\documentclass[useAMs,usenatbib,a4paper]{mn2e}
\usepackage{savesym}
\usepackage{graphicx}
\expandafter\let\csname equation*\endcsname\relax
  \expandafter\let\csname endequation*\endcsname\relax 
\usepackage{subfig}
\usepackage{amsmath}
\usepackage{verbatim}
\usepackage{array}
\usepackage{times}
\usepackage[total={17.8cm,24.0cm},centering]{geometry} 
\usepackage{rotating}

\newcommand{\be}{\begin{equation}}
\newcommand{\ee}{\end{equation}}
\newcommand{\eea}{\end{eqnarray}}
\newcommand{\bea}{\begin{eqnarray}}

\newcommand{\m}{\mathrm}

\title[New constraints on the structure and dynamics of black hole jets]{New constraints on the structure and dynamics of black hole jets}
\author[William J. Potter and Garret Cotter]{William J. Potter\thanks{E-mail:
will.potter@astro.ox.ac.uk (WJP)} and Garret Cotter
\\
Oxford Astrophysics. Denys Wilkinson Building, Keble Road, Oxford, OX1 3RH, United Kingdom}
\begin{document}

\date{}

\pagerange{\pageref{firstpage}--\pageref{lastpage}} \pubyear{2011}

\maketitle

\label{firstpage}

\begin{abstract}
Accreting black holes produce powerful relativistic plasma jets which emit radiation across all observable wavelengths but the details of the initial acceleration and confinement of the jet are uncertain. We apply an innovative new model that allows us to determine key properties of the acceleration zone via multi-frequency observations. The central component of the model is a relativistic steady-state fluid flow, and the emission from physically distinct regions can be seen to contribute to different energy bands in the overall spectrum.  By fitting with unprecedented accuracy to 42 simultaneous multiwavelength blazar spectra we are able to constrain the location of the brightest synchrotron emitting region, and show that there must be a linear relation between the jet power and the radius of the brightest region of the jet.  We also find a correlation between the length of the accelerating region and the maximum bulk Lorentz factor of the jet and find evidence for a bimodal distribution of accretion rates in the blazar population. This allows us to put constraints on the basic dynamical and structural properties of blazar jets and to understand the underlying physical differences which result in the blazar sequence.

\end{abstract}

\begin{keywords}
Galaxies: jets, galaxies: active, radiation mechanisms: non-thermal, radio continuum: galaxies, gamma-rays: galaxies.
\end{keywords}

\section{Introduction}

Accreting black holes can launch powerful relativistic plasma jets which are thought to be produced by the interaction of a rotating black hole with the magnetic field coupled to the accreting plasma. The rotation of the black hole induces rotation of inertial frames surrounding the black hole which twists the magnetic field close to the event horizon. This process transfers energy from the black hole's rotation into the energy contained in the magnetic field (\citealt{1977MNRAS.179..433B}). The enhanced pressure of the magnetic field is strong enough to drive and accelerate a plasma jet, however the details of the acceleration and confinement of the jet are uncertain (\citealt{1984RvMP...56..255B}, \citealt{2003ApJ...596.1080V}, \citealt{2006ApJ...641..103H} \citealt{2006MNRAS.367..375B}, \citealt{2006MNRAS.368.1561M}, \citealt{2007MNRAS.380...51K}, \citealt{2010ApJ...711...50T} and \citealt{2012MNRAS.423.3083M}). Radio observations of the jet in M87 show that it starts with a parabolic shape, transitioning to a conical jet at $10^{5}$ Schwarzschild radii, $r_{s}$ (\citealt{2012ApJ...745L..28A}). These observations are consistent with general relativistic magnetohydrodynamic (GRMHD) simulations in which the jet starts with a magnetically dominated parabolic accelerating base and transitions to a ballistic conical jet.  This transition occurs when the jet plasma comes into approximate equipartition between the energy contained in the magnetic field and particles. Here we use the first jet emission model which takes into account these state-of-the-art results from observation and theory in order to constrain the basic structure and dynamics of jets by fitting to a large sample of blazar spectra.

The most luminous class of supermassive black hole jets are blazars: jets which are pointing towards Earth so their emission is strongly Doppler-boosted. Evidence for their Doppler-boosting comes from the observed superluminal motion within their jets and short-timescale variability (\citealt{1995PASP..107..803U}, \citealt{1997ARA&A..35..445U}, \citealt{2001ApJS..134..181J} and \citealt{2010Natur.463..919A}). Non-thermal electrons are accelerated within the jets by processes which are currently not well understood (most likely a combination of shocks and magnetic reconnection e.g. \citealt{2012ApJ...745...63S, 2012ApJ...754L..33C, 2013arXiv1303.2569N}) and these electrons emit radiation across all observable wavelengths via synchrotron and inverse-Compton mechanisms. Blazars are particularly important to our understanding of jet properties because their emission is strongly Doppler-boosted and so the entire multiwavelength jet spectrum is clearly observed above the galactic emission. This makes it possible to constrain basic jet properties by using realistic jet emission models to fit to observed spectra. Previous models of blazar jet emission have focused either on understanding the high energy gamma-ray emission by modelling the jet as one or two fixed radius spherical blobs (e.g. \citealt{1998A&A...333..452K}, \citealt{2000ApJ...536..729L} \citealt{2002ApJ...581..127B}, \citealt{2007A&A...476.1151T}, \citealt{2009ApJ...692...32D}, \citealt{2011A&A...534A..86T}, \citealt{2013ApJ...768...54B}, \citealt{2014arXiv1412.3576T}, \citealt{2014A&A...571A..83P} and \citealt{2014ApJ...790...45P}), or using extended conical models to fit primarily to the flat radio spectrum (e.g. \citealt{1979ApJ...232...34B}, \citealt{1984ApJ...285..571M}, \citealt{1985A&A...146..204G}, \citealt{2000A&A...356..975K}, \citealt{2001A&A...372L..25M}, \citealt{2001MNRAS.325.1559S}, \citealt{2009ApJ...699.1919P}, \citealt{2010MNRAS.401..394J} and \citealt{2011arXiv1112.2560V}). 

In this paper we use one of the most sophisticated models for blazar jet emission currently available (\citealt{2012MNRAS.423..756P}, \citealt{2013MNRAS.429.1189P}, \citealt{2013MNRAS.431.1840P} and \citealt{2013MNRAS.436..304P}) to fit to the entire sample of Fermi blazars from \cite{2010ApJ...716...30A} with simultaneous multiwavelength observations and redshifts. Because this model takes into account the extended parabolic to conical geometry and acceleration of the jet fluid, we show that we are able to place meaningful constraints on the structure and dynamics of supermassive black hole jets by fitting to the blazar spectra across all observable wavelengths, with unprecedented accuracy. The structure of the paper is as follows: we start by introducing and explaining the assumptions of our jet model, we then show the results of fitting the model to the observed spectra and the constraints we obtain. Finally, we discuss these results in the context of current ideas on magnetic acceleration, the blazar sequence and AGN unification.

\section{Jet model}

Our model is motivated by the recent results from observations and simulations. The jet is modelled by a 1D time-independent relativistic fluid flow with a variable shape and bulk Lorentz factor \citep{2013MNRAS.429.1189P}. The total relativistic energy of the plasma is conserved via the equation for energy-momentum
\be
\nabla_{\mu}T^{\mu \nu}=0, \qquad T^{\mu \nu}=T^{\mu \nu}_{\m{Magnetic}}+T^{\mu \nu}_{\m{Particles}}+T^{\mu \nu}_{\m{Losses}},
\label{consE}
\ee
where $T^{\mu \nu}$ is the total energy-momentum tensor of the jet plasma which can be decomposed into magnetic and particle energy densities, and also a cumulative energy loss term which we include to conserve the total energy along the jet. The components of the energy-momentum tensor in the fluid rest frame, indicated by a prime, are given by
\bea
T'^{00}_{\m{Magnetic}}=\frac{B'^{2}}{2\mu_{0}}, \qquad T'^{00}_{\m{Particles}}=\int_{E_{\m{min}}}^{\infty} E_{e}n_{e}'(x,E_{e}) \m{d}E_{e}, \nonumber \\ T'^{00}_{\m{losses}}(x)=\int_{0}^{x} \frac{P'_{\m{synch}}(x)+P'_{\m{IC}}(x)+P'_{\m{ad}}(x)}{\pi R^{2}(x)}dx. \label{Tdecomp}
\eea
where we have set $c=1$, $x$ is the distance along the jet axis in the lab frame, $B'$ is the rest frame magnetic field strength, $E_{e}$ is the electron energy, $n_{e}$ the electron energy distribution and $P'_{\m{synch}}(x)+P'_{\m{IC}}(x)+P'_{\m{ad}}$ the sum of the synchrotron, inverse-Compton and adiabatic losses per unit length in the fluid rest frame (for a detailed calculation of these loss terms see sections 3--6 in \citealt{2012MNRAS.423..756P} and 5--6 in \citealt{2013MNRAS.429.1189P}). Making the assumption that the plasma is locally homogeneous and isotropic perpendicular to the jet axis, the energy-momentum tensor simplifies to that of a relativistic perfect fluid, $P=\rho/3$, in the fluid rest frame.
\be
T'^{\mu \nu}=\begin{pmatrix} \rho' & 0 &0 &0 \\ 0 & \frac{\rho'}{3} &0 &0\\ 0 &0 &\frac{\rho'}{3} &0\\ 0 &0 &0 &\frac{\rho'}{3}\end{pmatrix},
\ee
where $\rho'=T'^{00}$ is the total rest frame energy density. Using a Lorentz transformation to convert the energy momentum tensor from the rest frame to the lab frame (and assuming $\beta=v/c\approx 1$)
\bea
&&T^{\mu \nu}(x)=\Lambda^{\mu}_{\,\, \m{a}} T'^{\m{a} \m{b} } \Lambda _{\m{\,\,b}}^{ \nu}=... \nonumber \\  &&\hspace{1cm} \begin{pmatrix} \frac{4}{3}\gamma_{\m{bulk}}(x)^{2} \rho' & \frac{4}{3}\gamma_{\m{bulk}}(x)^{2} \rho' &0 &0\\\frac{4}{3}\gamma_{\m{bulk}}(x)^{2} \rho' &\frac{4}{3}\gamma_{\m{bulk}}(x)^{2} \rho' &0 &0\\0&0&\frac{\rho'}{3}&0\\0&0&0&\frac{\rho'}{3}\end{pmatrix} \label{LTEM},
\eea
where $\gamma_{\m{bulk}}$ is the bulk Lorentz factor of the jet plasma. The two factors of the Lorentz factor in the relation of the lab frame energy density to the rest frame energy density, $\rho \propto \rho' \gamma^{2}$, can be intuitively understood as one being due to an increase in energy by the Lorentz boost and one due to the increase in density because of the length contraction in the $x$-direction. Integrating (\ref{consE}) and using the 4-dimensional divergence theorem (see for example \citealt{1974ApJ...191..499P}) we find
\bea
&&\hspace{-1.0cm}\int \nabla_{\mu}T^{\mu \nu} d^{4}V=\int T^{\mu \nu} d^{3}S^{\mu}= \int_{x}^{x+dx}\int_{0}^{2\pi}\int_{0}^{R} T^{0\nu} RdRd\phi dx \nonumber\\
&&\hspace{-1.0cm}+\int_{t}^{t+dt}\int_{0}^{2\pi}\int_{0}^{R} T^{1\nu} RdRd\phi dt +\int_{t}^{t+dt}\int_{x}^{x+dx}\int_{0}^{2\pi} T^{2\nu} Rd\phi dxdt+ \nonumber\\ &&\hspace{-1.0cm}+\int_{t}^{t+dt}\int_{x}^{x+dx}\int_{0}^{R} T^{3\nu} dRdxdt=0.\nonumber\\ \label{divE}
\eea
where, $d^{4}V=\sqrt{|g|}dtdxdRd\phi$, is the invariant 4-volume, $g$, is the determinant of the metric tensor and $\sqrt{|g|}=R$, using cylindrical coordinates in Minkowski space (appropriate for the strongly radiating sections of the jet which occur many Schwarzschild radii from the black hole). The last three 3-surface integrals containing an integral over time are equal to zero due to the time-independence of the model and we have used the assumed radial and azimuthal symmetry of the jet plasma. The only non-zero components of (\ref{divE}) occur for $\nu=0$ or 1, in both cases this gives us our equation for conservation of energy-momentum
\be
\frac{\partial}{\partial x}\left(\frac{4}{3}\gamma_{\m{bulk}}(x)^{2}\pi R^{2}(x)\rho'(x)\right)=0.
\label{ce}
\ee
The particle flux $J^{\mu}$ is conserved along the jet by the equation
\be
\nabla_{\mu}J^{\mu}(E_{e},x)=0, \qquad J^{\mu}(E_{e})=n'_{e}(E_{e},x)U^{\mu}(x).
\label{consJ}
\ee
where $U^{\mu}(x)=\gamma(x)(1,\beta(x),0,0)$ is the jet fluid 4-velocity, $\beta(x)=v(x)/c$, $v(x)$ is the jet speed and $n'_{e}$ the electron number density in the rest frame. Integrating (\ref{consJ}) and using the divergence theorem as before we find
\be
\int \nabla_{\mu}J^{\mu} d^{4}V=\int J^{\mu} d^{3}S^{\mu}=\frac{\partial}{\partial x}(\pi R^{2}(x)n'_{\m{e}}(x)U^{0}(x))=0,
\label{cc}
\ee
where again the three 3-surface integrals which contain an integral over time vanish due to the time-independence of the model. These equations ensure that the total energy is conserved along the jet and naturally take into account the magnetic energy required to accelerate the jet and the internal energy gained when the jet decelerates via interactions with its environment. To illustrate this we shall explicitly show that in the case of an accelerating jet with minimal radiative and adiabatic losses the magnetic energy is converted into bulk kinetic energy of the plasma. Equations \ref{ce} and \ref{cc} have solutions
\be
\rho'(x)=\frac{3.\m{constant}}{4\pi R(x)^{2}\gamma_{\m{bulk}}(x)^{2}}, 
\ee
\be
n_{e}'(x)=\frac{\m{constant}}{\pi R(x)^{2}\gamma_{\m{bulk}}(x)}.
\ee
The electron energy density is given by (\ref{Tdecomp}) and we decompose $\rho'$ into the rest frame magnetic and particle energy densities. We shall temporarily neglect the energy loss term in this calculation for clarity.
\be
\rho'_{\m{Particles}}=\int_{E_{\m{min}}}^{\infty} E_{e}n_{e}'(x,E_{e}) \m{d}E_{e}\propto \frac{1}{R(x)^{2}\gamma_{\m{bulk}}(x)}
\ee
\be
\rho'\propto \frac{1}{R(x)^{2}\gamma^{2}_{\m{bulk}}(x)}, \qquad \rho'\approx \rho'_{\m{Particles}}+\rho'_{\m{Magnetic}}.
\ee
Since the total rest frame energy density is decreasing as $\gamma_{\m{bulk}}^{-2}$, whereas, the rest frame particle energy density is only decreasing as $\gamma_{\m{bulk}}^{-1}$, when the jet accelerates and the bulk Lorentz factor increases the jet plasma will become more and more particle dominated. This is because magnetic energy is being converted into particle energy as the bulk jet plasma is accelerated by magnetic forces and eventually, when the magnetic energy becomes smaller than the bulk particle energy, this acceleration ceases to be efficient. This demonstrates the benefit of using the relativistic fluid equations \ref{ce} and \ref{cc} as the basis of the jet model.

\begin{figure}
          \centering
          \includegraphics[width=8.5cm]{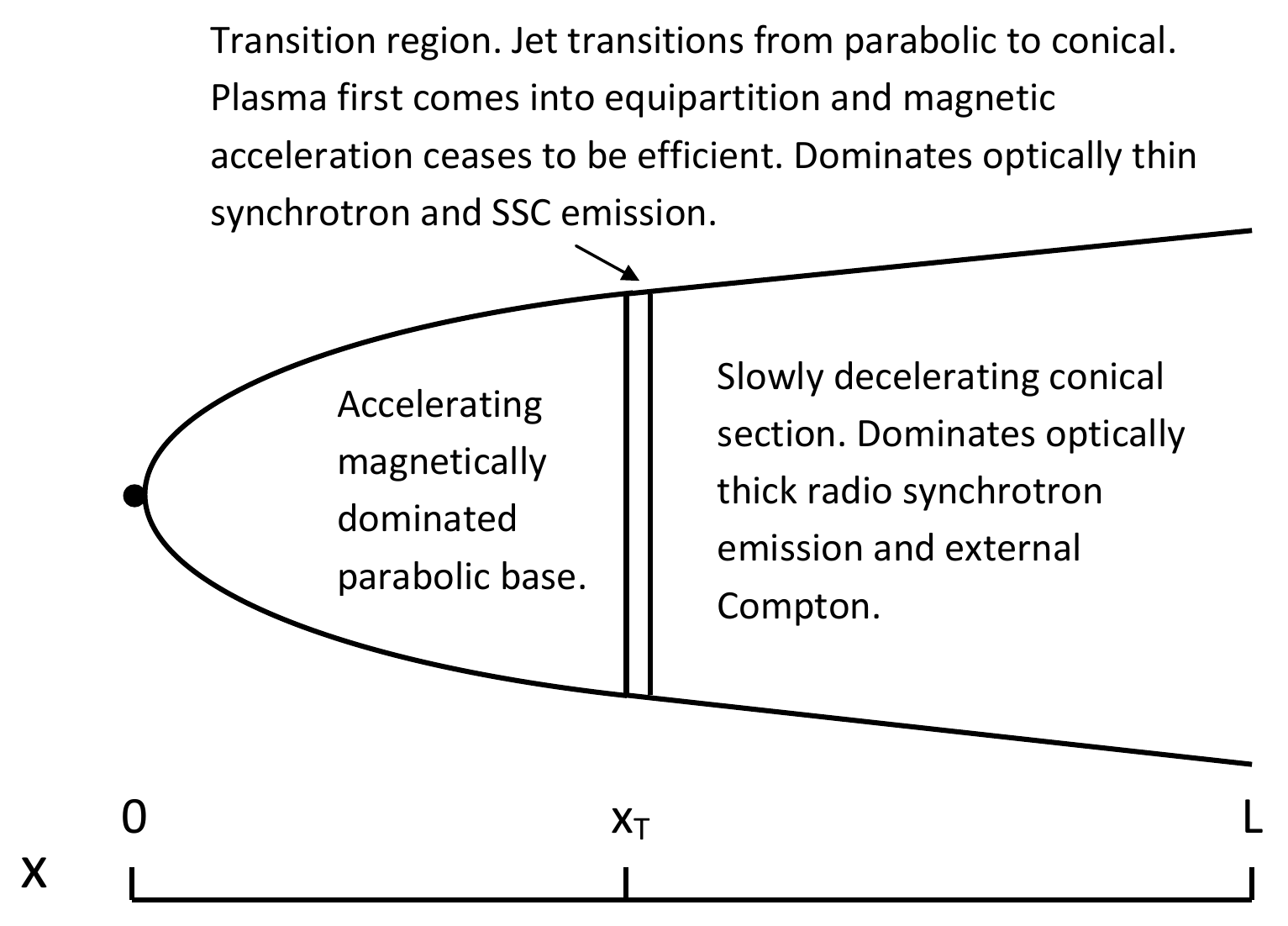}
          \caption{Jet schematic.}
          \label{schematic}
\end{figure}

\subsection{Model assumptions}

The best current constraints on the shape of a supermassive black hole jet come from radio observations of M87 which show a parabolic base extending out to a distance, $x_{T}=10^{5}r_{s}$, where the jet transitions to conical \citep{2012ApJ...745L..28A}. We use these observations as the basis for our jet model choosing our jet shape to be
\be
R(x)=C_{\m{par}}(r_{s}+x)^{A_{\m{par}}} \qquad \m{for} \,\,\,\, x\leq x_{T},
\label{radius}
\ee
\be
R(x)=C_{\m{par}}(r_{s}+x_{T})^{A_{\m{par}}}+(x-x_{T})\tan(\theta_{\m{opening}}) \,\,\, \m{for} \,\,\,\, x>x_{T},
\ee
where $r_{s}$ is the Schwarzschild radius and we estimate $A_{\m{par}}=0.58$ and $C_{\m{par}}=1.49r_{s}^{0.42}$ from the observations of M87 \citep{2012ApJ...745L..28A}, and $\theta_{\m{con}}$ is the lab frame conical jet opening angle (we assume that due to relativistic beaming $\theta_{\m{con}} \approx 1/\gamma_{\m{bulk}}$). Hereafter we shall refer to the region where the jet transitions from parabolic, accelerating and magnetically dominated to a slowly decelerating conical jet in equipartition, as the transition region of the jet and we shall use a subscript $T$ to denote quantities measured at this transition region. We choose to linearly scale the jet geometry from the observations of M87 by using an effective black hole mass, $M$, defined such that the transition from parabolic to conical occurs at $10^{5}r_{s}(M)=x_{T}$ for all fits, as observed in M87. This fixes the aspect ratio of the transition region radius to transition region distance ($R_{T}/x_{T}\sim 1/80$) to be fixed to that observed in M87, whilst allowing the absolute lengthscales to depend linearly on $M$. This seems a much more reasonable assumption than allowing the jet aspect ratio at the transition region to be a function of the absolute distance to the transition region (in this case the aspect ratio would vary as $R_{T}/x_{T}\propto x_{T}^{1-A_{\m{par}}}$ using equation \ref{radius}). The effective black hole mass, $M$, found by fitting the model to a blazar spectrum, represents the black hole mass that would be required for the transition region for this blazar to occur at $10^{5}r_{s}$ as in M87. In section 3.3 we discuss observational evidence which indicates that differences in the jet geometry are unlikely to be driven primarily by changes in the black hole masses and so the reader should consider the effective black hole mass, $M$, simply as a convenient parameter to vary the distance to, and radius of the transition region. 

In agreement with the observations of M87, GRMHD simulations find that the jet starts with a parabolic magnetically dominated accelerating base (e.g. \citealt{2006MNRAS.368.1561M}). In these simulations the jet is accelerated by a magnetic pressure gradient which converts magnetic energy into bulk kinetic energy and this process continues until the jet plasma approaches equipartition between the magnetic field and particle energies (where the magnetic energy is no longer large enough to significantly accelerate the plasma). We incorporate all of this information into our model, a schematic of which is shown in Fig. \ref{schematic}. In our model the jet starts with a magnetically dominated accelerating parabolic base and we use the relation between the jet shape and acceleration calculated by \citealt{2006MNRAS.367..375B} and consistent with simulations e.g. \citealt{2006MNRAS.368.1561M}
\be
\gamma_{\m{bulk}}(x) \propto x^{1/2}.
\ee
\be
\gamma{_\m{bulk}}(x)=\gamma_{0}+\left(\frac{\gamma_{\m{max}}-\gamma_{0}}{x_{\m{con}}^{1/2}}\right)x^{1/2}  \,\,\,\,\, \m{for} \,\,\,x \leq x_{\m{con}}.
\ee
Once efficient acceleration of the jet has ceased and it transitions to conical we assume the bulk Lorentz factor slowly decreases due to interaction of the jet with its environment i.e. entrainment. We choose a logarithmic deceleration so the conversion of bulk kinetic energy into in-situ non-thermal particle acceleration is continuous along the jet, as suggested by observations (e.g. \citealt{2005A&A...431..477J}). 
\be
\gamma_{\m{bulk}}(x)=\gamma_{\m{max}}-\left(\frac{\gamma_{\m{max}}-\gamma_{\m{min}}}{\log \left(\frac{L}{x_{\m{con}}}\right)}\right) \log \left(\frac{x}{x_{\m{con}}}\right), \,\,\,\,\,\, x \geq x_{\m{con}}. \label{dec}
\ee
In our previous work \citep{2013MNRAS.429.1189P} we found that accelerating non-thermal electrons in the parabolic base of the jet had little observable effect since the bulk Lorentz factor in the base is not as large as in the conical jet and so the emission from the parabolic regions is dominated by the  emission from the faster conical jet, which is more strongly Doppler-boosted. We also found that the average equipartition fraction in the base must be very low (i.e. almost completely magnetically dominated). This is because the timescale for electrons to lose energy by synchrotron emission was very short in the high magnetic fields present in the accelerating region and so the energy required to maintain a significant observable population of non-thermal electrons was unfeasibly large. For these reasons we assume that only a small amount of energy is contained in electrons at the base of the jet. We assume that the jet comes rapidly into equipartition at the transition region between parabolic and conical regions where the jet stops accelerating efficiently. This rapid particle acceleration at the transition between parabolic and conical sections could be caused by a combination of a recollimation shock if the jet becomes underpressured relative to its environment (e.g. \citealt{2002MNRAS.336..328L}, \citealt{2009ApJ...699.1274B}) and enhanced magnetic reconnection due to the jet compression but a detailed consideration of these mechanisms is beyond the scope of this paper. In our previous work we found that BL Lac type blazars were well modelled by jets which maintain equipartition between non-thermal electrons and the magnetic field in the conical section, whereas flat spectrum radio quasars (FSRQs) favoured an equipartition fraction which began in equipartition at the start of the conical jet and became progressively more particle dominated at large distances. This is consistent with the theoretical interpretation of jet morphologies by \cite{1999ApJ...522..753M}, who suggested that FRII jets should be particle dominated at large distances while FRI jets remain highly magnetised.   

\begin{figure}
	\centering
		\includegraphics[width=8cm, clip=true, trim=1.5cm 1cm 2.0cm 1.5cm]{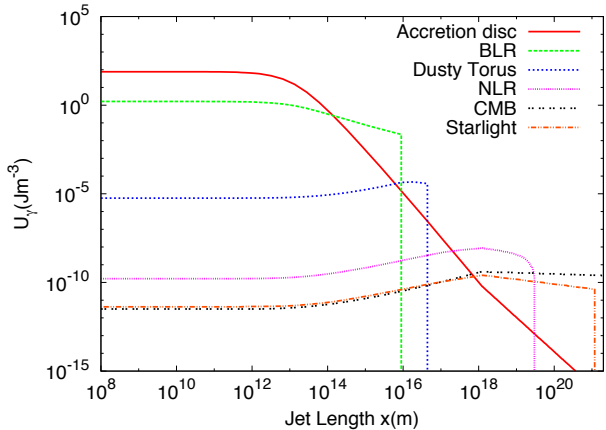} 
                 \caption{The energy density of the different external photon sources as a function of jet length, calculated for the fit to the FSRQ J1512 in the plasma rest frame. The rest frame energy densities account for the Doppler boosting and beaming of the external photon sources into the plasma rest frame. The acceleration of the bulk Lorentz factor (up to $x\sim10^{18}$m) of the jet leads to the increase in the energy density of the CMB as measured in the plasma rest frame and the slow deceleration of the conical jet leads to the decrease in the CMB energy density beyond this distance. The synchrotron break constrains the transition region of FSRQs to occur outside the expected radii of the BLR and dusty torus, at large distances along the jet the CMB is the dominant external photon field. We find that inverse-Compton scattering of CMB photons at large distances is able to fit very well to the high energy X-ray and gamma-ray observations for FSRQs.  }
                 \label{Uphot}
\end{figure}

We have chosen to use a simple electron-positron jet composition in this work to increase the constraining power of our model. We know that high-energy non-thermal electrons exist in jets since we see polarised synchrotron emission \citep{2005AJ....130.1418J} and so it is sensible to first try to fit observations with an electron-positron jet before adding additional components which carry extra free parameters. Including hadronic emission processes would add additional free parameters to the model making it more difficult to constrain the model. Hadronic emission models currently seem to require highly super-Eddington jet powers to match to data. For example, in \citealt{2013ApJ...768...54B}, the proton kinetic luminosities range from $2.1\times10^{39}-4.4\times10^{43}\m{W}$, even for a very large $10^{10}M_{\odot}$ black hole, $4.4\times10^{43}$W is $350$ times the Eddington Luminosity, which seems implausibly large. Adding additional components to our model such as a cold proton component would simply increase the jet power. We can view our model as providing an estimate of the power contained in the magnetic and non-thermal electrons in the jet and thus an estimate of the lower limit of the jet power. 

We calculate the inverse-Compton emission by Lorentz transforming the external radiation fields into the plasma rest frame and using the full Klein-Nishina cross-section. We comprehensively account for the sources of external radiation as functions of distance along the jet \citep{2013MNRAS.429.1189P} these include photons from: the accretion disc, the broad line region (BLR), the dusty torus, the narrow line region (NLR), starlight and the cosmic microwave background (CMB). FSRQs are defined observationally as blazars in which broadened emission lines from the accretion disc have been observed and so for these sources we fit a standard composite blackbody thin accretion disc model to the spectra (\citealt{1973A&A....24..337S}). The distribution of these different external photon fields is shown in Figure \ref{Uphot}. It is worth emphasising the importance of including the CMB photons which we find contribute significantly to the gamma-ray emission of FSRQs at large distances. We also calculate the amount of $\gamma$-ray absorption due to photon-photon collisions which produce electron positron pairs (see \citealt{2013MNRAS.436..304P}).

\subsubsection{Acceleration of non-thermal electrons}

The mechanism which accelerates non-thermal electrons in jets is not well understood currently, the most popular physical processes are acceleration by shocks and magnetic reconnection. These mechanisms tend to produce non-thermal electron distributions which have a power law component at low energies and a cutoff at high energies \citep{1987ApJ...322..256K, 2007ApJ...670..702Z, 2012ApJ...745...63S} and we have found that this is compatible with observed emission spectra of blazars \cite{2013MNRAS.431.1840P}. Our model takes into account the radiative and adiabatic energy losses to the electron population travelling along the jet and also the acceleration of additional electrons which is required to keep the plasma close to equipartition along the jet (as found from observations e.g. \citealt{2005ApJ...626..733C} and \citealt{2006ApJ...642L.115H}). The energy for this in-situ acceleration comes from the internal energy gained from the deceleration of the jet, where bulk energy of the plasma is converted into enhancing the magnetic field strength and accelerating non-thermal electrons via shocks, and also from magnetic reconnection whereby internal magnetic energy is converted into accelerating non-thermal electrons. Since the precise electron distributions originating from these processes are far too computationally expensive to calculate within our model (they require dedicated simulations in their own right) we assume that the initial electron distribution and the distribution of any additional non-thermal electrons accelerated in-situ is given by
\be
N_{\m{injected}}(x,E_{e})=AE_{e}^{-\alpha} e^{-E_{e}/E_{\m{max}}}, \label{Ch2loss2}
\ee
where $\alpha$ is the electron energy distribution spectral index. Of course, due to the energy losses from the electron population (which are a complicated function of the electron energy and jet parameters) the total electron distribution will quickly deviate from the simple form in (\ref{Ch2loss2}) as the plasma flows along the jet and radiates. 

This model incorporates current observations and results from simulations and theory. We have made the model flexible and rigorous but at the same time chosen to make simplifying assumptions in order to limit the number of free parameters. The model has only 12 free parameters which is in fact far fewer than a model with two independent spherical emission zones, however, we argue that our model is considerably more realistic and than these models.

\subsubsection{Adiabatic losses in a steady jet}

The external adiabatic energy losses which transfer internal energy from a thin cylindrical volume element of jet plasma to the confining medium occur at a rate
\be
\frac{dE_{\m{int}}}{dt}=-\int_{x=0}^{x=L}\int_{R({x,t_{0})}}^{R(x,t_{0}+ \delta t)} \frac{p_{\m{ext}}(x)}{\delta t}2\pi R(x,t)dx dR(x,t),
\ee
where $p_{\m{ext}}$ is the pressure of the jet's external environment, $E_{\m{int}}=E_{\m{P}}+E_{B}$ is the internal energy contained in the cylindrical volume element, as measured in the lab frame and $E_{\m{P}}$ and $E_{B}$ are the components of energy contained in particles and magnetic fields. We decompose the change in the volume of the cylindrical element, $dV(x,t)$ defined via
\be
dV(x,t)=\int_{R(t_{0})}^{R(t)}2\pi R(x,t)dxdR(x,t),
\ee
 into a long-term time average $d\bar{V}(x)$ and a short-term fluctuating part $d\tilde{V}(x,t)$. If the time average converges for suitably long timescales then this is equivalent to saying that the time average of the fluctuations will tend to zero (i.e. at a fixed distance along the jet its radius will not systematically expand or shrink with time). Our model assumes a jet with a steady long-term average volume and so over suitably long timescales the work done by the jet expanding the external medium will equal the work done by the external medium compressing the jet. This means that such a steady-state jet will not do any net work on the external medium.

Whilst the lab frame volume occupied by the jet is likely to be well-described by a time-independent jet model, clearly individual fluid elements will expand in radius as they propagate along the jet. In this case the plasma will do work on itself as it expands radially by accelerating the bulk radial velocity of the jet (analogous to the adiabatic radial expansion and cooling in a supernova explosion). The rate of change of relativistic particle energy due to internal adiabatic losses is given by (e.g. \citealt{2006MNRAS.367.1083K})
\be
\frac{d\ln E_{\m{P}}\m{(V)}}{d\ln V}=-\frac{1}{3},
\ee
\be
E_{\m{P}}(2R)=0.63E_{\m{P}}(R).
\label{adiabatic}
\ee
where $V$ is the volume of a thin cylindrical fluid element and $R$ is the jet radius, both measured in the plasma rest frame. For a jet which undergoes internal adiabatic losses and has a fixed equipartition ratio $A_{\m{equi}}$. The rate of loss of internal energy is modified by the store of available magnetic energy to become
\be
d\ln E_{\m{int}}=-\frac{1}{3(1+A_{\m{equi}})}d\ln V, \qquad A_{\m{equi}}=\frac{E_{B}}{E_{\m{P}}},
\ee
\be
E_{\m{int}}\propto R^{\frac{-2}{3(1+A_{\m{equi}})}},
\ee
Where again $E_{\m{int}}=E_{B}+E_{\m{P}}$.  In our model our primary concern is to conserve the total energy contained in the jet plasma. It is not currently known how the total energy of the plasma is distributed between magnetic, particle and bulk kinetic energies along the jet. Observations show continuous injection of energy into the acceleration of non-thermal electrons along the jet which are observed via optical synchrotron emission (\citealt{2001A&A...373..447J} and \citealt{2005A&A...431..477J}). The main candidates for this in-situ acceleration are internal shocks, which convert bulk kinetic energy into non-thermal particle energy, and magnetic reconnection which converts magnetic energy into particle energy via resistive dissipation. Importantly, synchrotron observations favour a continuous acceleration mechanism since they do not observe a series of discrete, bright, strong shock fronts (\citealt{2001A&A...373..447J}). The only processes which lead to a net decrease in the total energy contained in the jet plasma are the emission of radiation and external adiabatic losses (external adiabatic losses are zero in a time-independent jet as we have argued above). 

Since the energy in the non-thermal electrons requires constant replenishment (\citealt{2001A&A...373..447J} and \citealt{2005A&A...431..477J}) this leads to three possibilities for the source of this energy: resistive dissipation of the magnetic field, conversion of the bulk kinetic energy of both the non-thermal, and possibly an additional component of cold particles or the conversion of the radial kinetic energy into internal energy. The problem with powering the in-situ particle acceleration with purely magnetic dissipation is that the jet would need to be magnetically dominated over the majority of its length and so would not remain close to equipartition as suggested by observations (e.g. \citealt{2005ApJ...626..733C}). A dominant cold component would allow equipartition to be maintained but at the expense of increasing the total jet power to account for this extra source of energy (the main effect of this in our current work would be to increase our estimation of the total jet power). The only limitation on this process is that it should not require an unreasonably large jet power. The final possibility is that the work done by the non-thermal particles expanding the jet (accelerating the bulk radial velocity) is efficiently converted back into non-thermal particle energy. This could occur, for example, if the jet undergoes recollimation shocks in which an initially overpressured jet expands into the external medium until it is underpressured, the pressure of the environment then accelerates material radially towards the jet axis where it eventually collides supersonically. This forms a recollimation shock which converts some of the radial kinetic energy back into internal energy via non-thermal shock acceleration and magnetic field enhancement. 

In this paper we have chosen this final mechanism in which adiabatic losses are replenished by the conversion of radial kinetic energy into internal energy. This is because this scenario is the simplest to implement, provides a lower bound estimate of the jet power and does not require the introduction of an additional unknown parameter (the ratio of the energy in cold particles compared to non-thermal electrons). We shall investigate the effect of the introduction of a cold particle component of the fluid flow in a future paper. 

It is worth commenting that the detailed question of whether adiabatic losses are really appropriate for a diffuse magnetised and non-thermal jet plasma is far from certain. The prolonged existence of non-thermal electrons demonstrates that the plasma is collisionless (particle-particle collisions are infrequent) and so the usual assumptions which go into adiabatic expansion are not fulfilled (the system does not move slowly and reversibly between adjacent states of thermal equilibrium). Since the number density of non-thermal electrons is low, and their energies very large, scattering of the particles from the field lines is stochastic and is not likely to be well described by a fluid approximation. To determine whether adiabatic losses apply in their usual form in these extreme conditions might be best tackled through particle in cell simulations. 

\begin{figure*}
	\centering
		\subfloat[]{ \includegraphics[width=15cm, clip=true, trim=1cm 1cm 0.5cm 2cm]{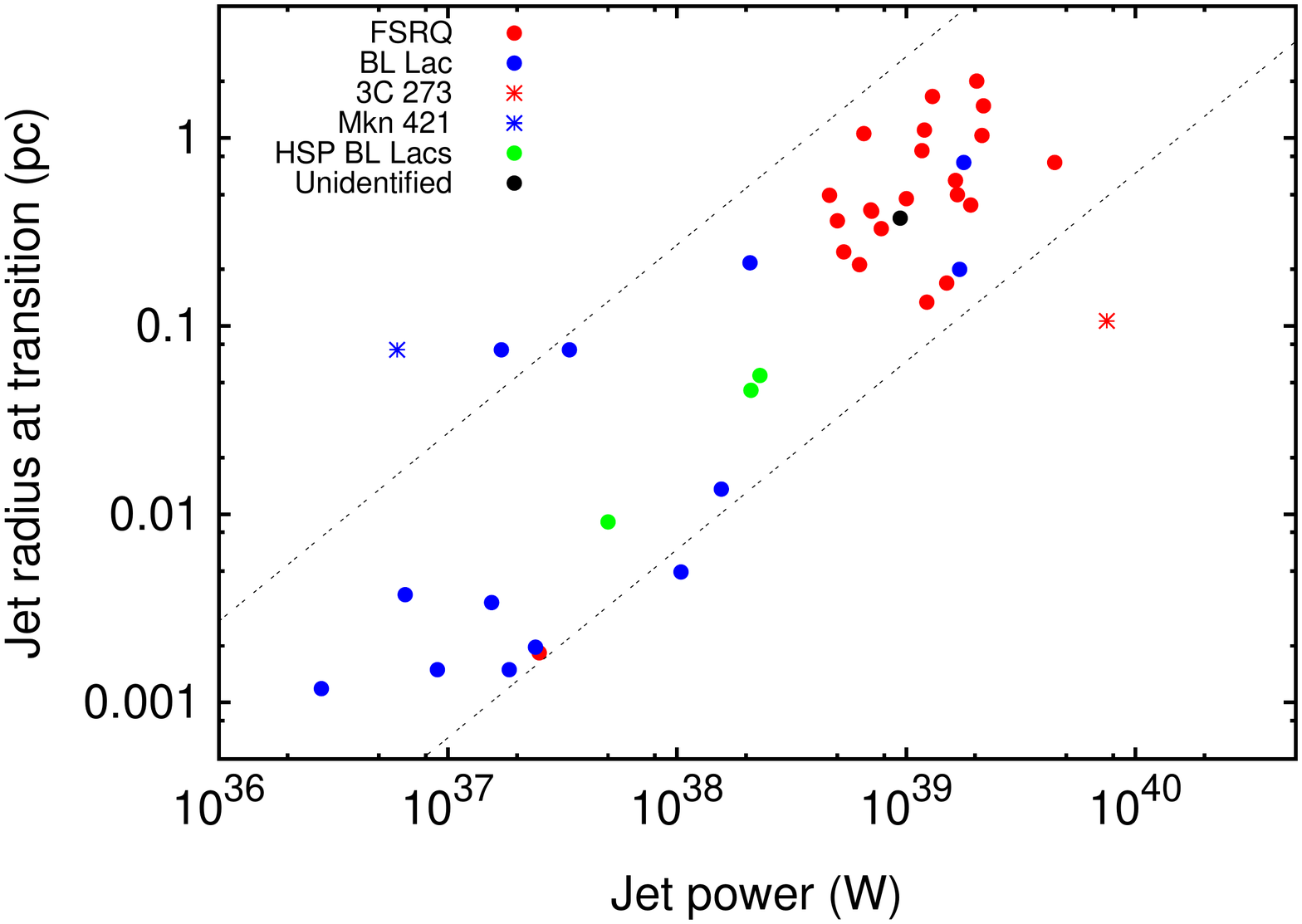} }
\\
		\subfloat[]{ \includegraphics[width=15cm, clip=true, trim=1cm 1cm 0.5cm 2cm]{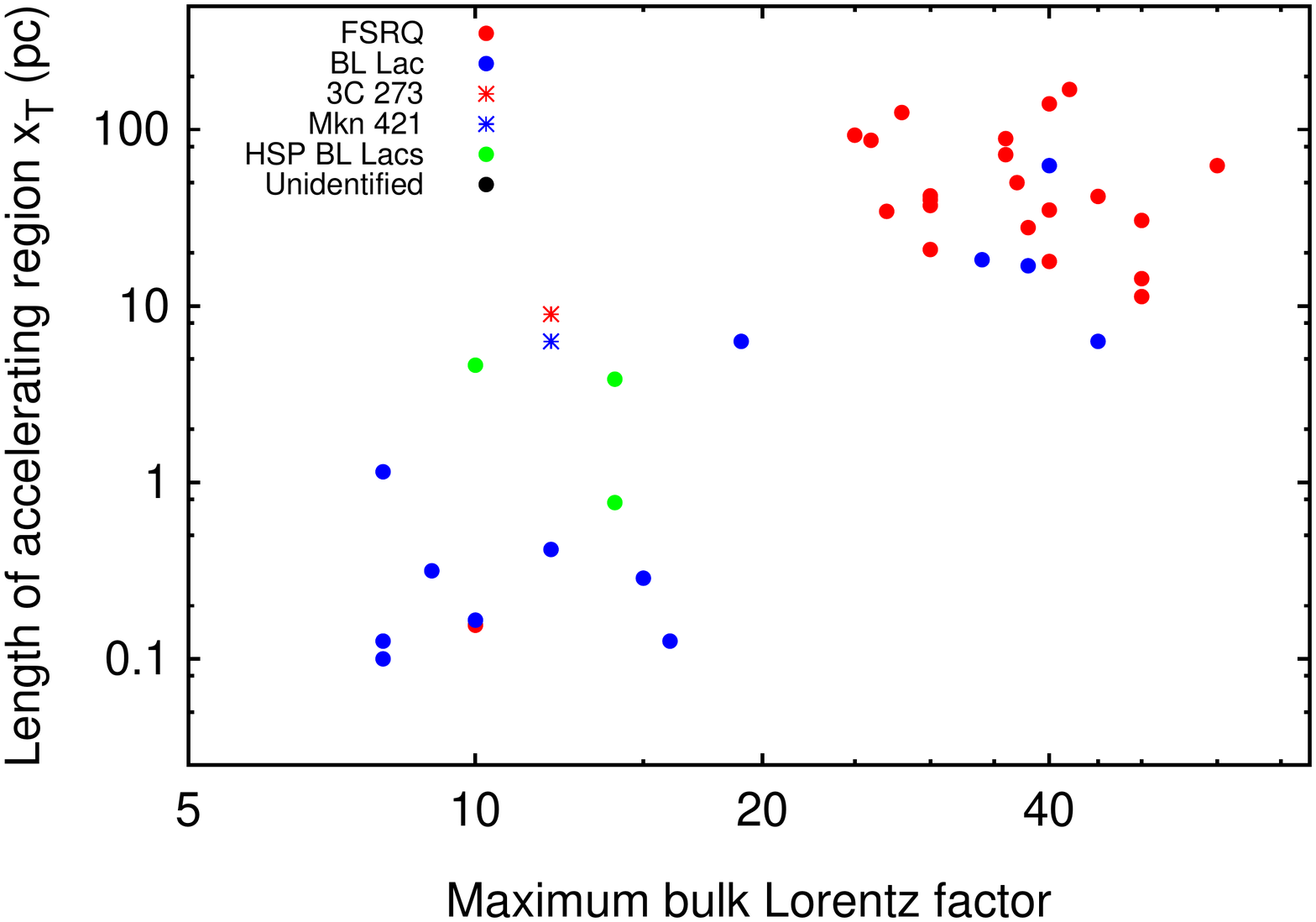} }	
		
	\caption{The results of fitting our model to all 38 blazars with simultaneous multiwavelength spectra and redshifts from Abdo et al. 2010 and the 4 quasi-simultaneous HSP BL Lac SEDs from Padovani et al. 2012. Figure a shows the radius of the jet transition region plotted against the jet power for all fits (the transition region is where the jet transitions from parabolic to conical, stops accelerating and first approaches equipartition due to strong non-thermal particle acceleration). We see a clear correlation between the jet power and radius of the transition region as suggested in our previous work. BL Lacs have systematically smaller radius transition regions than FSRQs, with an approximately linear correlation between the two as illustrated by the dashed lines. We have highlighted the two outliers to the relation, Markarian 421 and 3C273. We argue that these two blazars are atypical since Mkn421 is one of the closest known blazars and 3C273 is the one of the most powerful (and seems to be misaligned). Figure b shows the maximum bulk Lorentz factor plotted against the distance of the transition region from the black hole (the ratio of transition region radius to length is fixed at $\sim 1:80$). We see that the bulk Lorentz factor is systematically larger for FSRQs than BL Lacs and increases with the distance over which the jet accelerates as expected for steady magnetic acceleration. The relationship between the length of the accelerating region and the maximum bulk Lorentz factor is approximately $\gamma_{max}\propto x_{T}^{1/4}$, or equivalently in terms of the transition region radius $\gamma_{max}\propto R_{T}^{1/4}$. }
	\label{corr}
\end{figure*}

\section{Results}

Our model is able to fit to all 38 Fermi blazars with simultaneous multiwavelength spectra and redshifts from \cite{2010ApJ...716...30A} and the 4 high power, high synchrotron peak frequency (HSP) BL Lacs from \cite{2012MNRAS.422L..48P} with unprecedented precision. The spectral fits are shown in Figure \ref{spectra} and their parameters in Figure \ref{table}. One of the main achievements of this work is to show that an extended fluid model can fit to both the radio and high energy emission simultaneously. Fitting to the radio data simultaneously with all other wavelengths allows tight constraints to be placed upon the radius of the brightest synchrotron emitting region (in our model this is the transition region where the jet first becomes conical, stops accelerating and comes into equipartition, see Fig. \ref{schematic}). We have summarised our main results in Figure \ref{results_schem}.

\subsection{Constraining the radius of the transition region}

The radius of the brightest emitting synchrotron region is tightly constrained by modelling the break frequency where the synchrotron emission goes from being self-absorbed and close to flat in flux at low frequencies, to optically thin at high frequencies (the synchrotron break has also been used to constrain the properties of a small number of X-ray binary systems with observed jet spectra e.g. \citealt{2013MNRAS.429..815R}). This break at low frequencies $\sim10^{-4}$eV ($\sim20$GHz) is clearly seen in many of the spectra e.g. J0423, J0531, J1057, J1159, J1221, J1256 and J1512 in Fig. \ref{spectra} (for a simplified derivation of the break frequency and an explicit plot illustrating the sensitive dependence on the jet radius see equation 8 and Fig. 2 in \citealt{2013MNRAS.431.1840P}) . The determined radius of the transition region is shown for all 42 blazars in Figure \ref{corr}a. There is a clear linear correlation between the radius of the transition region and the jet power. This is significant because it shows that there is a simple underlying relation between the structural properties of jets and their power. In this work we have assumed that the transition region is the location where the jet stops accelerating and first transitions from a parabolic to a conical jet based on the results of GRMHD simulations. This means that the initial power of the jet determines the radius at which it will stop accelerating and start to brightly emit synchrotron radiation. This result can be used to inform and constrain the acceleration mechanism of the jet. Currently, simulations tend to find that jets stop accelerating at roughly $~10^{3}r_{s}$ (\citealt{2006MNRAS.368.1561M} and \citealt{2007MNRAS.380...51K}), however, the transition from parabolic to conical in M87 occurs at $10^{5}r_{s}$. These observations combined with our findings suggest that real jets can accelerate over much larger distances than currently found in simulations. This difference could be a result of the interaction of the jet with its host environment and its initial mass-loading combined with the effects of radiative energy losses and drag, which are difficult to include properly in simulations.

\subsection{A relation between the length of the accelerating region and jet speed}

Figure \ref{corr}b shows the length of the accelerating region of the jet (our jet shape is linearly scaled from M87 and so the radius of the transition region is $\approx1/80$ of the jet length at that point) plotted against the maximum bulk Lorentz factor at this radius. There is a clear correlation between the length of the accelerating region and the maximum jet speed, with evidence for bimodality. This suggests that a steady acceleration process is operating in the jet. The longer the distance over which the jet accelerates, the higher its bulk Lorentz factor becomes. This is consistent with steady magnetic acceleration in the parabolic section of the jet, however, it also shows that the final Lorentz factor of the jet depends on its initial jet power. We also see a clear association of FSRQs with fast jets and BL Lacs with slower jets. In simulations the maximum bulk Lorentz factor depends primarily on the initial mass loading or equipartition ratio of the magnetic field to particle energies. This is because the jets only accelerate efficiently up to equipartition, where approximately half of the initial magnetic energy is converted into bulk kinetic energy in the particles i.e.
\be
\epsilon_{ML}=\frac{E_{p}}{E_{B}}, \qquad \gamma_{\m{max}}\approx \frac{1+\epsilon_{ML}}{2\epsilon_{ML}}
\ee
where $\epsilon_{ML}$ is the mass loading fraction and $\gamma_{\m{max}}$ is the terminal maximum bulk Lorentz factor. From Fig. \ref{corr}b we find a relationship $\gamma_{\m{max}}\propto R_{T}^{1/4}$ between the radius of the transition region and the maximum bulk Lorentz factor of the jet. This is a weaker dependence of the maximum bulk Lorentz factor with jet radius than typically found in simulations $\gamma_{\m{max}}\propto R^{1/2}$ \cite{2006MNRAS.368.1561M, 2007MNRAS.380...51K}, although often the chosen prescription for the confining wind or rigid jet boundary will determine this relationship in simulations. This difference could be due to the entrainment of material surrounding the jet or radiative drag on the jet plasma as electrons lose bulk momentum by inverse-Compton scattering external radiation fields. 

This is the first attempt to constrain the relation between the length of the accelerating region of the jet and the maximum bulk Lorentz factor using a large sample of blazars, so these results are very useful to compare the results of simulations and observations. One potential problem in interpreting our results is our assumption that the brightest synchrotron emitting region occurs when the jet first comes into equipartition and stops accelerating. This scenario seems to us the most likely, especially given the close proximity of the bright feature HST1 in M87 with the transition from parabolic to conical \cite{2012ApJ...745L..28A}. However, it is possible that the region where substantial particle acceleration occurs is not close to the transition between parabolic and conical. It is worth emphasising that the tight constraints our model places on the radius of the brightest synchrotron emitting region are independent of the assumptions we have made about the accelerating region of the jet. Irrespective of the validity of our individual assumptions our results show the importance of using extended fluid jet emission models to understand the relationships between basic structural and dynamical quantities, since these are crucial to informing, constraining and testing our theoretical models and the results of GRMHD simulations. 

\begin{figure*}
	\centering
		\subfloat[]{ \includegraphics[height=5.5cm, clip=true, trim=2cm 1cm 1cm 2cm]{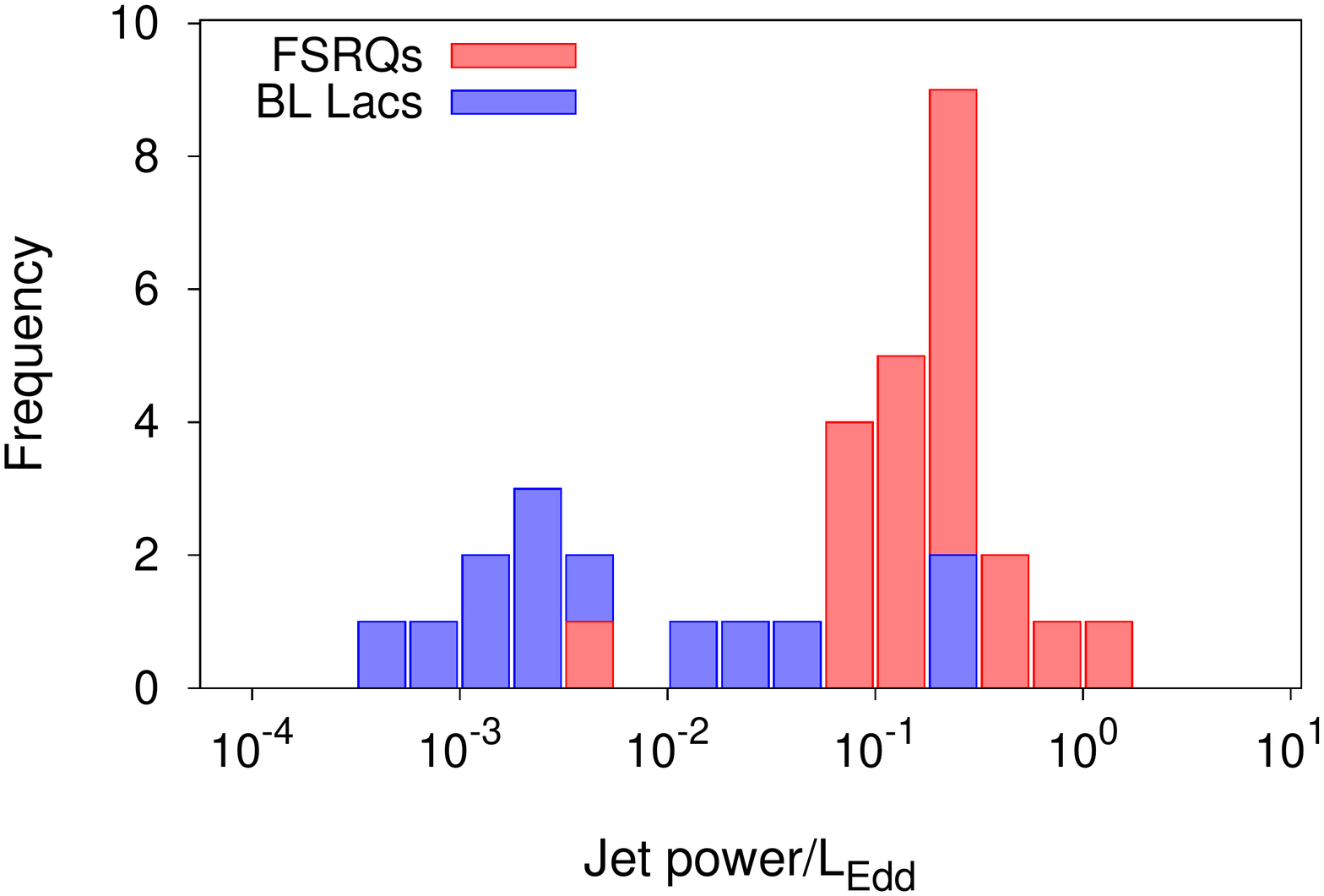} }
		\subfloat[]{ \includegraphics[height=5.5cm, clip=true,  trim=2cm 2cm 1cm 2cm]{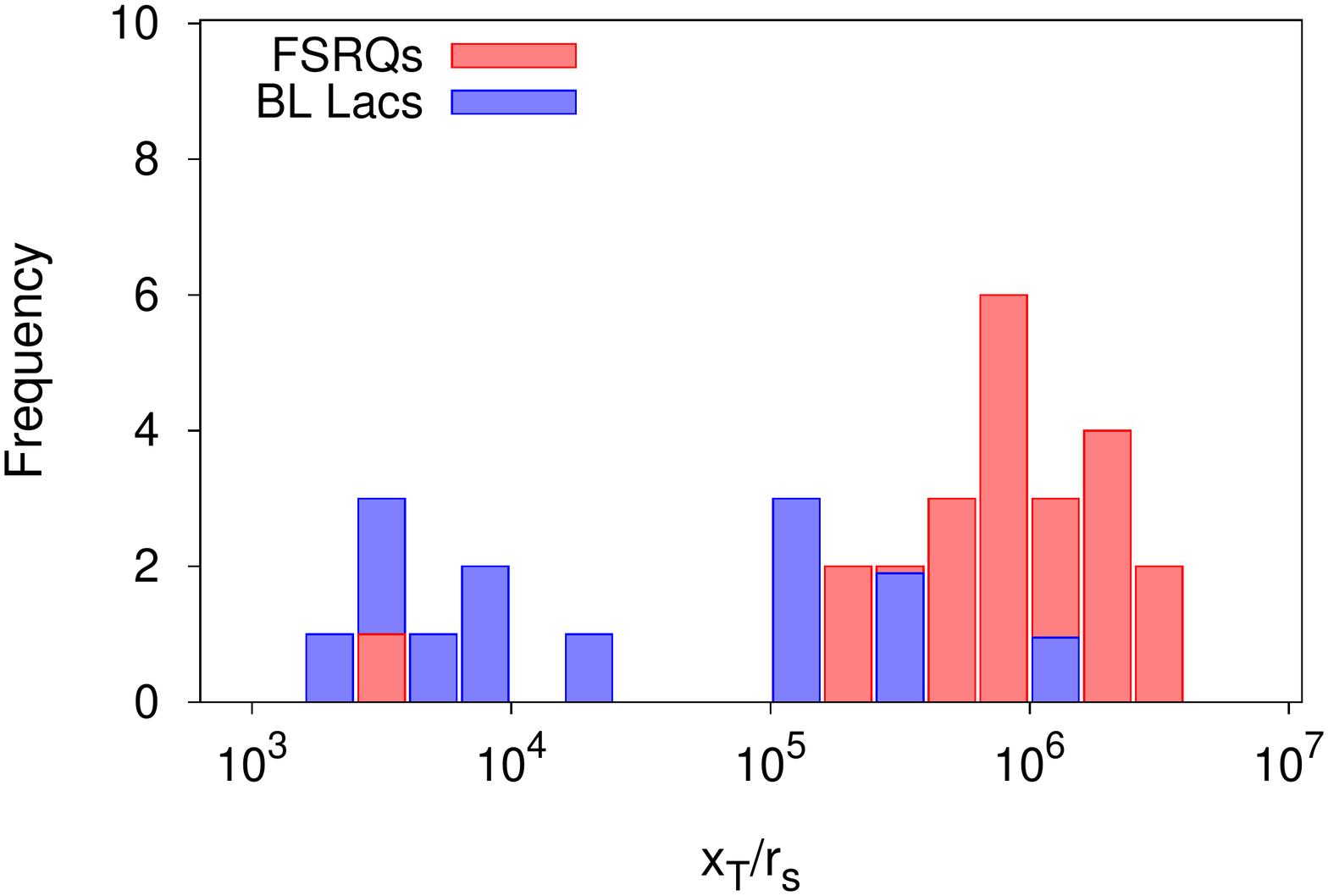} }	
	\caption{The distribution of fractional Eddington luminosity and the distance to the jet transition region under the assumption of a constant black hole mass of $5\times 10^{8}M_{\odot}$ for all 37 blazars with simultaneous multiwavelength {\it Fermi} observations and classifications. Figure a shows a clear dichotomy in the fractional Eddington luminosity between BL Lacs and FSRQs, suggestive of two accretion modes. Figure b shows that the length of the accelerating parabolic region of the jet, $x_{T}$, extends out to larger distances in FSRQs than BL Lacs. This could be responsible for the different maximum bulk Lorentz factors shown in Figure \ref{corr}b. }
	\label{distributions}
\end{figure*}

\subsection{Evidence for a bimodal accretion rate and AGN unification}

In order to understand and interpret our results in an astrophysical context it is necessary to make assumptions about the black hole mass distribution in the blazar population. Although black hole masses are notoriously difficult to determine accurately, evidence suggests that the black hole masses of BL Lacs and FSRQs (generally associated with low and high power jets) are similar, with the typical variation in the black hole masses being roughly an order of magnitude \citep{2013ApJ...764..135S}. These results suggest that there is no large systematic variation in the black hole masses of blazars and therefore that the black hole mass is unlikely to be driving the correlations shown in Fig. \ref{corr}, which exist over many orders of magnitude. In order to interpret our results in terms of the Eddington luminosity we shall therefore assume a fiducial blazar black hole mass of $5\times10^{8}M_{\odot}$ \citep{2012ApJ...748...49S, 2013ApJ...764..135S}. Under this assumption we find evidence for a bimodal distribution of accretion rates in the blazar population (Fig. \ref{distributions}). We find that FSRQs are high power jets close to the Eddington luminosity whilst BL Lacs are much lower power jets, consistent with previous work \cite{2011MNRAS.414.2674G}. These results agree remarkably well with the results of AGN luminosity distributions (see Fig. 14 in \citealt{2014MNRAS.440..269M}) with FSRQs corresponding to high excitation radio galaxies (HERGs) with $0.05L_{\m{EDD}}-1L_{\m{EDD}}$ and BL Lacs corresponding to low excitation radio galaxies (LERGs) with $2\times 10^{-4}L_{\m{EDD}}-0.05L_{\m{EDD}}$. This close agreement between our model and the observations of AGN acts as an important, independent verification of the results and assumptions of our model. This strongly suggests that there is a correspondence between FSRQs and HERGs (closely associated with FRII sources) and BL Lacs and LERGs (closely associated with FRI sources) supporting the idea of AGN unification. 

Under the same assumption of a fiducial black hole mass we find a bimodal distribution of the length of the accelerating region of a jet and its power in terms of the Eddington luminosity. This suggests that for a given black hole mass more powerful jets will accelerate over larger distances and reach higher terminal velocities. This could be interpreted in terms of a more powerful jet travelling a larger distance before being disrupted by its environment (see e.g. \citealt{1995ApJS..101...29B}). We also find a close association of FSRQs with fast jets and BL Lacs with slower jets. Together Figures \ref{distributions}a and \ref{distributions}b provide strong evidence for the existence of bimodal accretion rates and association of FSRQs with powerful, fast FRII type jets and BL Lacs with less powerful and slower FRI type jets.

\begin{figure*}
	\centering
		\includegraphics[width=14cm, clip=true, trim=0cm 0cm 0.0cm 0.0cm]{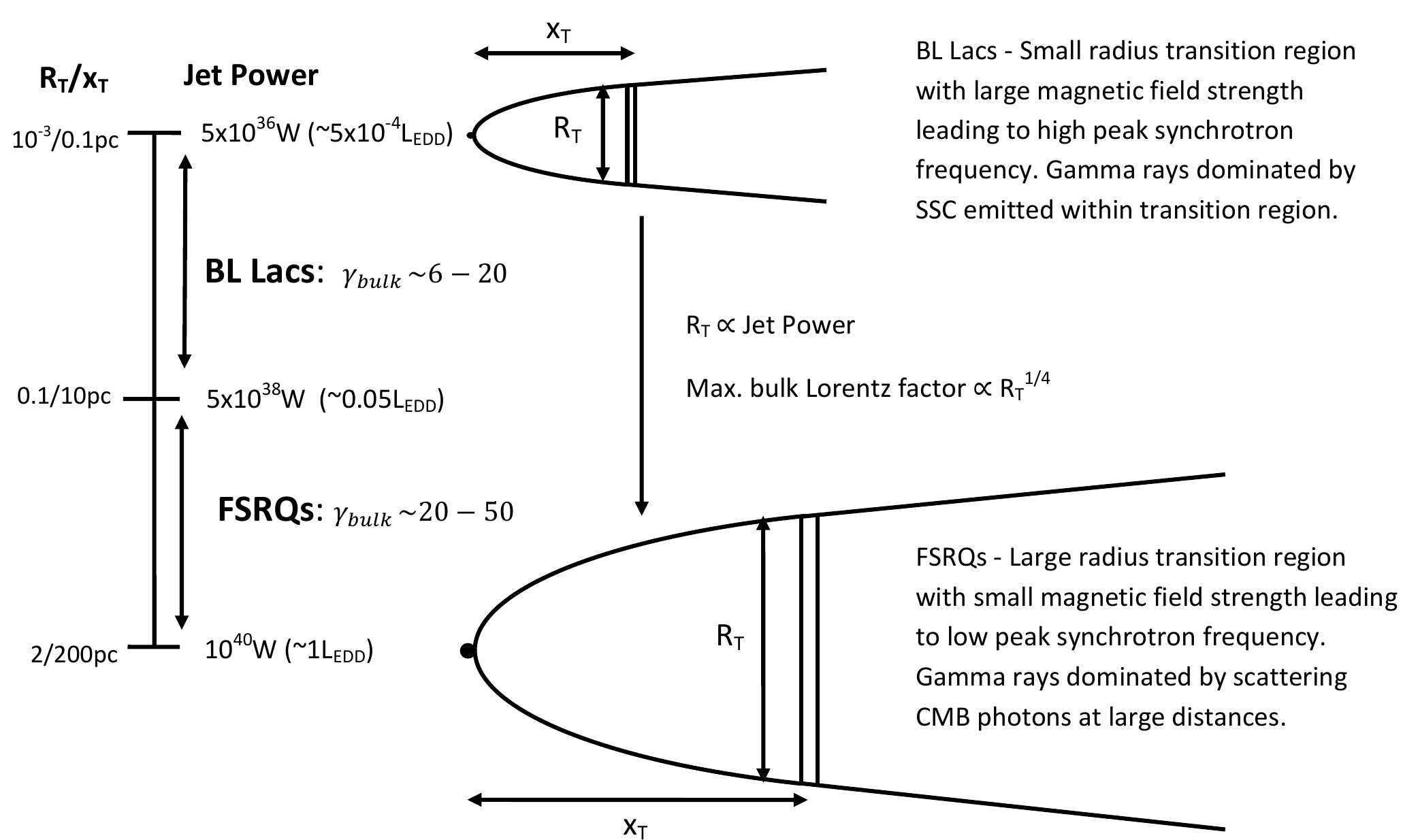} 
                 \caption{A summary of our main results. The reader may find it useful to refer to the jet schematic in Fig. \ref{schematic} to understand the jet geometry and basic assumptions of our model. }
                 \label{results_schem}
\end{figure*}

\subsection{A physical understanding of the blazar sequence}

The results of this work allow for a physical interpretation of the blazar sequence. The blazar sequence, originally discovered by \cite{1998MNRAS.299..433F}, describes the combination of an observed anti-correlation between the jet power and the synchrotron peak frequency in the blazar population, and a correlation between the jet power and the Compton dominance (the ratio of synchrotron to inverse-Compton luminosity). This result has been the subject of considerable discussion and speculation in recent years (e.g. \citealt{1998MNRAS.301..451G, 2008MNRAS.387.1669G, 2008MNRAS.391.1981M, 2011ApJ...740...98M}), however, our work is the first systematic attempt to simultaneously model both the radio and high energy emission of a large sample of blazars to understand their structural and dynamical properties and their relation to the blazar sequence. We find that BL Lacs are low power blazars, with jets that accelerate over shorter distances and reach lower maximum Lorentz factors than more powerful FSRQs. This has several implications for their spectra: firstly the brightest synchrotron emitting region in BL Lacs is smaller in radius than in FSRQs and this means that they have larger magnetic field strengths. This is easily calculated by equating the rest frame magnetic energy contained in a thin slab of the jet plasma in equipartition, to half the total energy of the slab 
\be
\frac{4\gamma^{2}}{3}\frac{B'^{2}}{2\mu_{0}}\pi R^{2}l_{s}=\frac{W_{j}}{2}, \qquad B'_{T}\propto \frac{\sqrt{W_{j}}}{\gamma_{\m{max}}R_{T}}
\ee
where a subscript $T$ indicates the value of a quantity at the transition region, $W_{j}$ is the lab frame jet power and $l_{s}$ is one light-second $l_{s}=c\times 1\m{s}$. Since BL Lacs have both lower maximum bulk Lorentz factors and smaller transition region radii than FSRQs (from Fig. \ref{corr}a we find $W_{j}\propto R_{T}$) the magnetic field strength at the transition region is greater (whilst the jet power is smaller, this factor is suppressed by a square-root). The synchrotron peak frequency is directly proportional to the magnetic field strength and so we find that as the jet power and jet radius increase the synchrotron peak frequency decreases. This offers a physical explanation for the anticorrelation between jet power and peak synchrotron frequency.

The inverse-Compton emission from BL Lacs is dominated by synchrotron self-Compton (SSC) emission due to density of synchrotron photons in the compact, bright transition region with a large magnetic field strength. We find that the power emitted by SSC is naturally limited to being comparable to, or smaller than, the power emitted by synchrotron emission and this prevents SSC dominated sources (i.e. most BL Lacs) from becoming Compton-dominant. This is because the highest energy electrons are both responsible for emitting the majority of the synchrotron power ($\propto E_{e}^{2}$) and SSC power ($\propto E_{e}^{2}U_{\m{synch}}\sim E_{e}^{4}$). This means that as the spectrum becomes Compton-dominant, the SSC radiative cooling lifetimes for these high energy electrons decreases rapidly and so it is difficult to replenish this high energy population and maintain Compton-dominance via SSC (unless the electron spectrum, $N_{e}(E_{e})\propto E_{e}^{-\alpha}$, has a very low spectral index, i.e. $\alpha<1$, so that the majority of the energy in the electron population is concentrated in the highest energy electrons).

In FSRQs we find that the radius of the brightest synchrotron emitting region, constrained by fitting the synchrotron break, is too large, and the magnetic field strength too small, for substantial SSC emission to occur. This is because SSC emission requires a large energy density of synchrotron seed photons to Compton up-scatter and the power emitted by synchrotron radiation is proportional the the square of the magnetic field strength. For FSRQs we find the radius of the transition region to be between $0.1$pc-$2$pc and the corresponding distance to be $\approx10$pc-$200$pc. At these large distances which are well outside of the BLR ($R_{\m{BLR}}<1$pc, e.g. \citealt{2005ApJ...629...61K}) and outside the dusty torus ($R_{\m{DT}}<10$pc, e.g. \citealt{2006NewAR..50..728E} and \citealt{2008ApJ...685..160N}), we find the high energy emission is well fitted by scattering CMB photons. Since inverse-Compton scattering of external photon sources is generally required to fit to most low-peak frequency, Compton-dominant FSRQ spectra, it is no surprise that most previous studies have required the brightest emitting region to occur within the BLR or dusty torus, since these were the only external photon sources which were modelled (e.g. \cite{2000ApJ...545..107B}, \cite{2008MNRAS.387.1669G}, \cite{2009ApJ...704...38S},  \cite{2011A&A...534A..86T} and \cite{2014ApJ...785..132J}). However, crucially, these works neglected to model the radio observations. We find that by simultaneously modelling both the radio and high energy emission we are able to constrain the brightest region in FSRQs, to occur at much larger distances than the BLR or dusty torus.

\subsection{Limitations of a time-independent model}

In this and our preceding papers we have developed the first relativistic fluid jet model which calculates the emission from all sections of the jet. We decided to initially study a time-independent model in order to understand the average properties of the jet fluid flow before adding time-dependent perturbations. Whilst this has the benefit of allowing us to understand correlations between the long-term average jet properties it does not allow us to investigate the dramatic short-term variability of blazars. Bearing in mind the very short timescales for gamma-ray flares (e.g. \citealt{2007ApJ...669..862A}, \citealt{2007ApJ...664L..71A} and \citealt{2010Natur.463..919A}) and the range in flaring behaviour, we expect that whilst the long-term steady gamma-ray emission seems to be well fitted by scattering CMB photons at large distances, it seems very unlikely that flares at such large distances could be responsible for the very short timescale gamma-ray variability observed due to the long radiative cooling and light-crossing timescales. We therefore expect that short timescale flaring events are likely to occur at much shorter distances along the jet, probably within the parabolic accelerating region. We shall investigate variability and flaring by adding time-dependent perturbations to the steady fluid flow in the near future.      

\section{Conclusion}

In this work we have used one of the most realistic jet emission models to date to fit with unprecedented accuracy to the spectra of the entire sample of Fermi blazars with simultaneous multiwavelength observations and redshifts from \cite{2010ApJ...716...30A}. From this we have been able to place constraints on the basic structure and dynamics of blazar jets. Our main results are as follows:

\begin{itemize}
\item We find a linear correlation between the radius of the brightest synchrotron emission region and the jet power, which we constrain via the optically thick to thin synchrotron break. In most FSRQs this radius ($>$0.1pc) is large enough to place this region, containing the most intense particle acceleration, well beyond the broad line region and dusty torus.
\item We find a correlation between the final radius of the accelerating region of the jet and the maximum bulk Lorentz factor of the jet given approximately by $\gamma_{\m{bulk}}\propto R_{T}^{1/4}$. 
\item There is a bimodal distribution of jet power in the blazar population with FSRQs having large jet powers ($5\times 10^{38}\m{W}-10^{40}\m{W}$) and BL Lacs ($5\times 10^{36}\m{W} - 5\times 10^{38}\m{W}$). Assuming a fiducial black hole mass of $5\times10^{8}M_{\odot}$ we find excellent agreement with the distribution of AGN luminosities from \cite{2014MNRAS.440..269M}, with a close correspondence of FSRQs with high excitation radio galaxies (HERGs, $0.05L_{\m{EDD}}-1L_{\m{EDD}}$) and BL Lacs with Low excitation radio galaxies (LERGs, $2\times 10^{-4}L_{\m{EDD}}-0.05L_{\m{EDD}}$) .
\item There is a bimodal distribution of maximum bulk Lorentz factors in the blazar population with FSRQs having large bulk Lorentz factors (20-50) and BL Lacs lower bulk Lorentz factors (6-20). Together with the differences in jet power, this provides evidence for AGN unification with FSRQs corresponding to powerful, fast FRII type jets and BL Lacs corresponding to weaker, slower FRI type jets.
\item FSRQs have larger maximum bulk Lorentz factors and a larger radius when the jet first comes into equipartition and intense particle acceleration occurs. This means they have lower magnetic field strengths in this bright region, which leads to sub-dominant SSC emission and low peak synchrotron emission. We find that for FSRQs, their large bulk Lorentz factors, maintained to large distances, allow the Compton-dominance and low inverse-Compton peak frequency to be explained via inverse-Compton scattering of CMB photons. 
\item BL Lacs have lower bulk Lorentz factors and smaller radius jets at the location of the brightest synchrotron emission, leading to higher magnetic field strengths in this region than FSRQs. This means that the peak synchrotron emission from BL Lacs occurs at higher frequencies and SSC dominates the inverse-Compton emission. These dynamical and structural constraints allow an understanding of the physics underlying the famous blazar sequence.
\end{itemize} 

These results are summarised in Fig. \ref{results_schem} and demonstrate the merit of using realistic, extended fluid jet emission models to fit to blazar spectra and constrain their basic properties. 

\section*{Acknowledgements}

WJP acknowledges funding from the University of Oxford. GC acknowledges support from STFC rolling grant ST/H002456/1. We would like to thank Alexander Tchekhovskoy, Lance Miller, Martin Hardcastle, Roger Blandford, Richard Booth, Jim Hinton, Steven Balbus, Masanori Nakamura, Sera Markoff, Markus Bottcher, Rob Fender, Mitch Begelman and Jonathan McKinney for many valuable discussions.

\bibliographystyle{mn2e}
\bibliography{Jetpaper2refs}
\bibdata{Jetpaper2refs}

\begin{figure*}
	\centering
		\subfloat[J0035]{ \includegraphics[height=8cm, clip=true, trim=1.5cm 1cm 1cm 3.5cm,angle=-90]{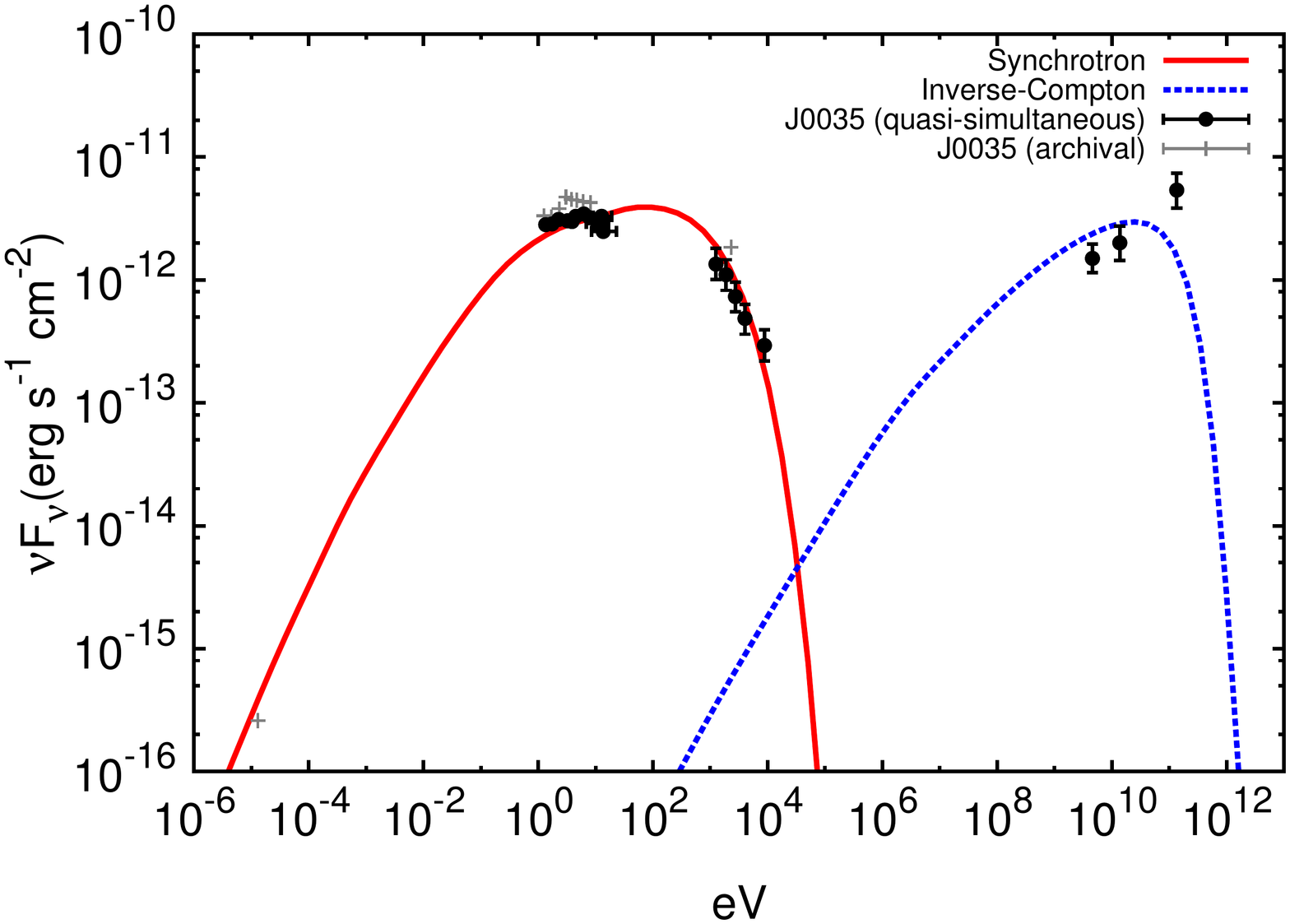} }
		\subfloat[J0137]{ \includegraphics[height=8cm, clip=true, trim=1.5cm 1cm 1cm 3.5cm,angle=-90]{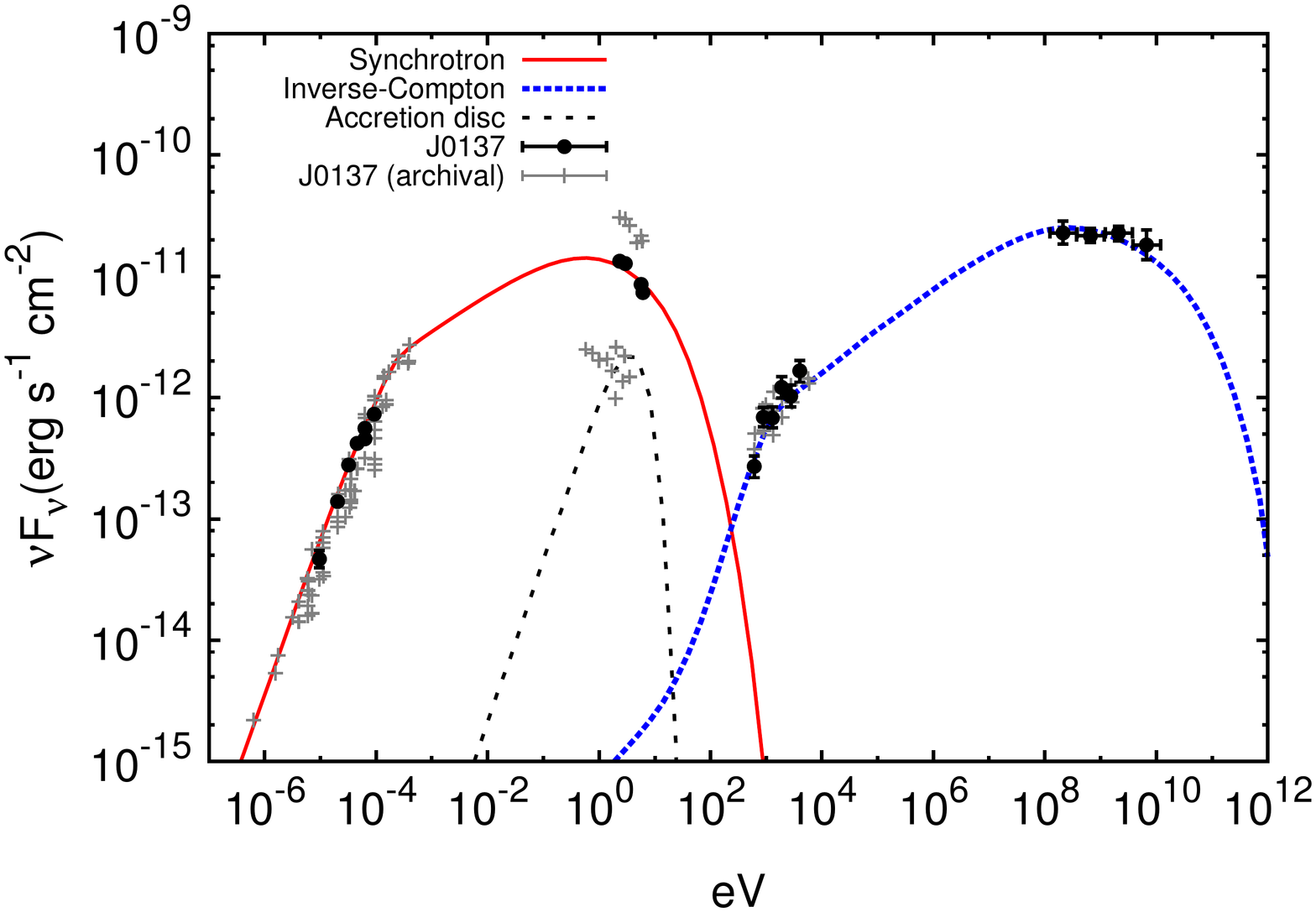} }
\\
		\subfloat[J0210]{ \includegraphics[height=8cm, clip=true, trim=1.5cm 1cm 1cm 3.5cm,angle=-90]{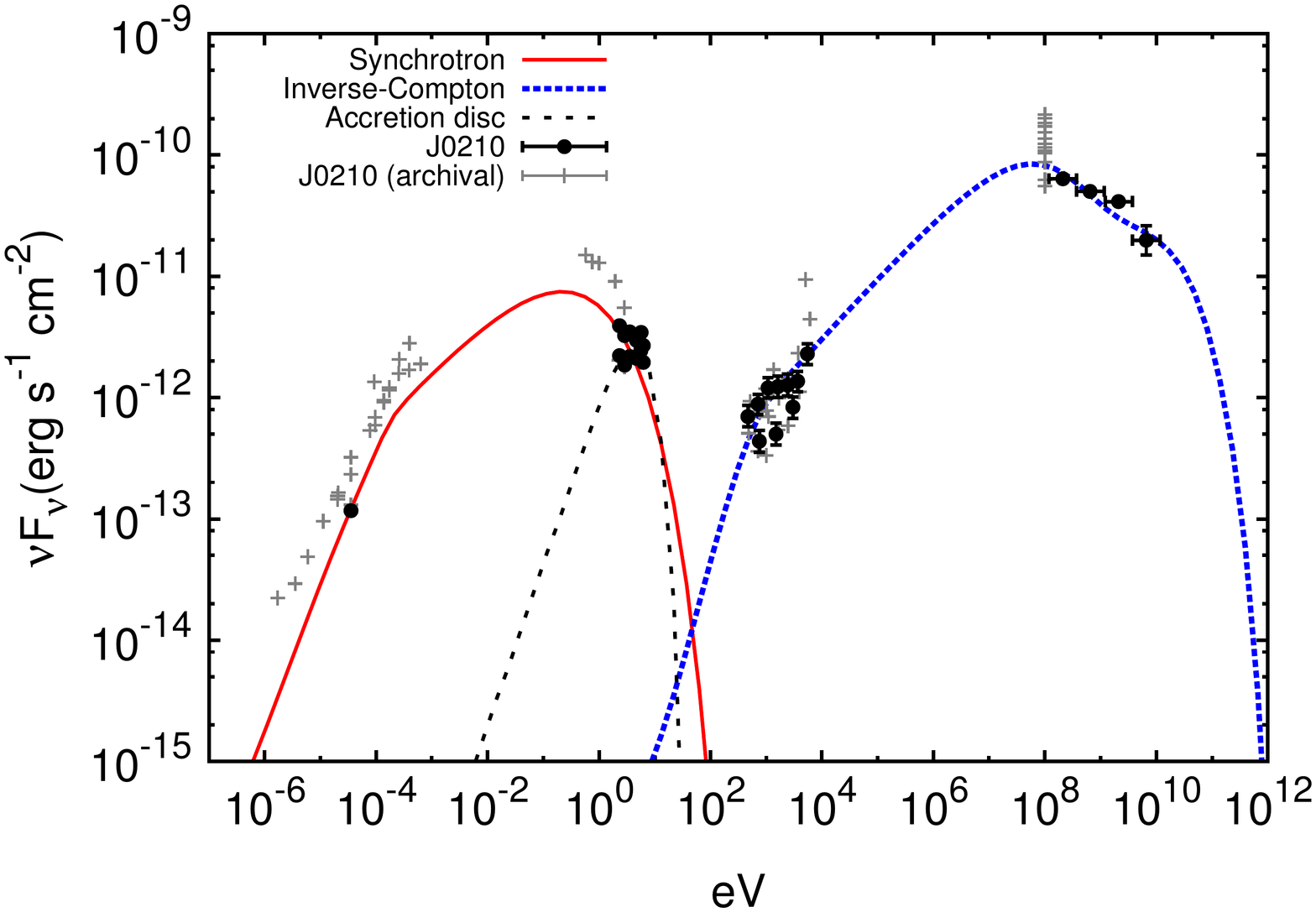} }	
		\subfloat[J0222]{ \includegraphics[height=8cm, clip=true, trim=1.5cm 1cm 1cm 3.5cm,angle=-90]{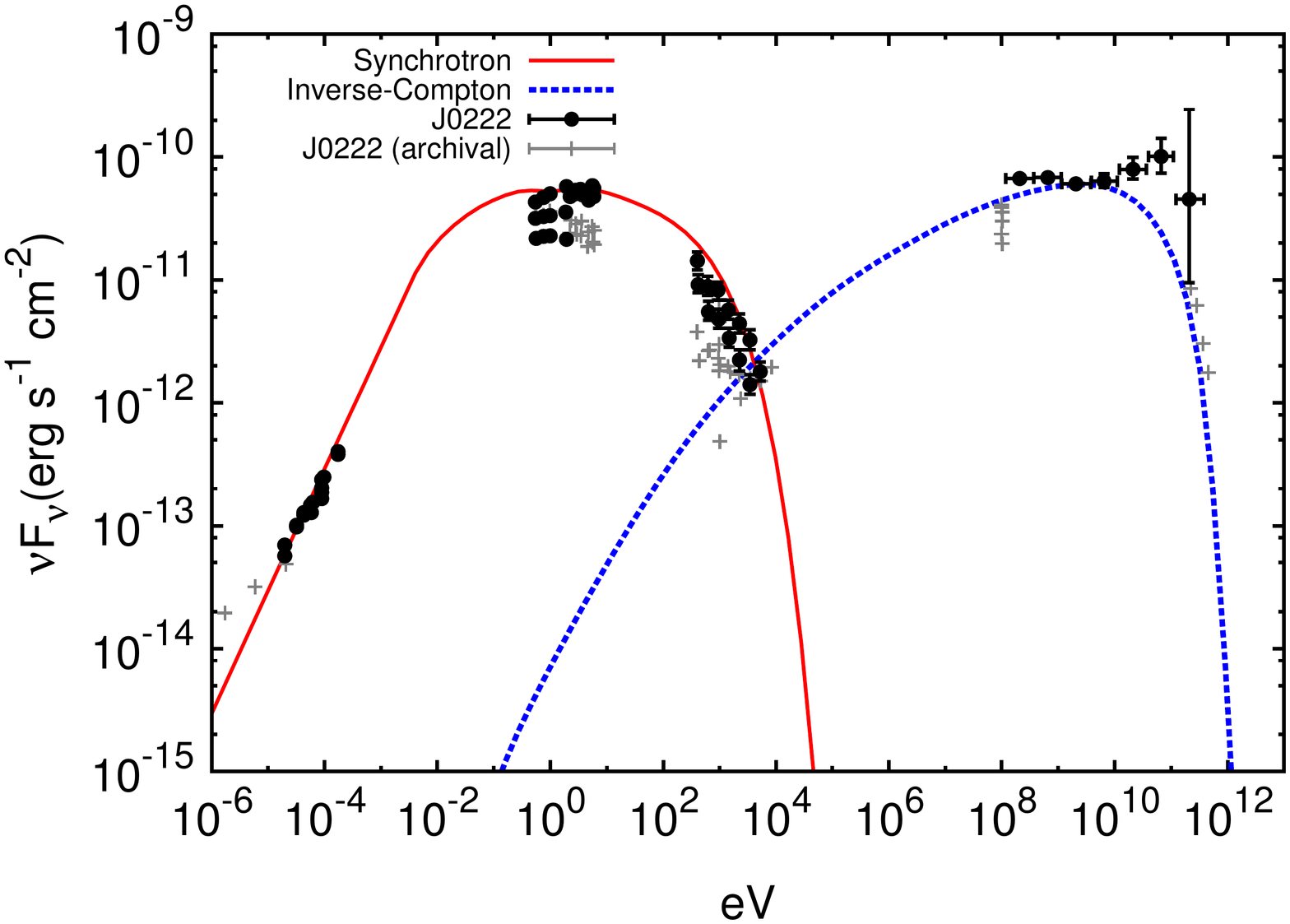} }
\\
                 \subfloat[J0229]{ \includegraphics[height=8cm, clip=true, trim=1.5cm 1cm 1cm 3.5cm,angle=-90]{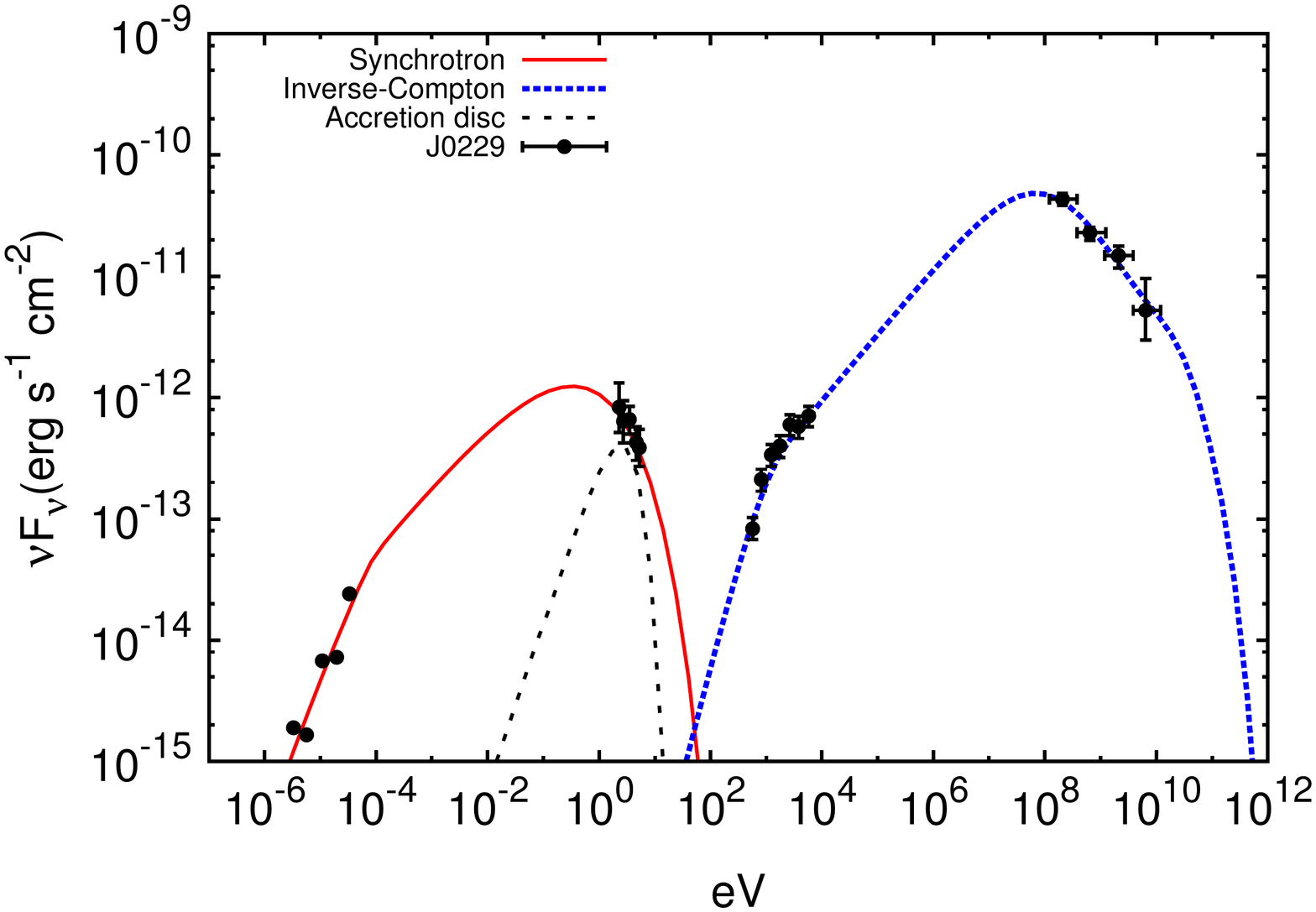} }
		\subfloat[J0238.4]{ \includegraphics[height=8cm, clip=true, trim=1.5cm 1cm 1cm 3.5cm,angle=-90]{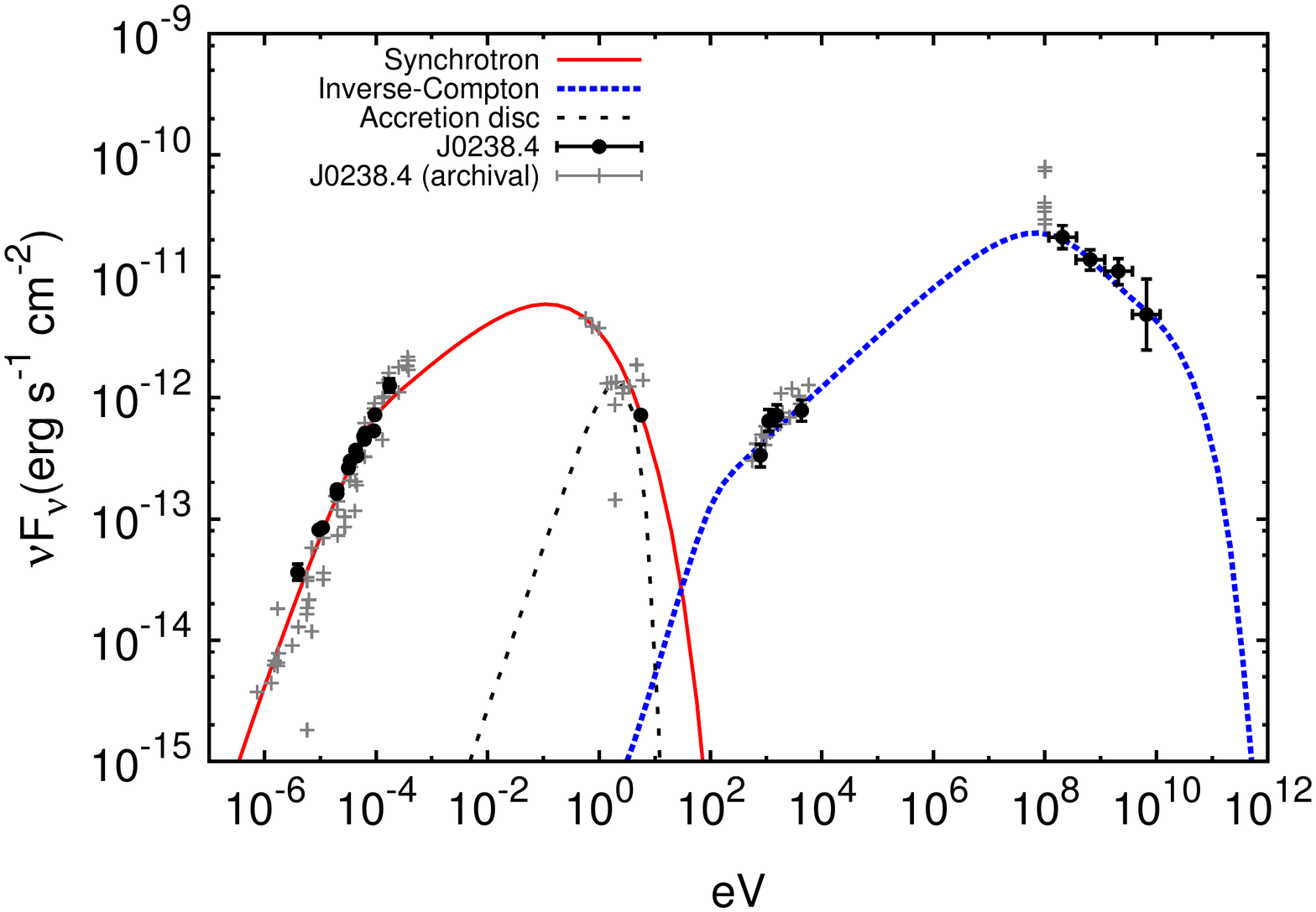} }	
		
	\caption{The fits to the SEDs of all 38 {\it Fermi} blazars with simultaneous multiwavelength observations and redshifts from Abdo et al. 2010, and the 4 non-simultaneous spectra of the high synchrotron peak frequency FSRQs from Padovani et al. 2010 (J0035, J0537, J0630 and J1312). The model is able to fit very well to all of the spectra including radio observations. Simultaneous (and quasi-simultaneous) data points are shown as solid circles whilst archival data are included as crosses. Due to computational time constraints fits were performed by eye.}
\label{spectra}
\end{figure*}

\begin{figure*}
	\centering
		\subfloat[J0238.6]{ \includegraphics[height=8cm, clip=true, trim=1.5cm 1cm 1cm 3.5cm, angle=-90]{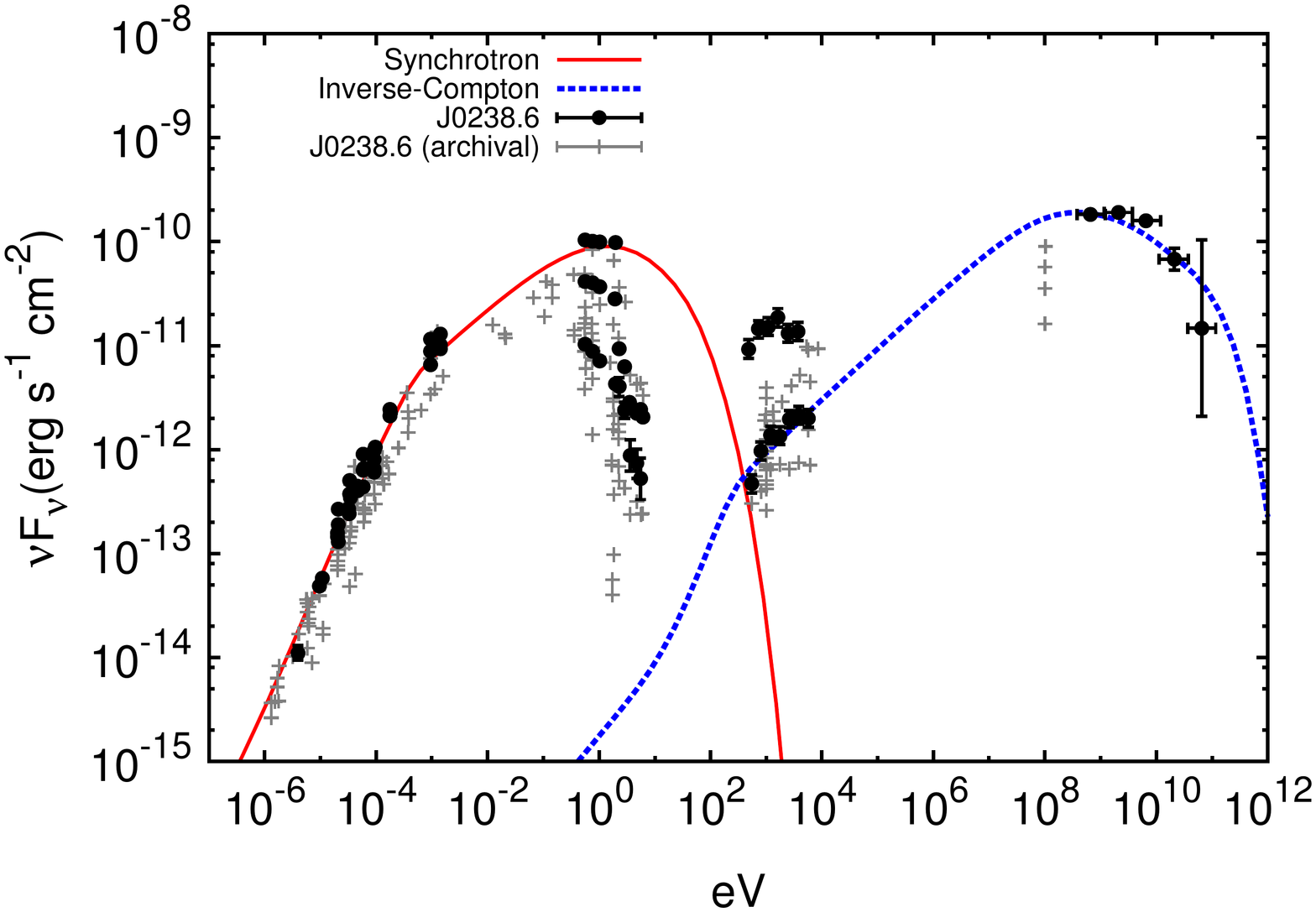} }
		\subfloat[J0349]{ \includegraphics[height=8cm, clip=true, trim=1.5cm 1cm 1cm 3.5cm,angle=-90]{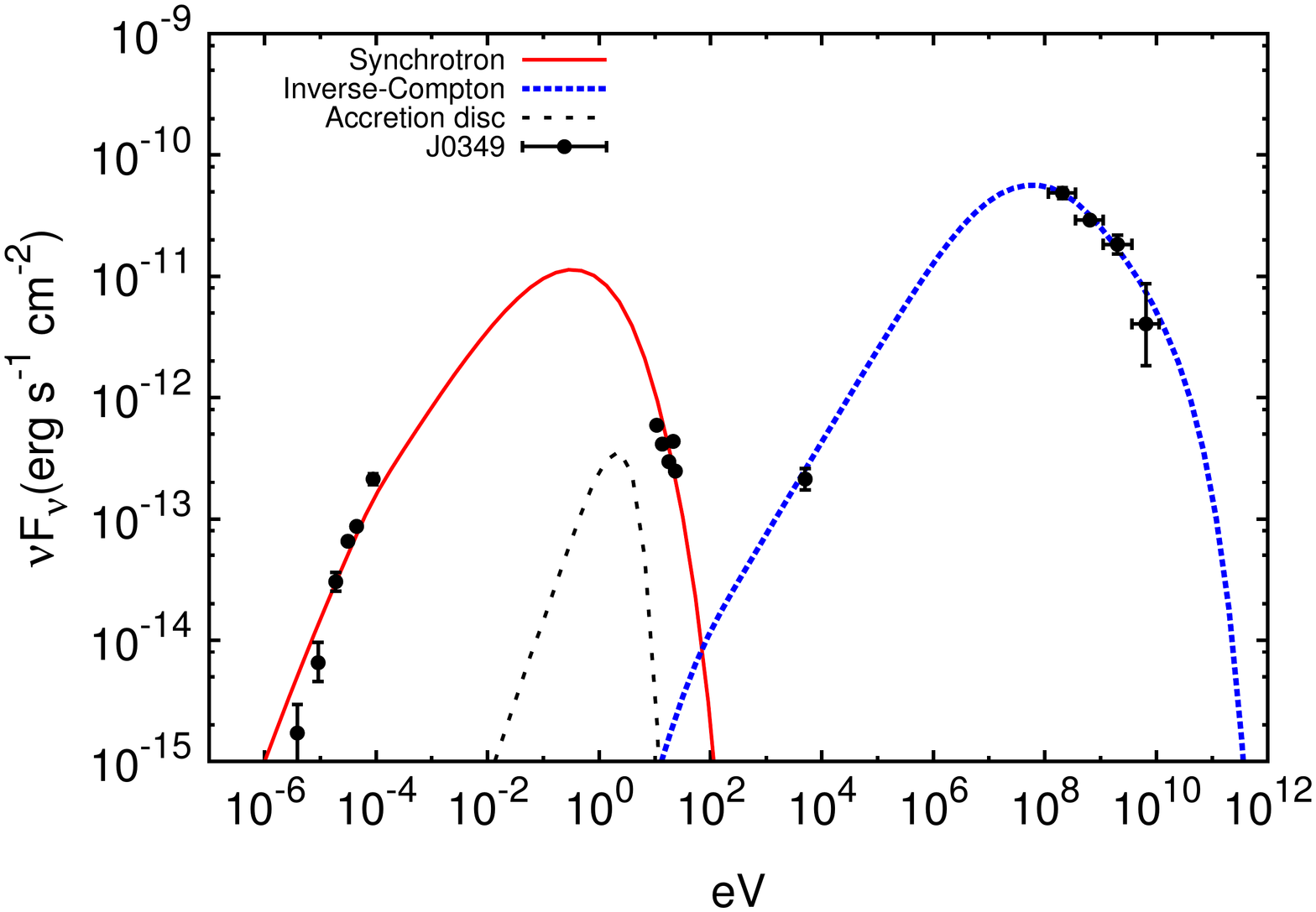} }	
\\
		\subfloat[J0423]{ \includegraphics[height=8cm, clip=true, trim=1.5cm 1cm 1cm 3.5cm,angle=-90]{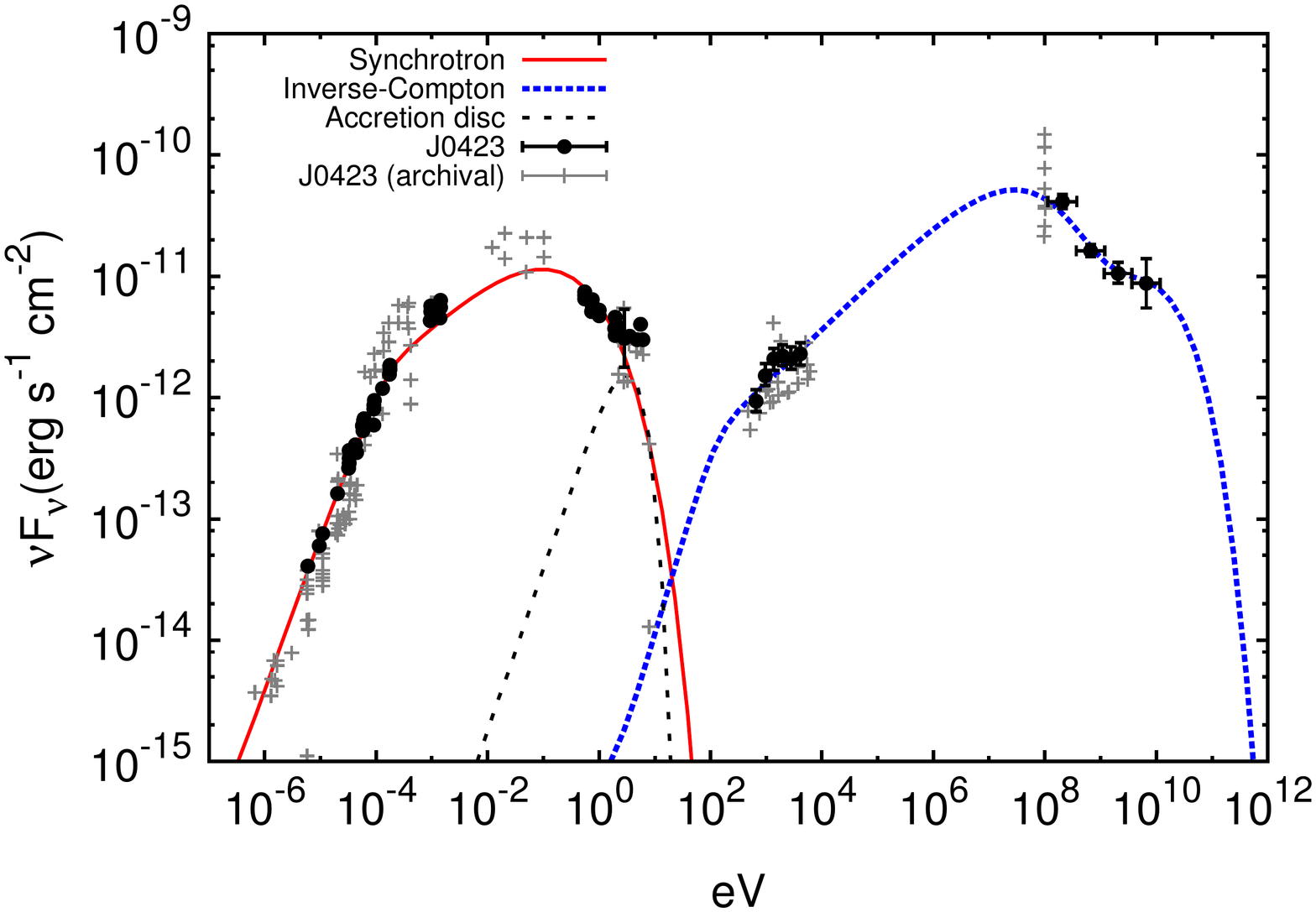} }
		\subfloat[J0457]{ \includegraphics[height=8cm, clip=true, trim=1.5cm 1cm 1cm 3.5cm,angle=-90]{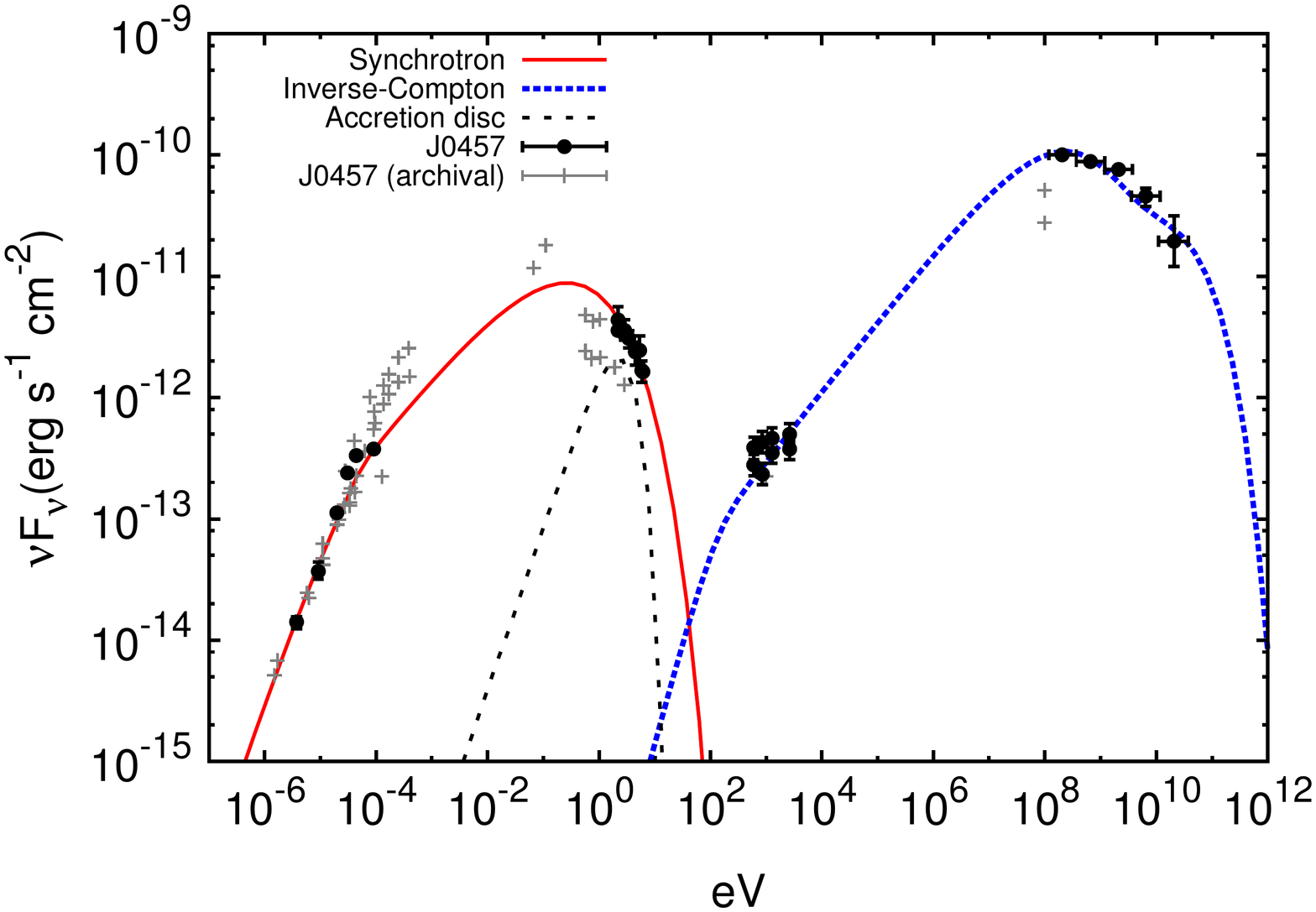} }
\\	
		\subfloat[J0507]{ \includegraphics[height=8cm, clip=true, trim=1.5cm 1cm 1cm 3.5cm,angle=-90]{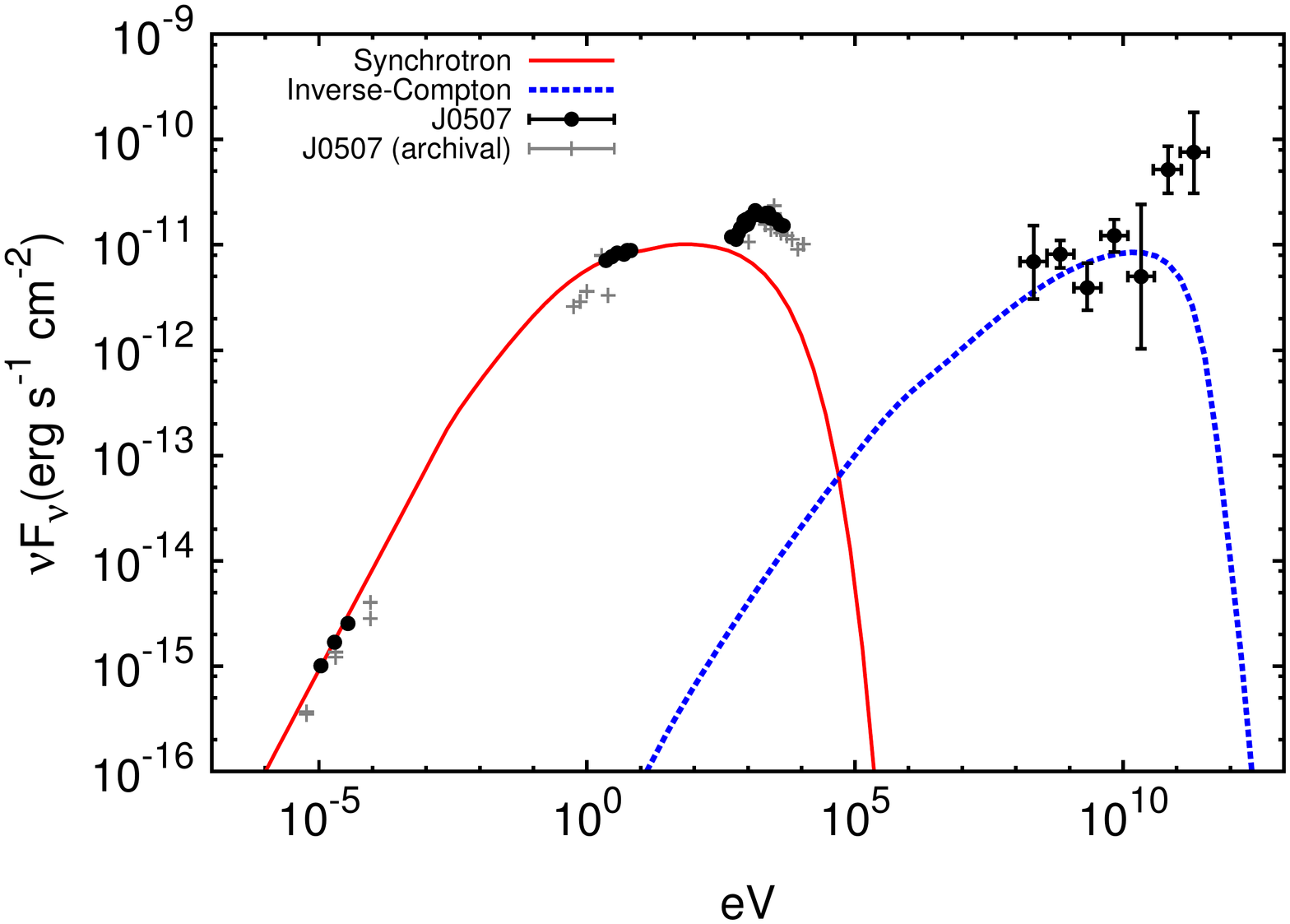} }
		\subfloat[J0531]{ \includegraphics[height=8cm, clip=true, trim=1.5cm 1cm 1cm 3.5cm,angle=-90]{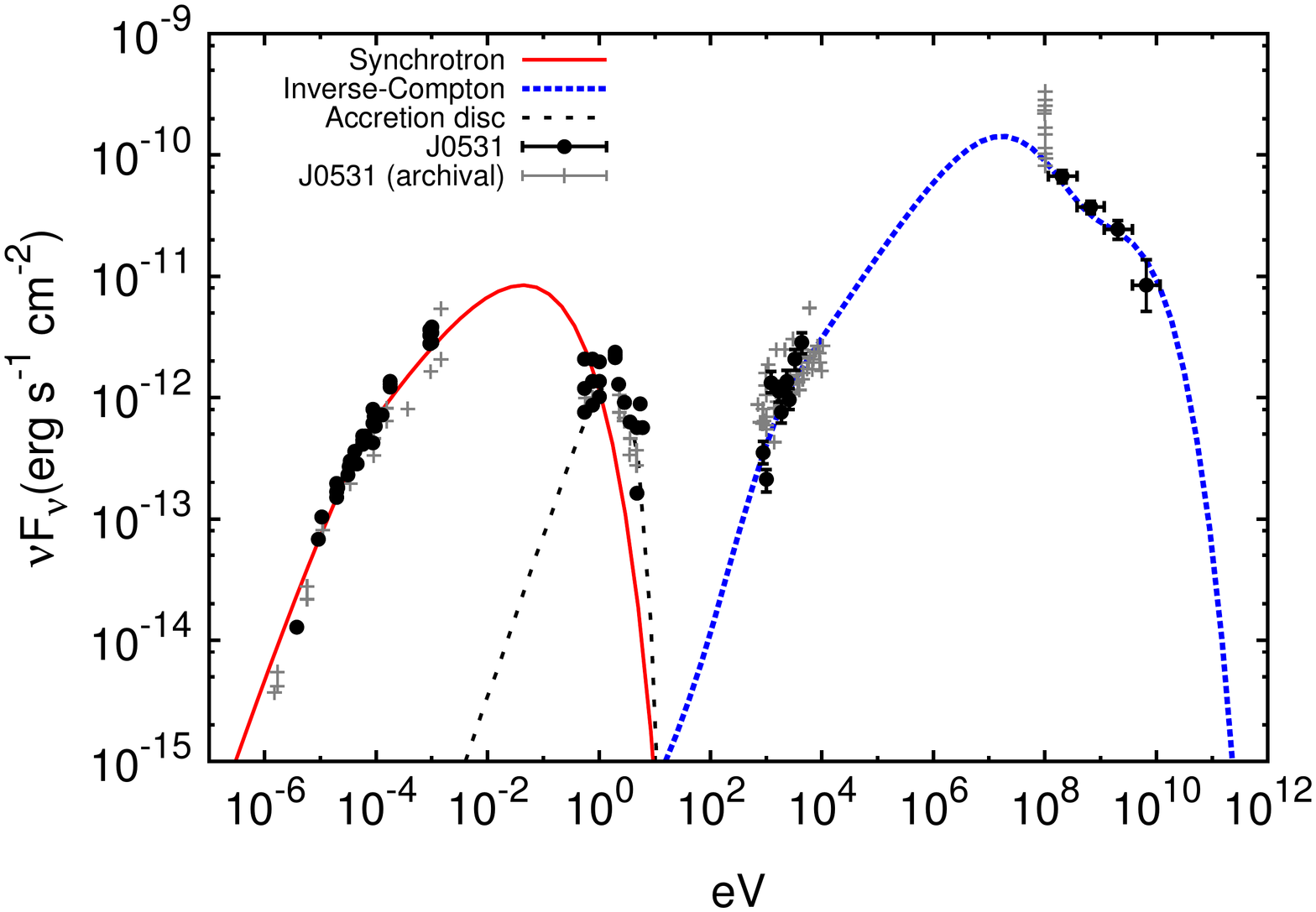} }	

\end{figure*}

\begin{figure*}
	\centering
		\subfloat[J0537]{ \includegraphics[height=8cm, clip=true, trim=1.5cm 1cm 1cm 3.5cm,angle=-90]{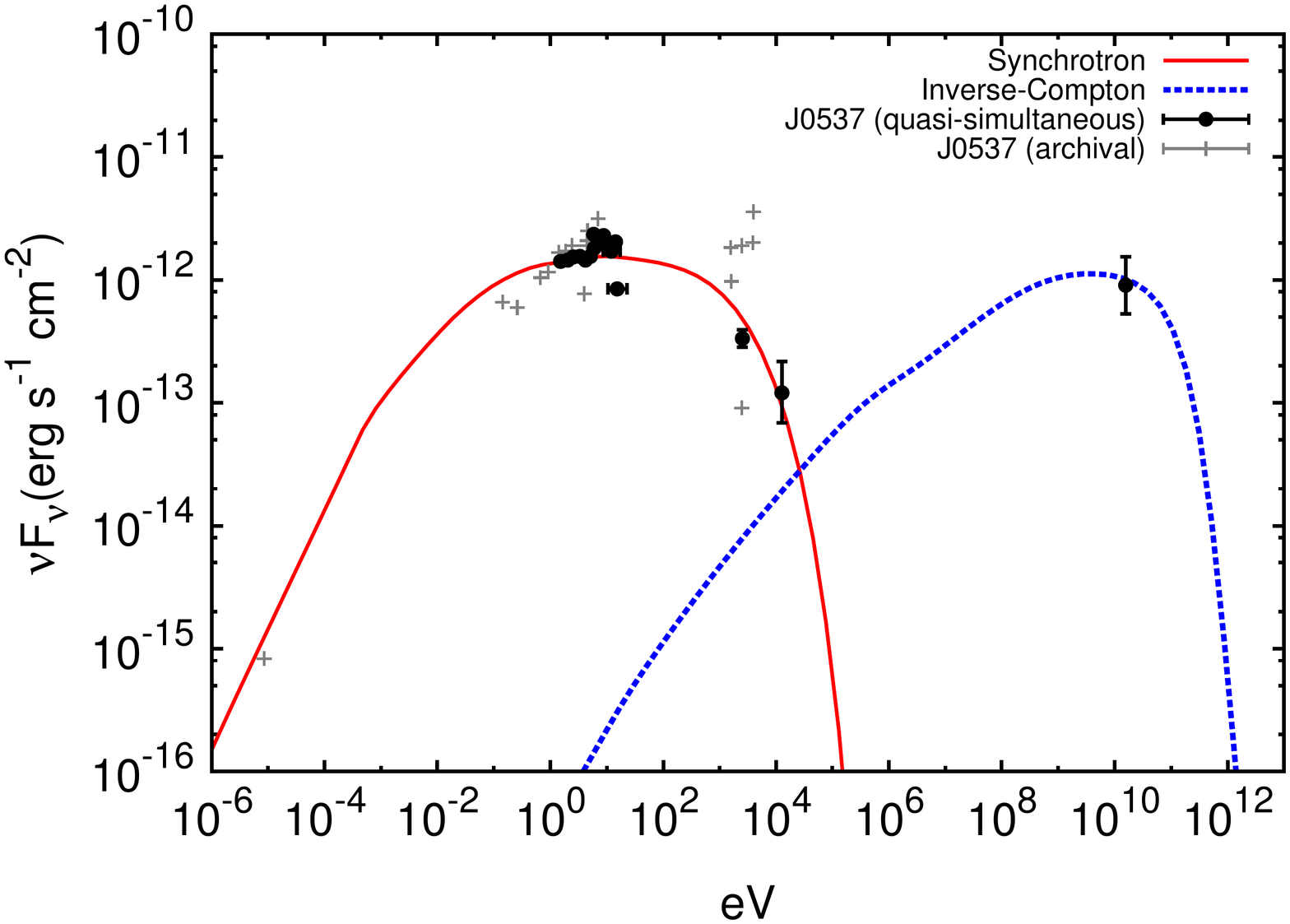} }
		\subfloat[J0538]{ \includegraphics[height=8cm, clip=true, trim=1.5cm 1cm 1cm 3.5cm,angle=-90]{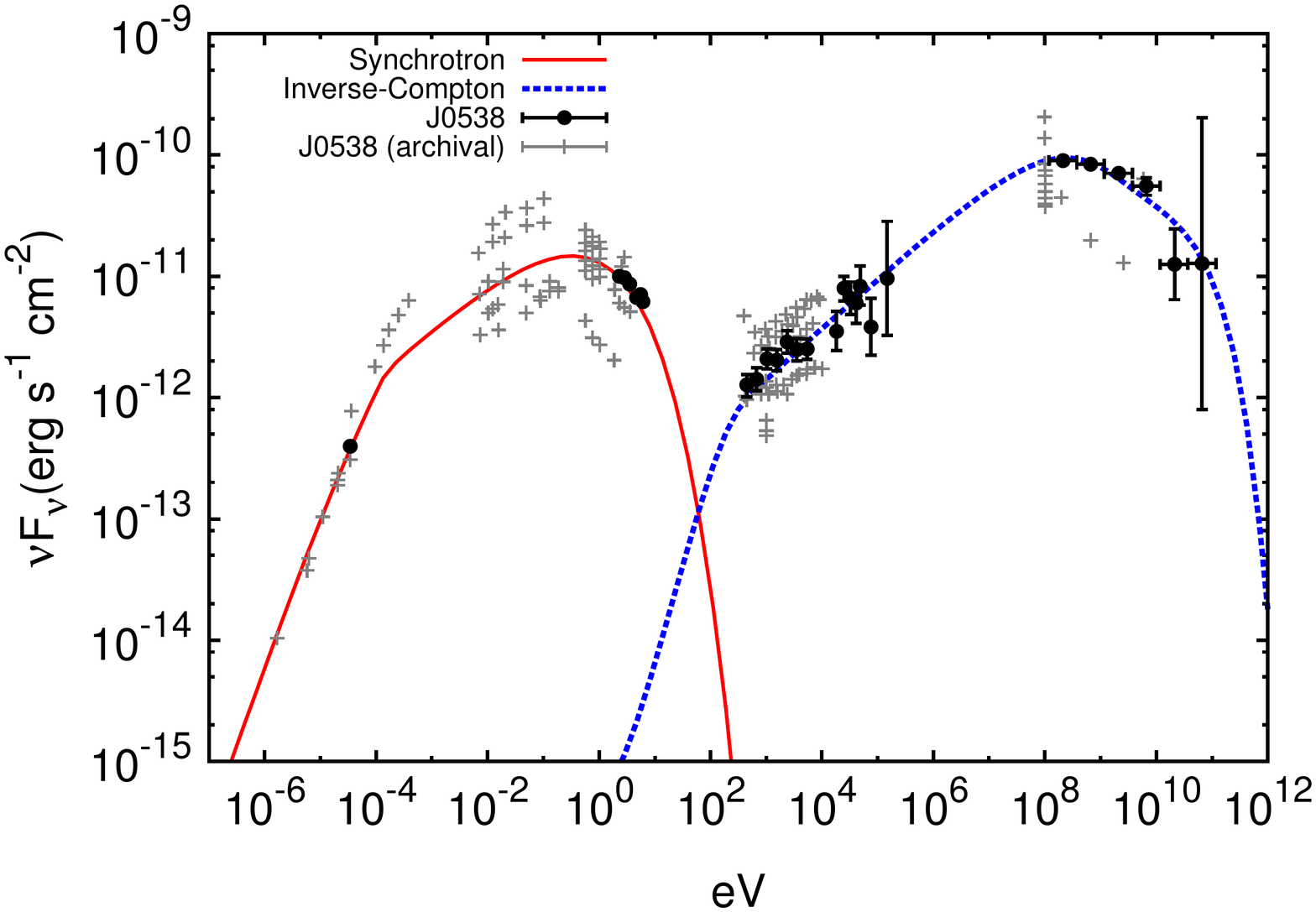} }
\\
		\subfloat[J0630]{ \includegraphics[height=8cm, clip=true, trim=1.5cm 1cm 1cm 3.5cm,angle=-90]{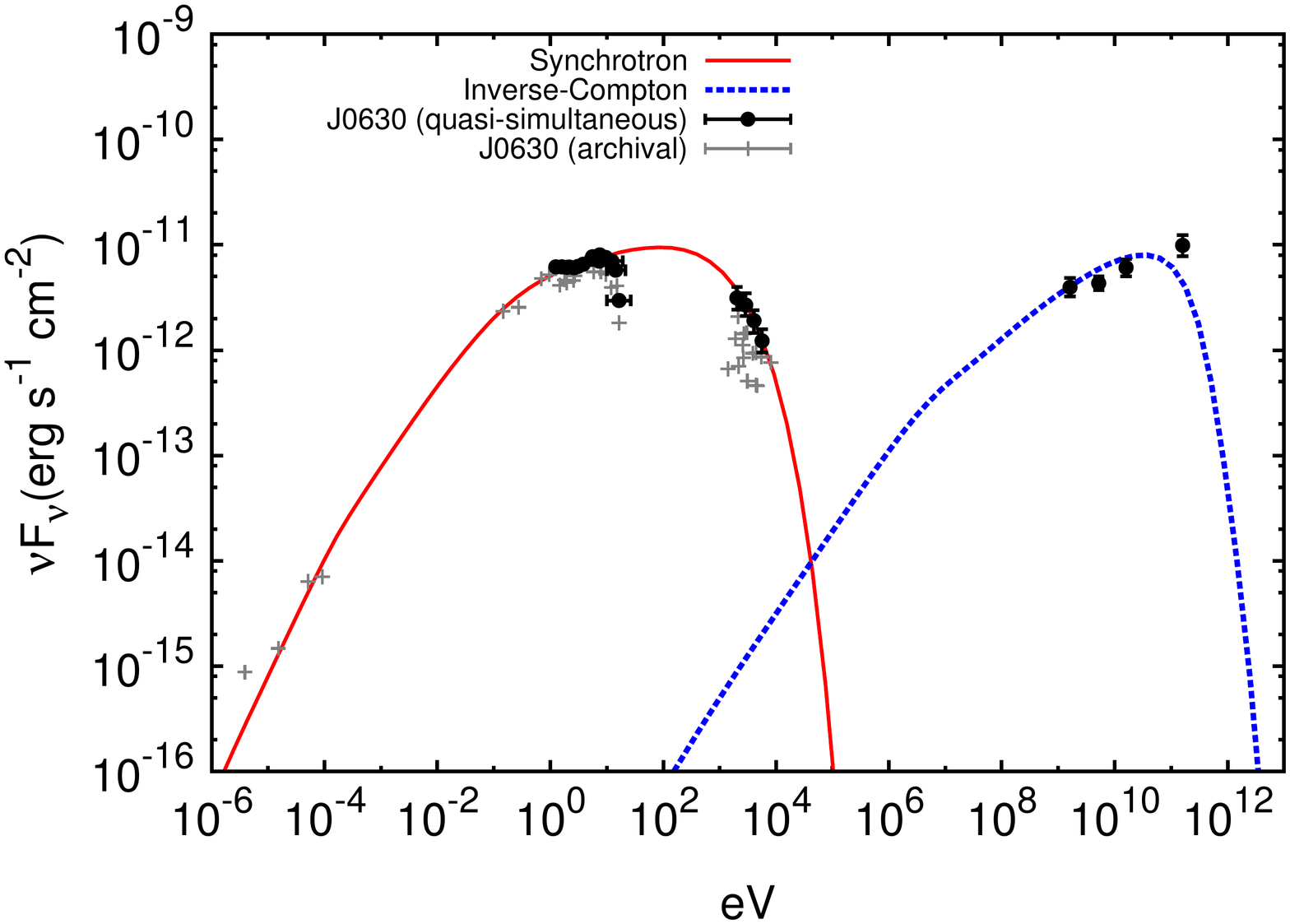} }
		\subfloat[J0730]{ \includegraphics[height=8cm, clip=true, trim=1.5cm 1cm 1cm 3.5cm,angle=-90]{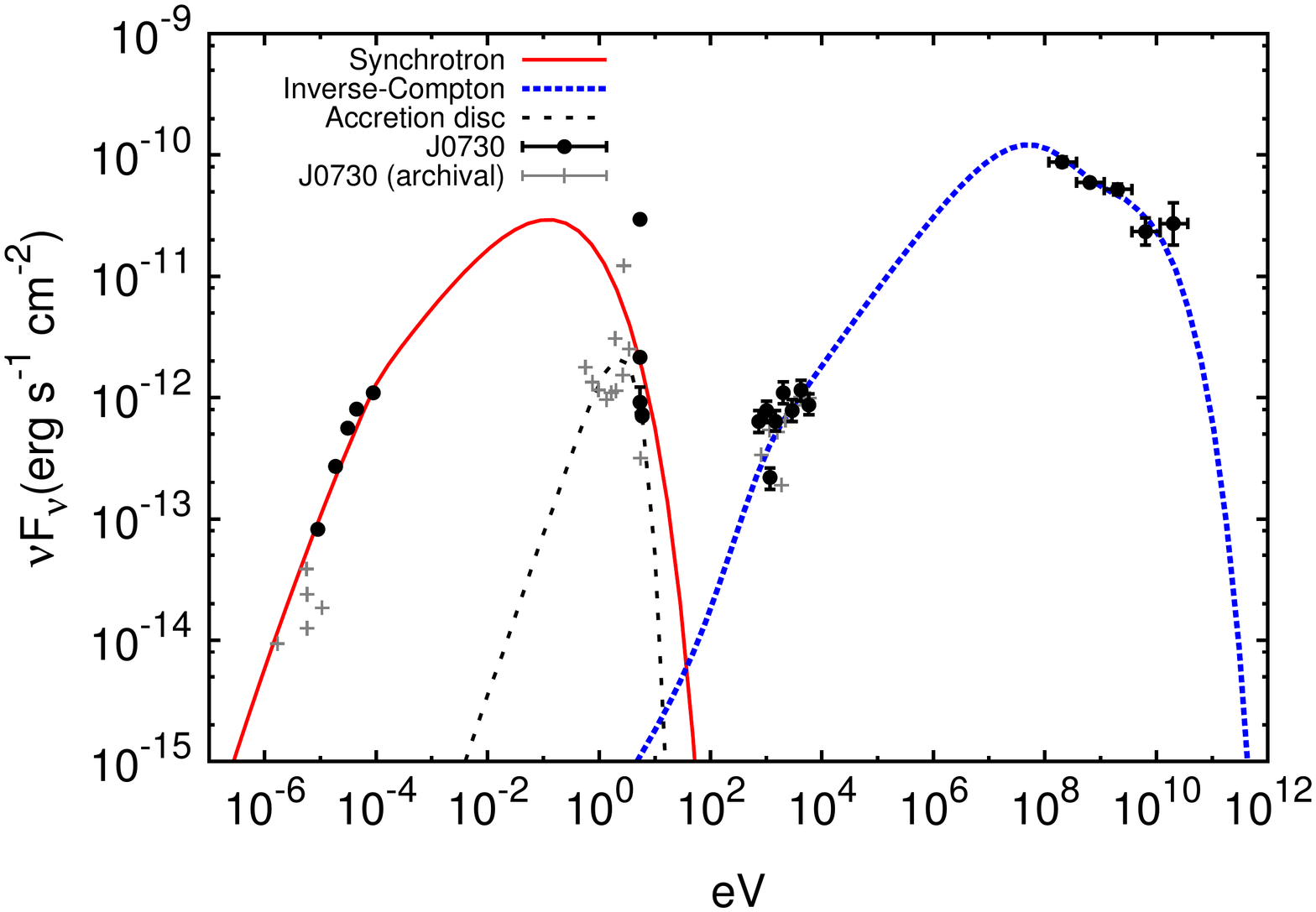} }	
\\	
		\subfloat[J0855]{ \includegraphics[height=8cm, clip=true, trim=1.5cm 1cm 1cm 3.5cm,angle=-90]{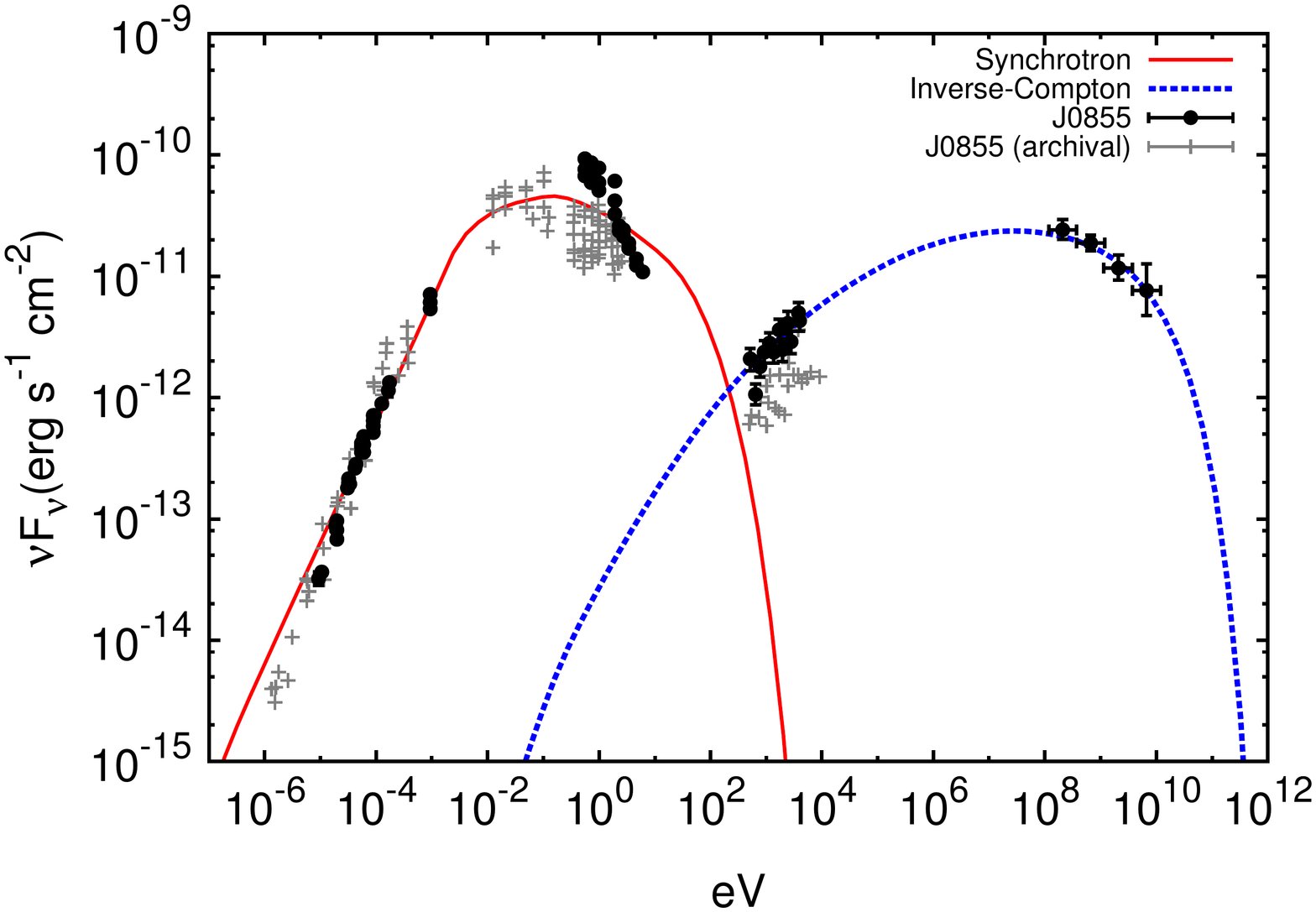} }
		\subfloat[J0921]{ \includegraphics[height=8cm, clip=true, trim=1.5cm 1cm 1cm 3.5cm,angle=-90]{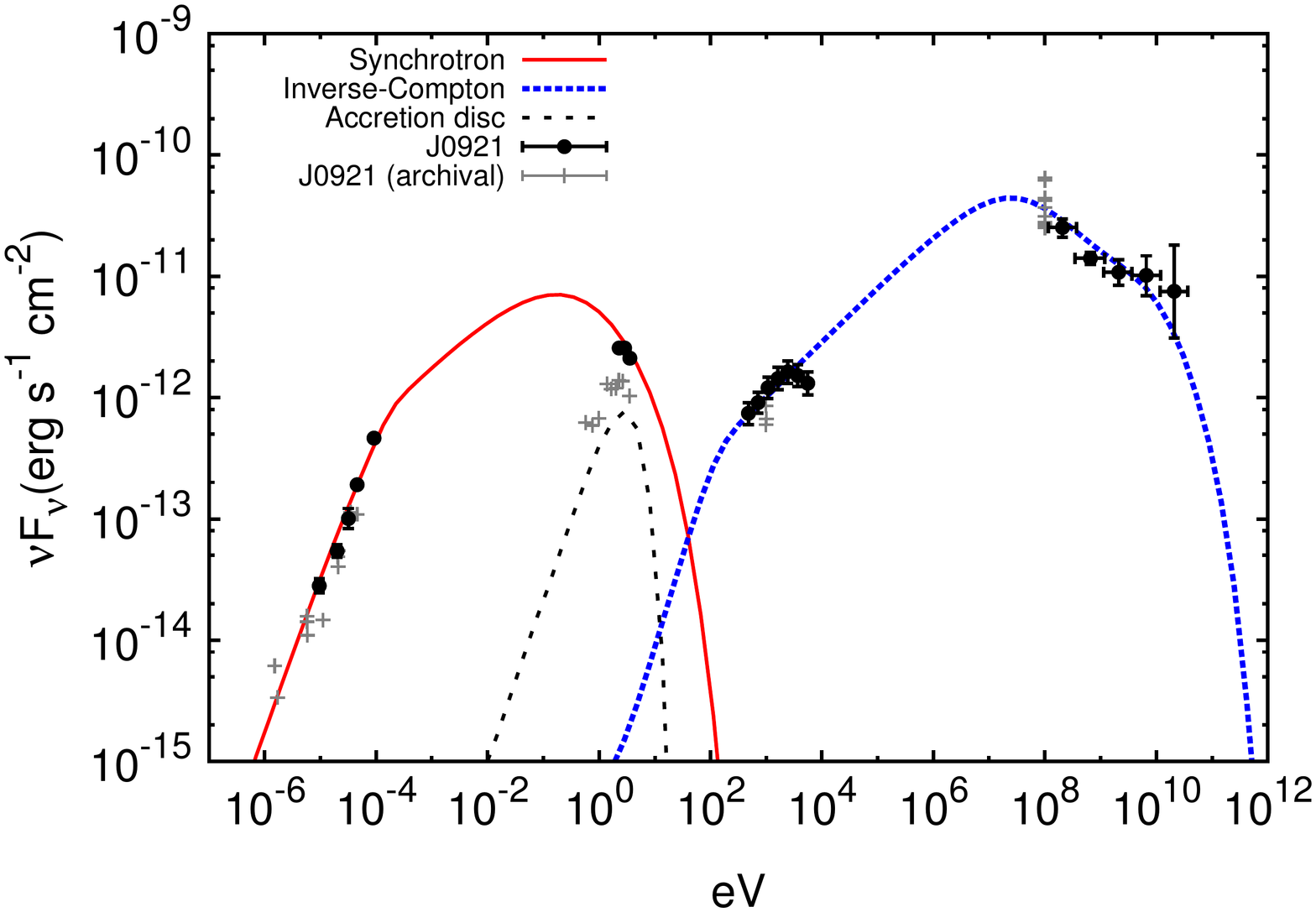} }

\end{figure*}

\begin{figure*}
	\centering
		\subfloat[J1057]{ \includegraphics[height=8cm, clip=true, trim=1.5cm 1cm 1cm 3.5cm,angle=-90]{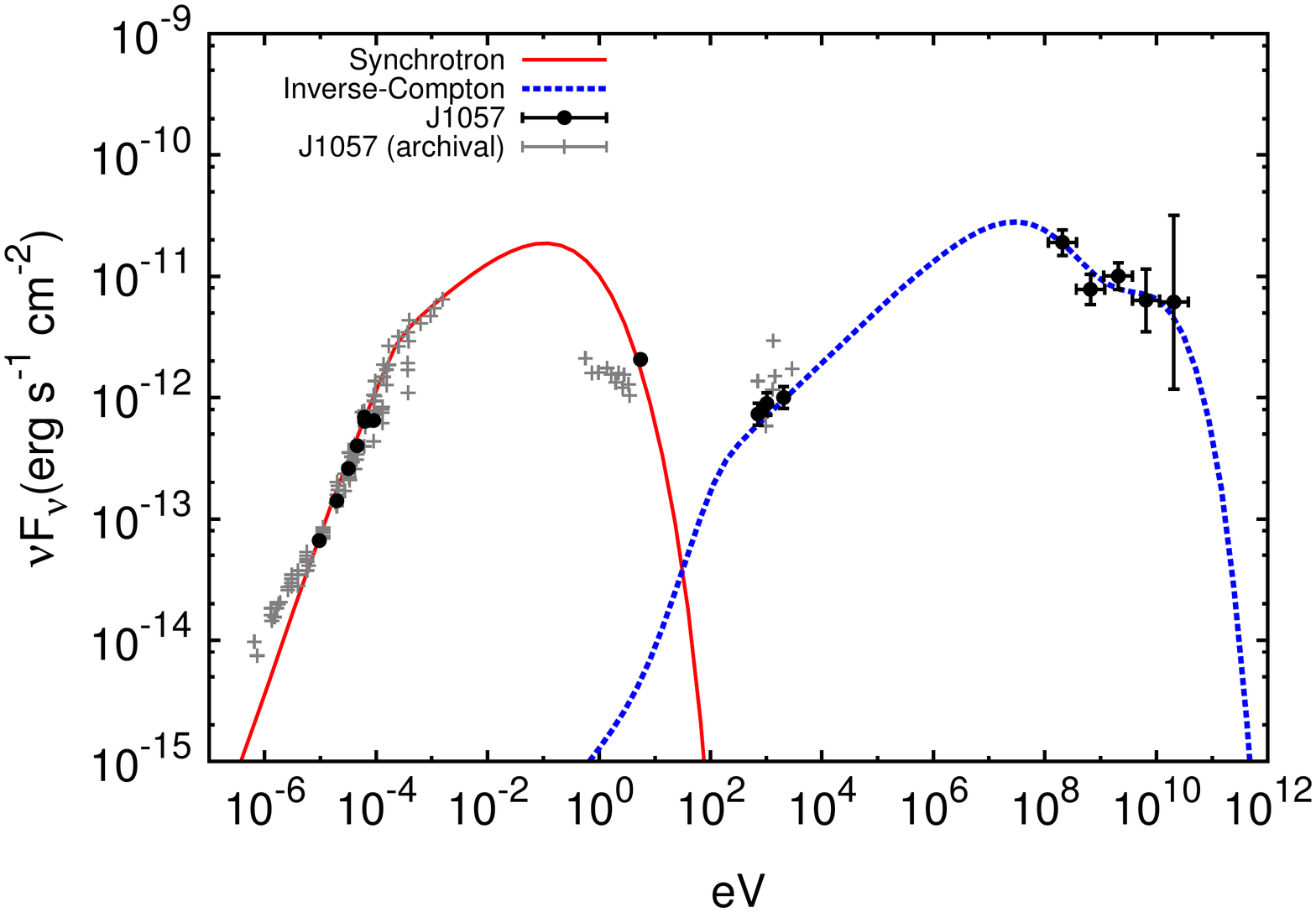} }	
		\subfloat[J01015]{ \includegraphics[height=8cm, clip=true, trim=1.5cm 1cm 1cm 3.5cm,angle=-90]{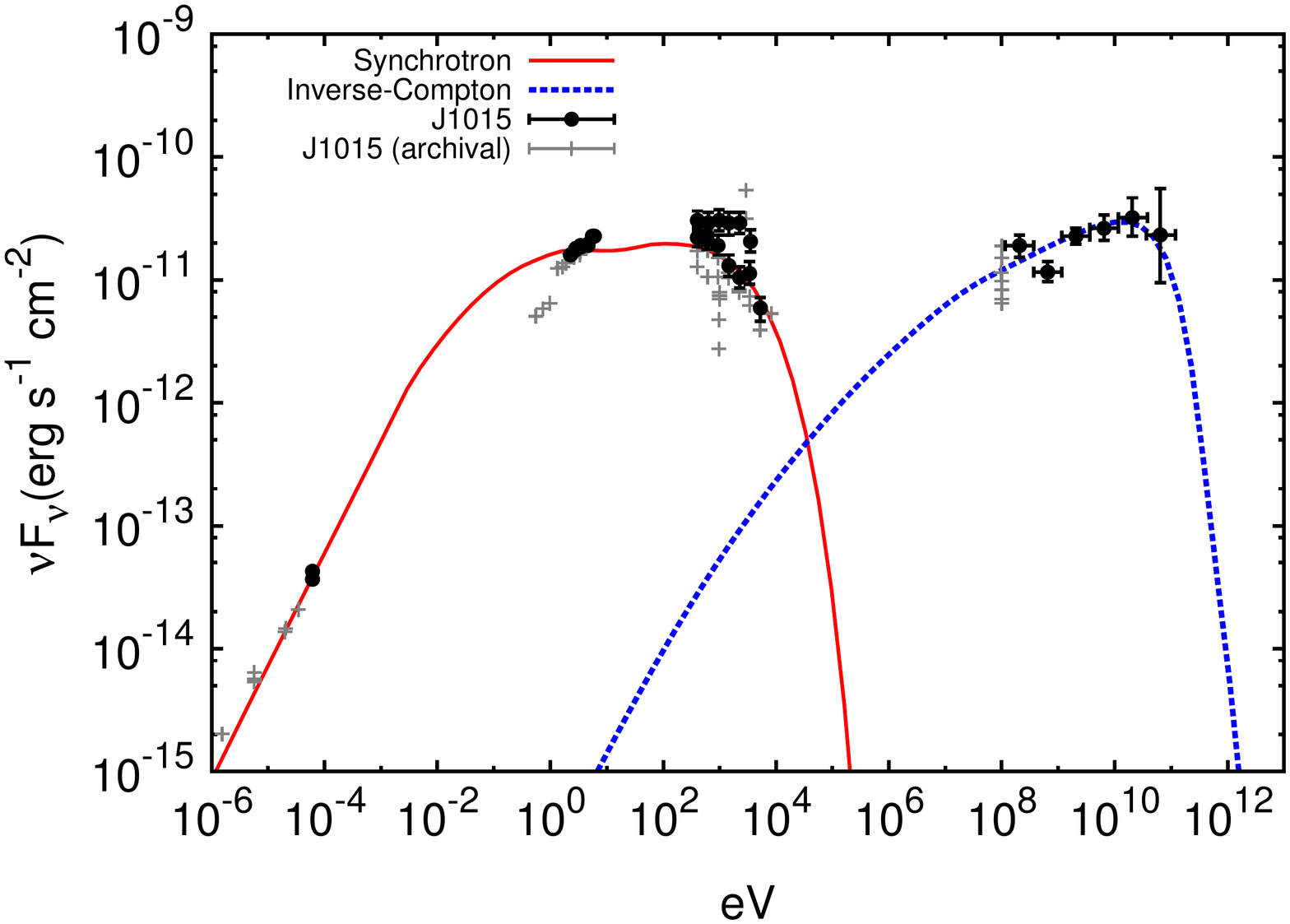} }
\\
		\subfloat[Mkn421]{ \includegraphics[height=8cm, clip=true, trim=1.5cm 1cm 1cm 3.5cm,angle=-90]{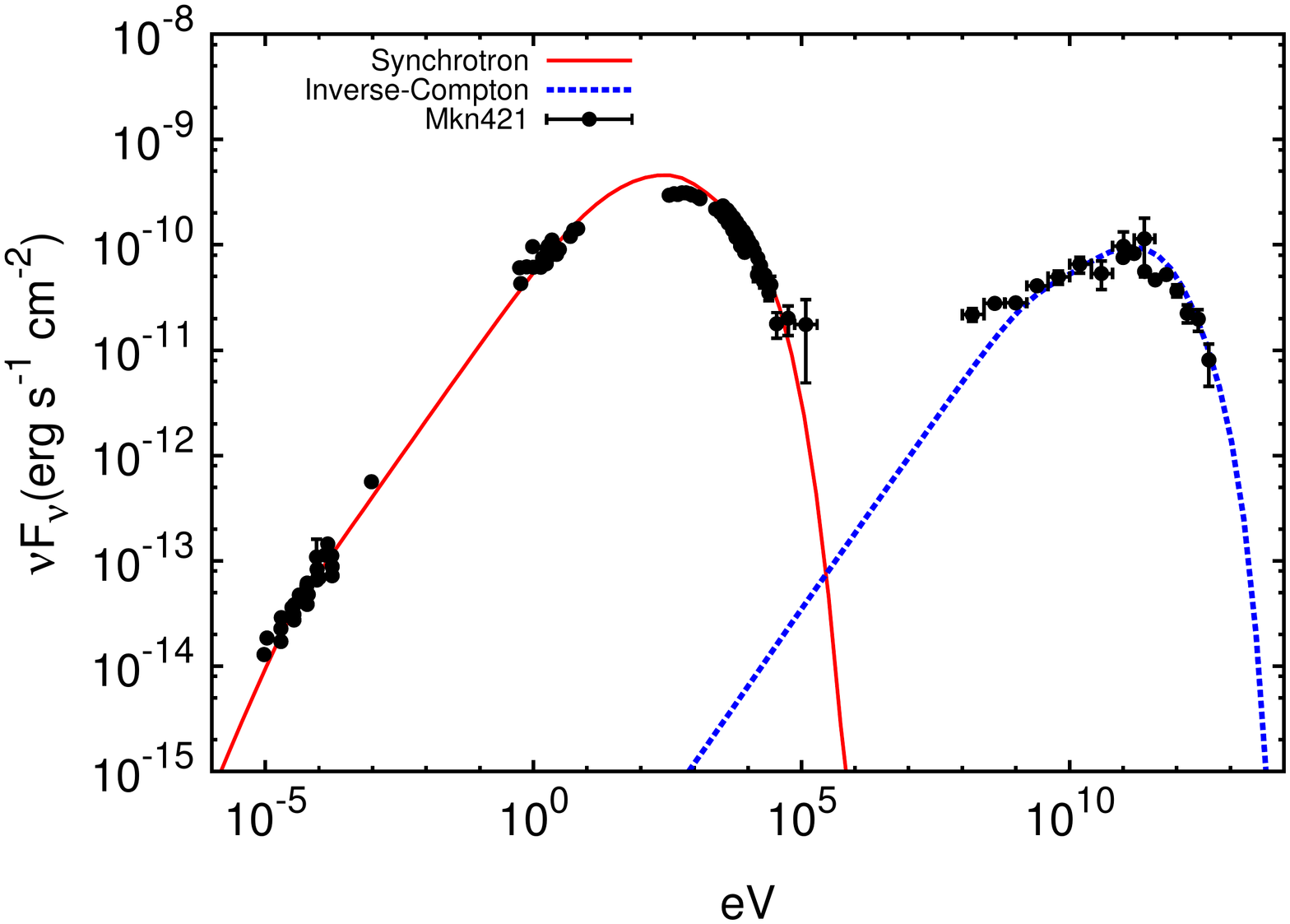} }
		\subfloat[J1159]{ \includegraphics[height=8cm, clip=true, trim=1.5cm 1cm 1cm 3.5cm,angle=-90]{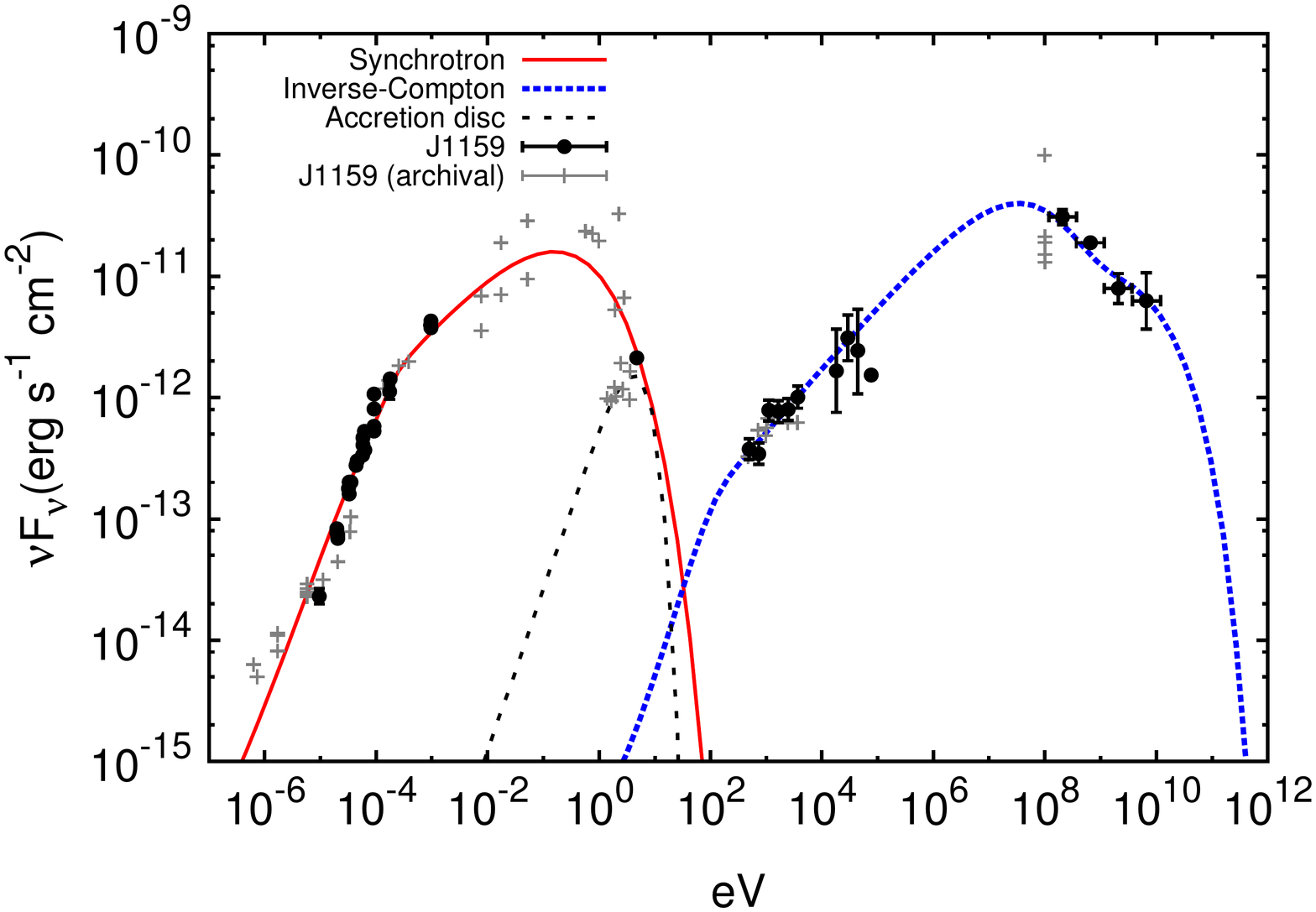} }
\\	
		\subfloat[J1221]{ \includegraphics[height=8cm, clip=true, trim=1.5cm 1cm 1cm 3.5cm,angle=-90]{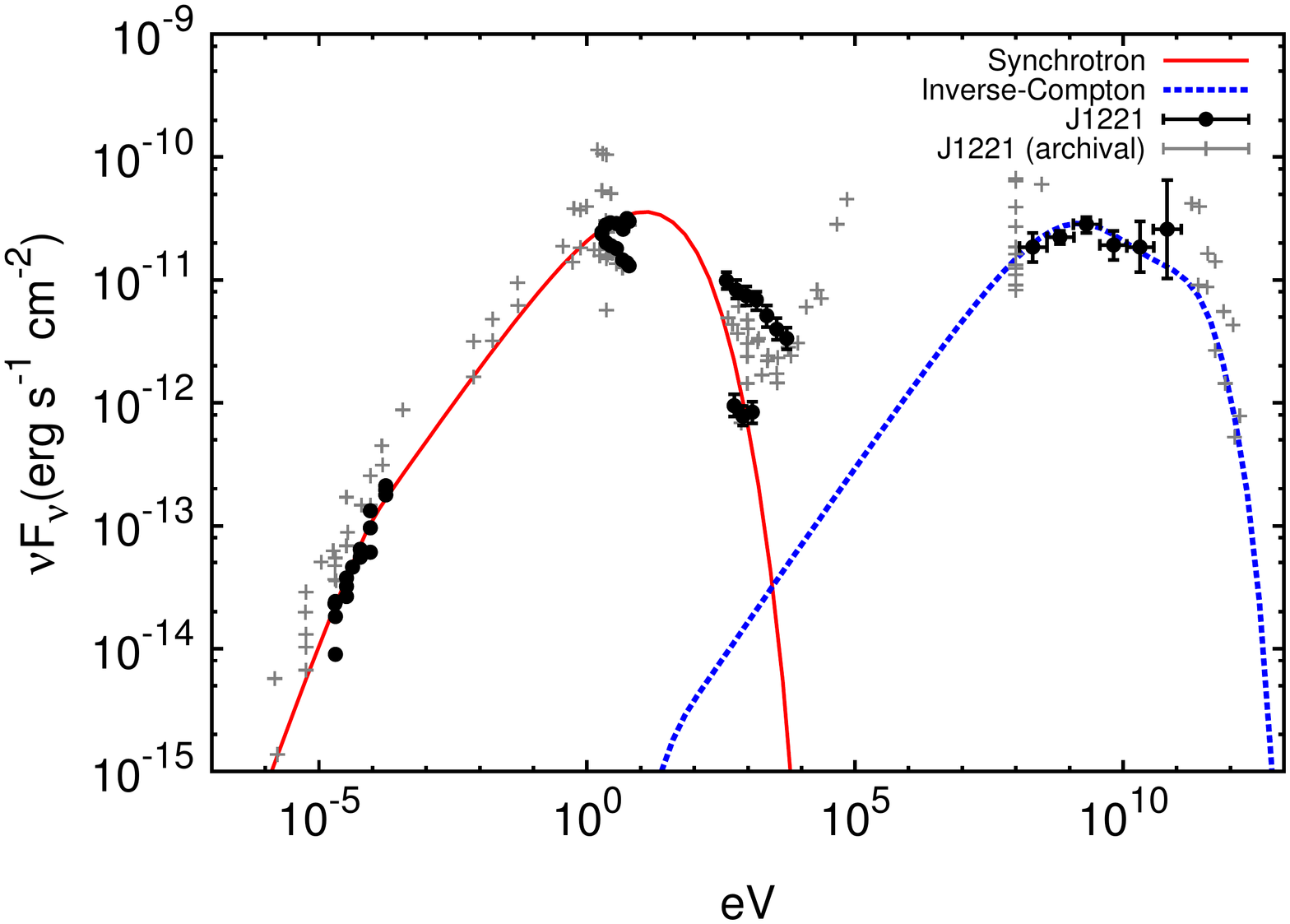} }			
		\subfloat[J1229]{ \includegraphics[height=8cm, clip=true, trim=1.5cm 1cm 1cm 3.5cm,angle=-90]{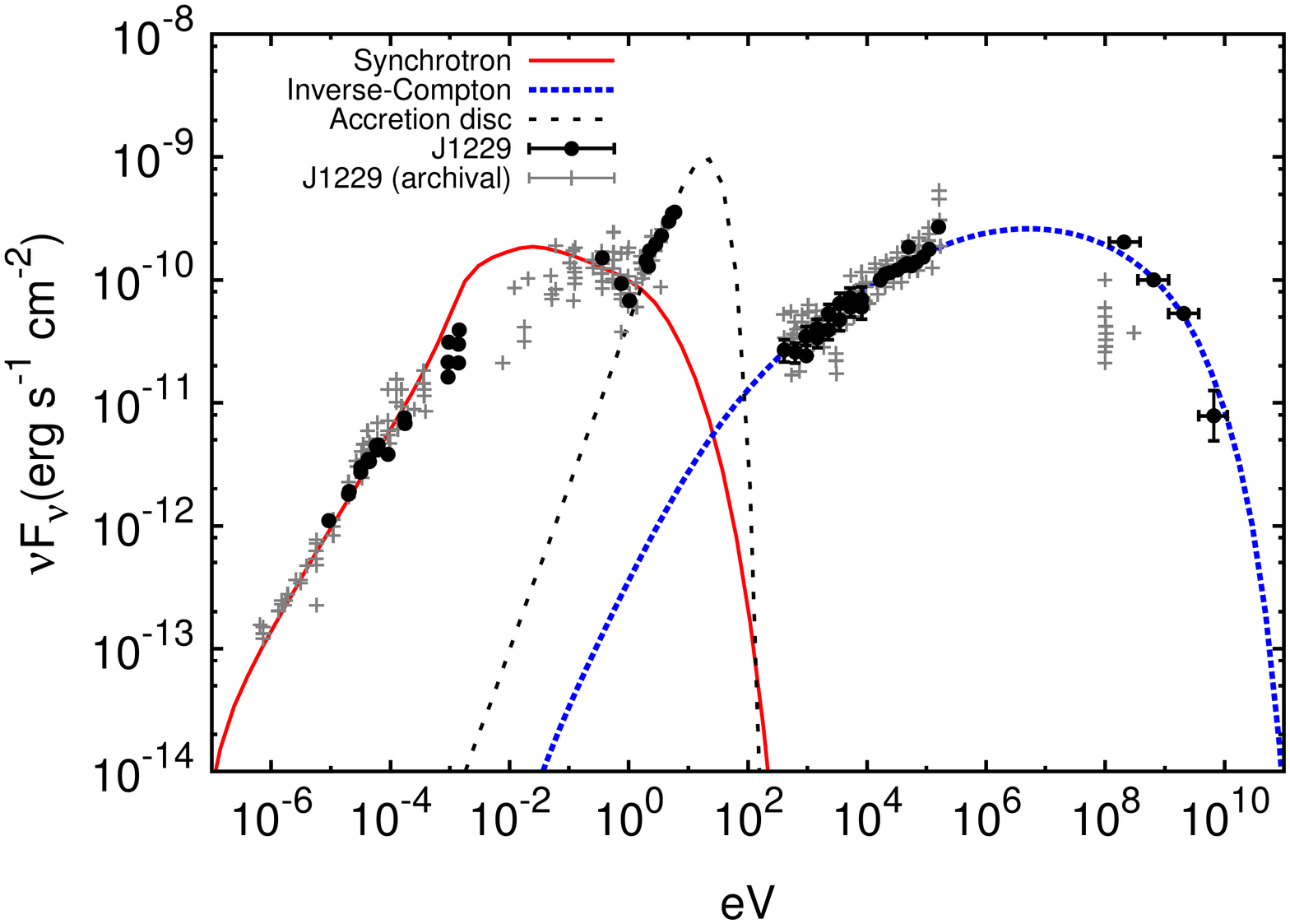} }		

\end{figure*}

\begin{figure*}
	\centering
		\subfloat[J1256]{ \includegraphics[height=8cm, clip=true, trim=1.5cm 1cm 1cm 3.5cm,angle=-90]{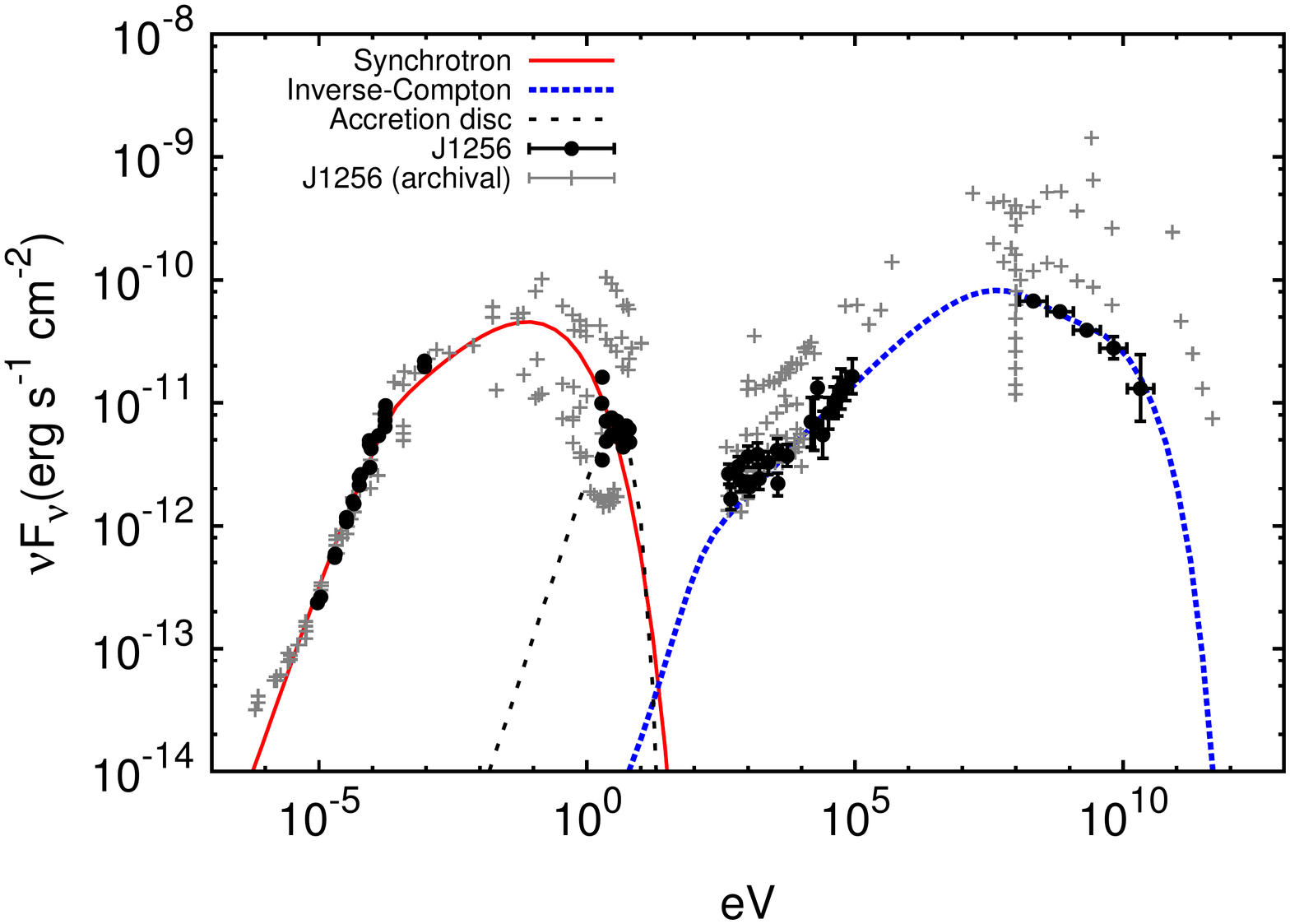} }
		\subfloat[J1310]{ \includegraphics[height=8cm, clip=true, trim=1.5cm 1cm 1cm 3.5cm,angle=-90]{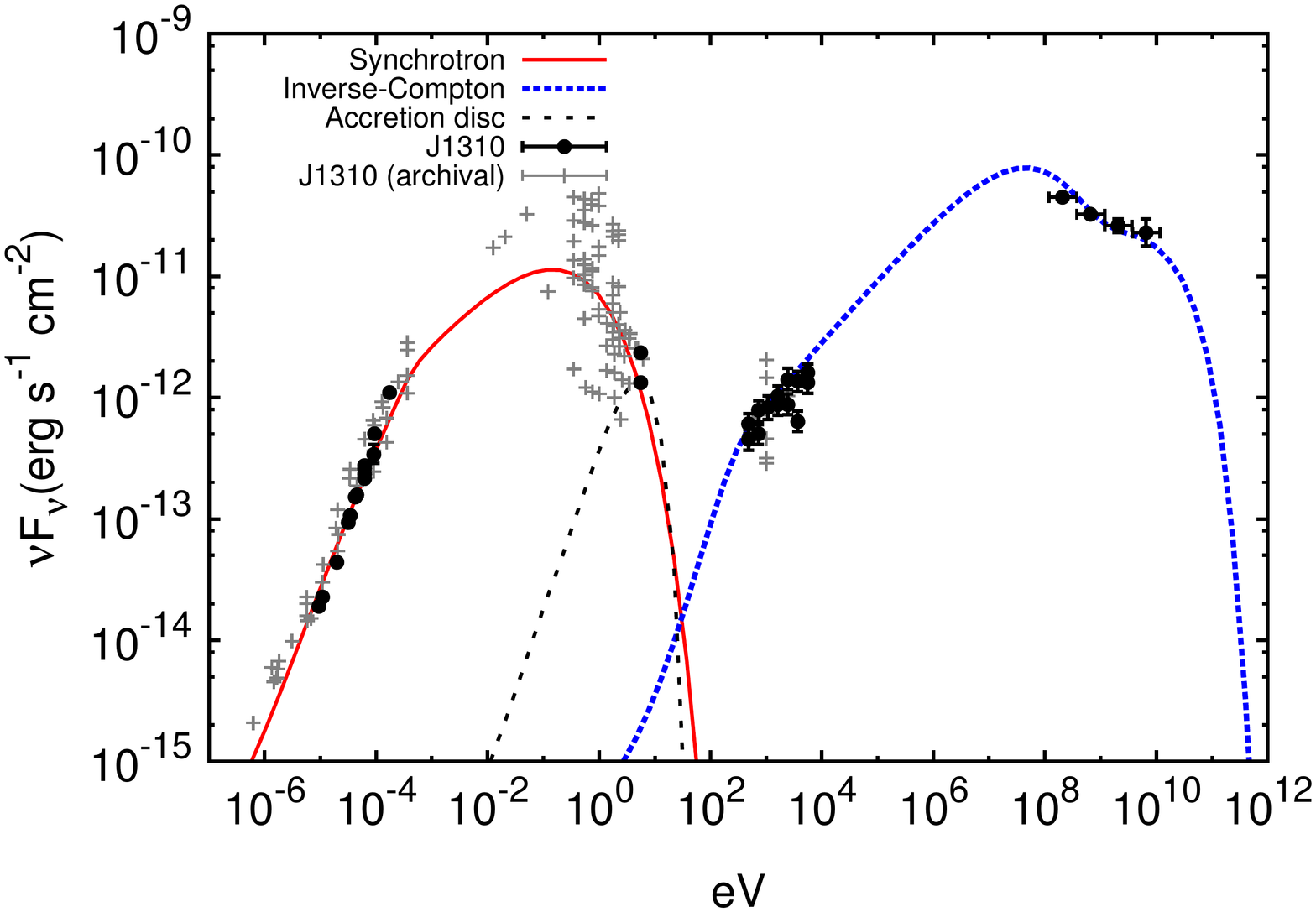} }	
\\
		\subfloat[J1312] {\includegraphics[height=8cm, clip=true, trim=1.5cm 1cm 1cm 3.5cm,angle=-90]{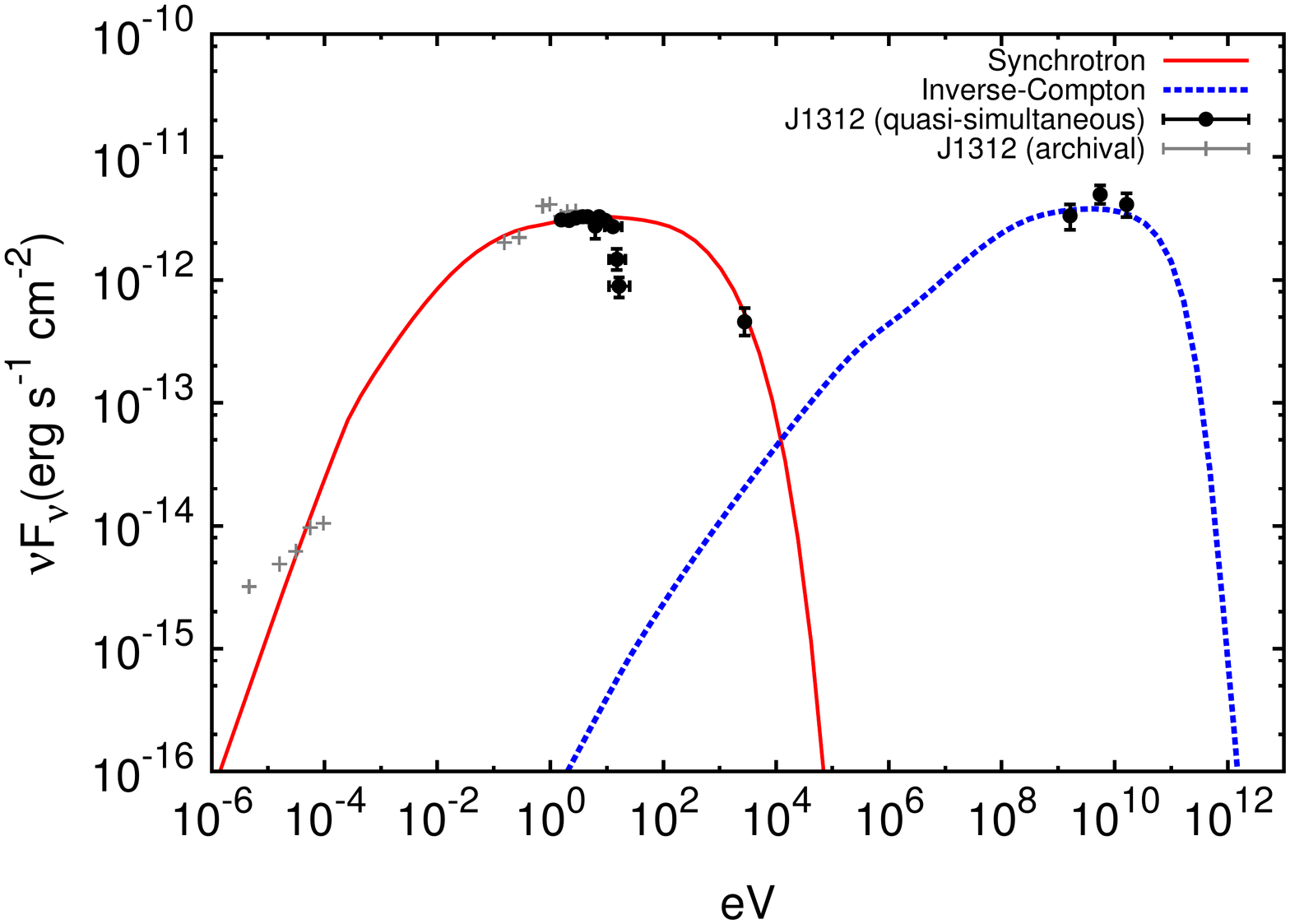} }
		\subfloat[J1457]{ \includegraphics[height=8cm, clip=true, trim=1.5cm 1cm 1cm 3.5cm,angle=-90]{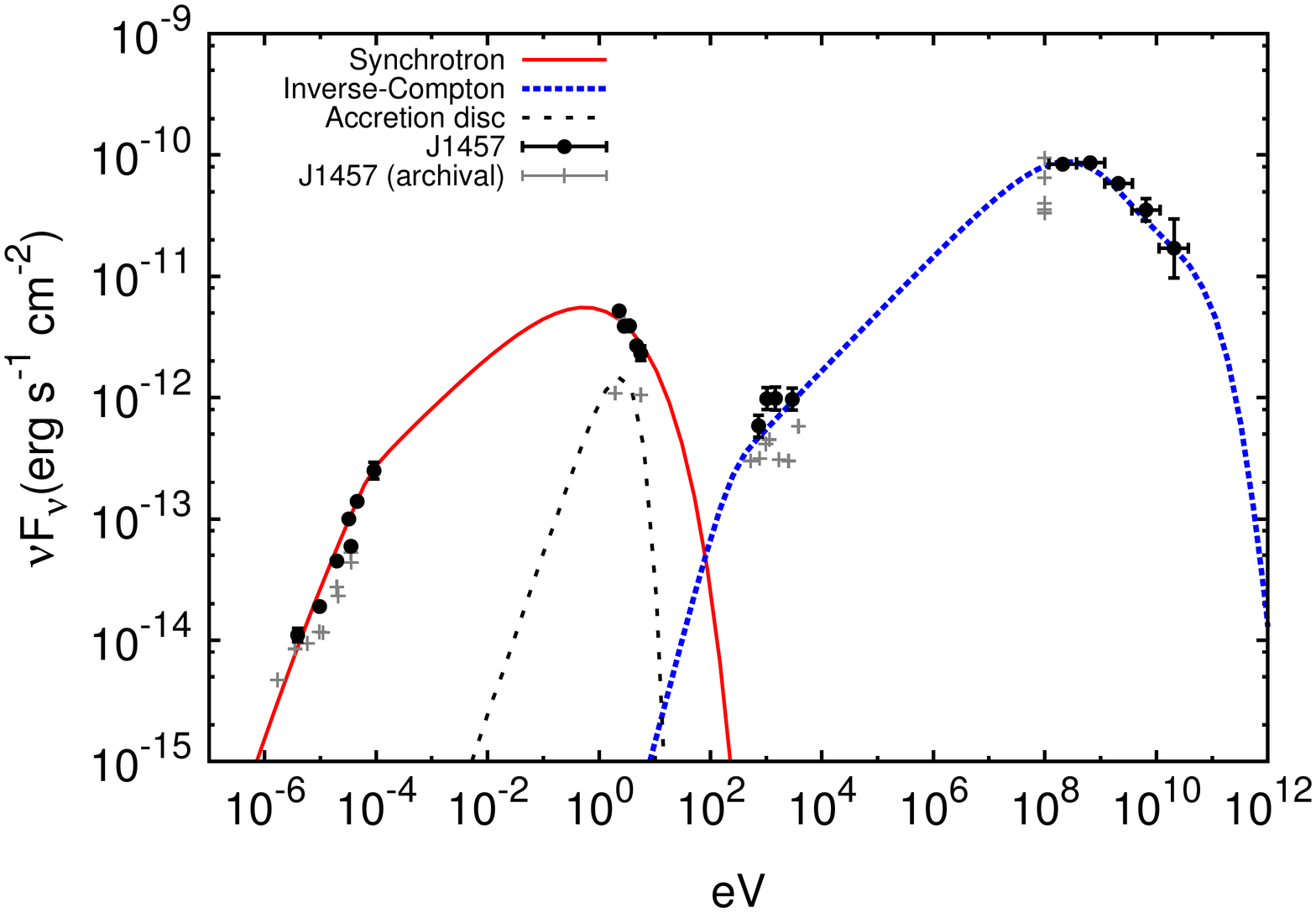} }		
\\
		\subfloat[J1504]{ \includegraphics[height=8cm, clip=true, trim=1.5cm 1cm 1cm 3.5cm,angle=-90]{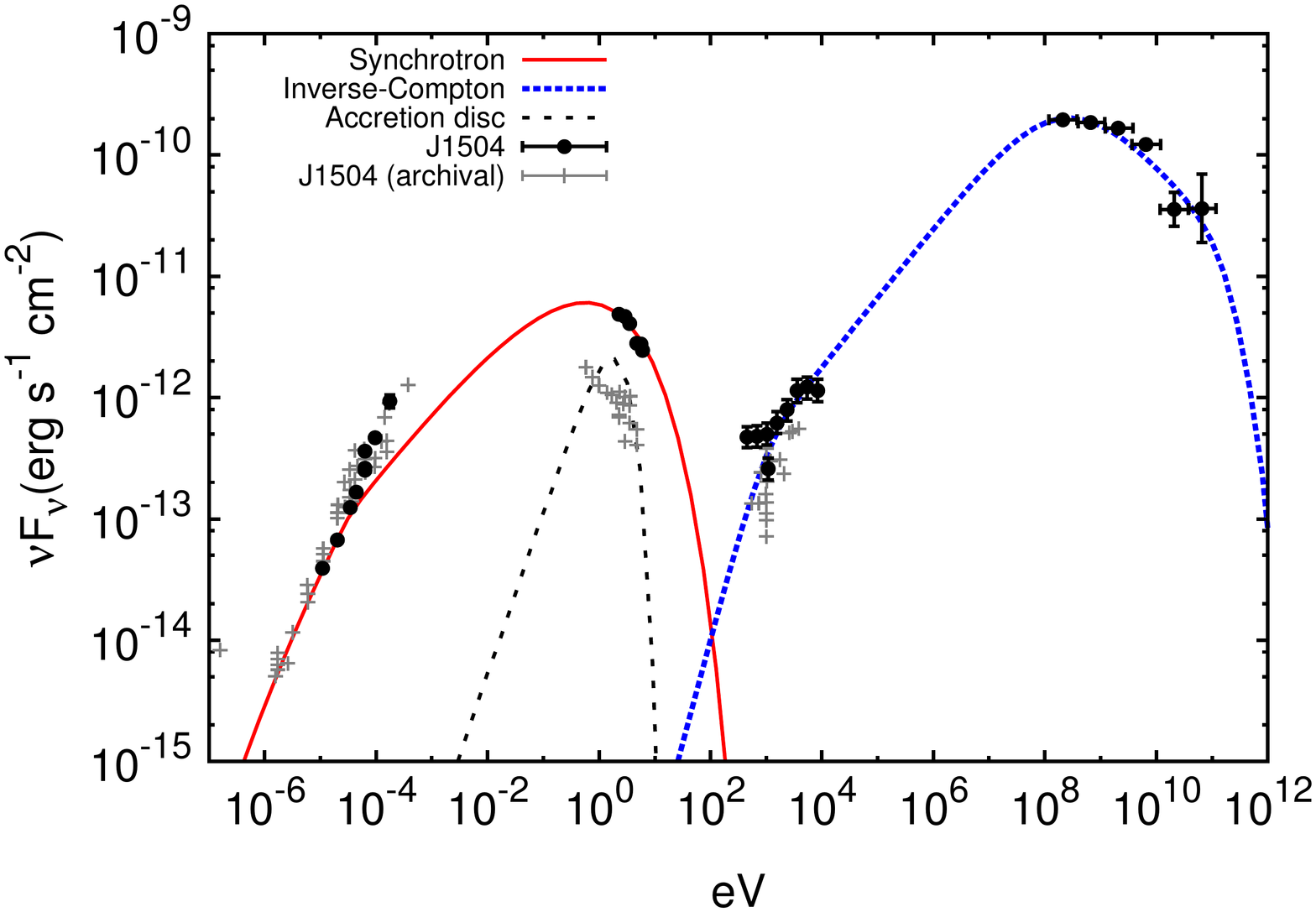} }	
		\subfloat[J1512]{ \includegraphics[height=8cm, clip=true, trim=1.5cm 1cm 1cm 3.5cm,angle=-90]{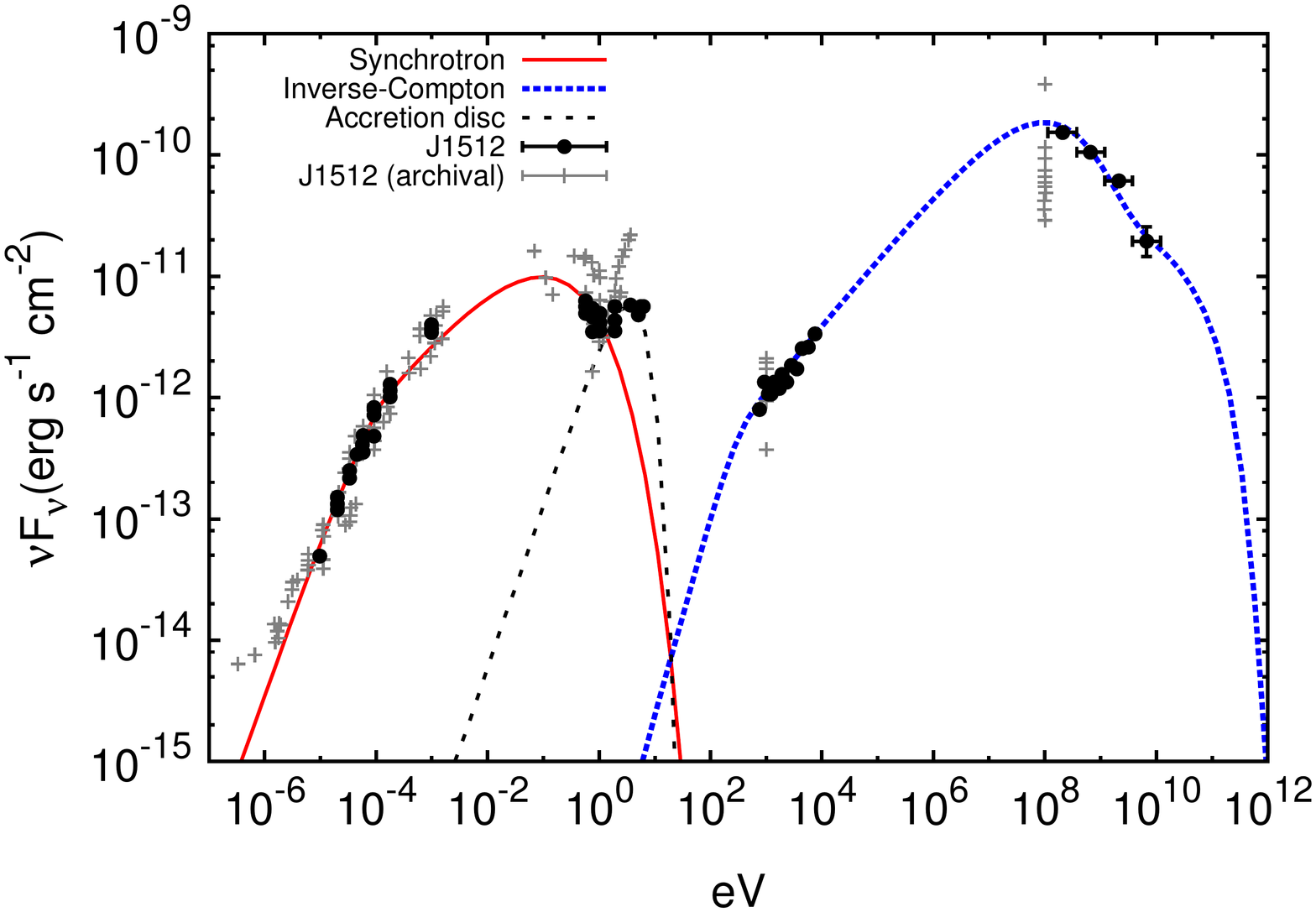} }				

\end{figure*}

\begin{figure*}
	\centering
		\subfloat[J1522]{ \includegraphics[height=8cm, clip=true, trim=1.5cm 1cm 1cm 3.5cm,angle=-90]{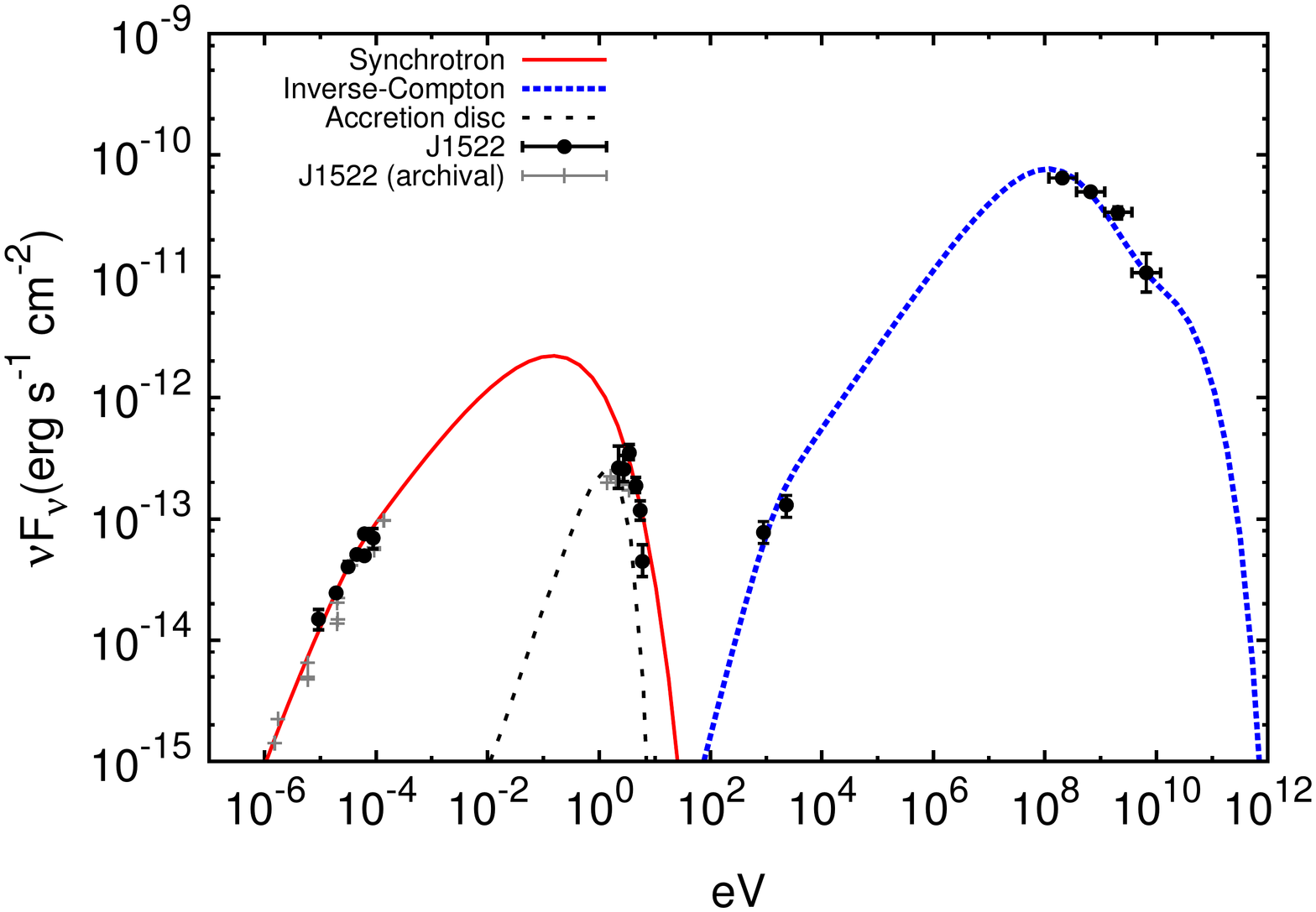} }	
		\subfloat[Mkn501]{ \includegraphics[height=8cm, clip=true, trim=1.5cm 1cm 1cm 3.5cm,angle=-90]{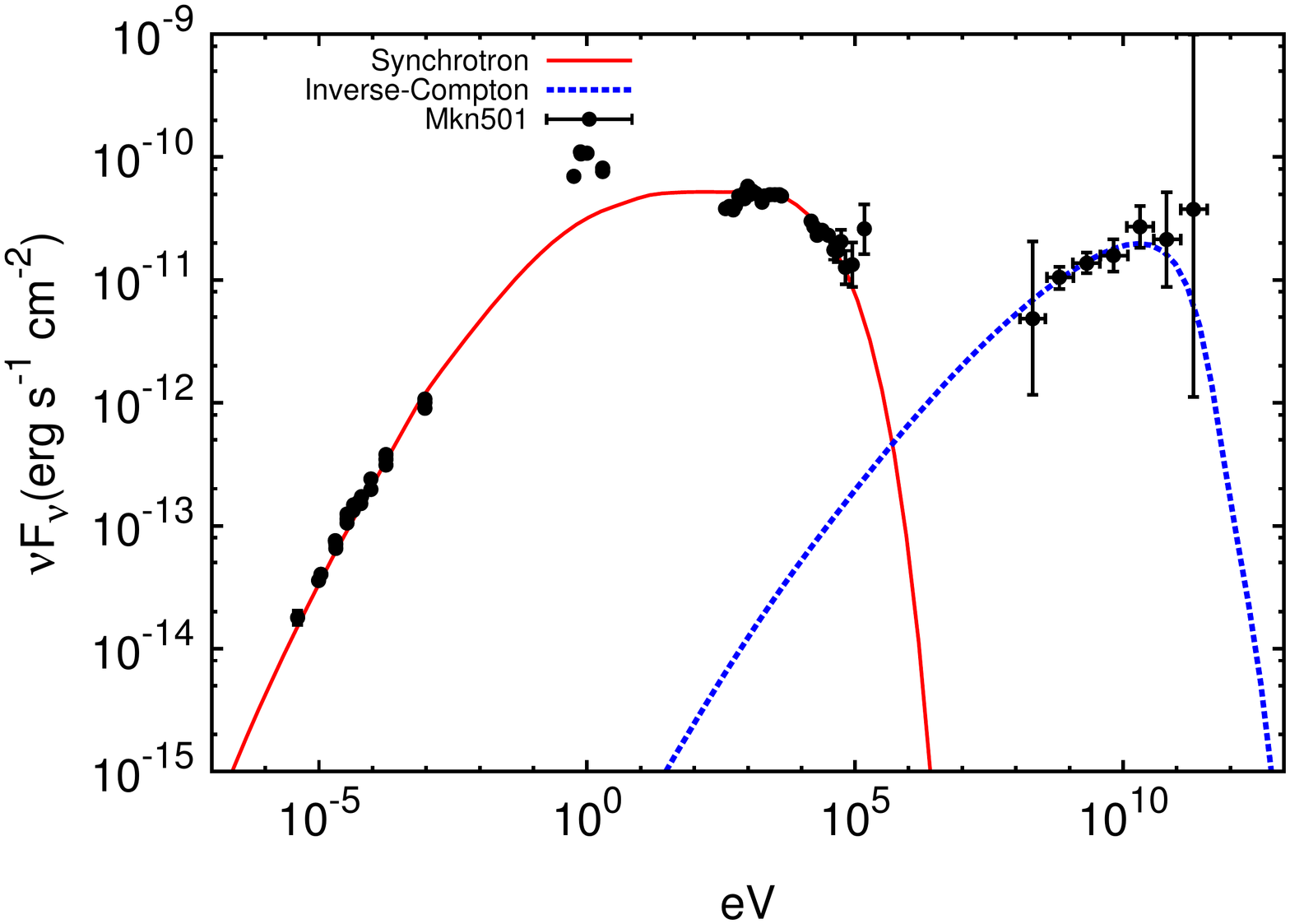} }	
\\
		\subfloat[J1719]{ \includegraphics[height=8cm, clip=true, trim=1.5cm 1cm 1cm 3.5cm,angle=-90]{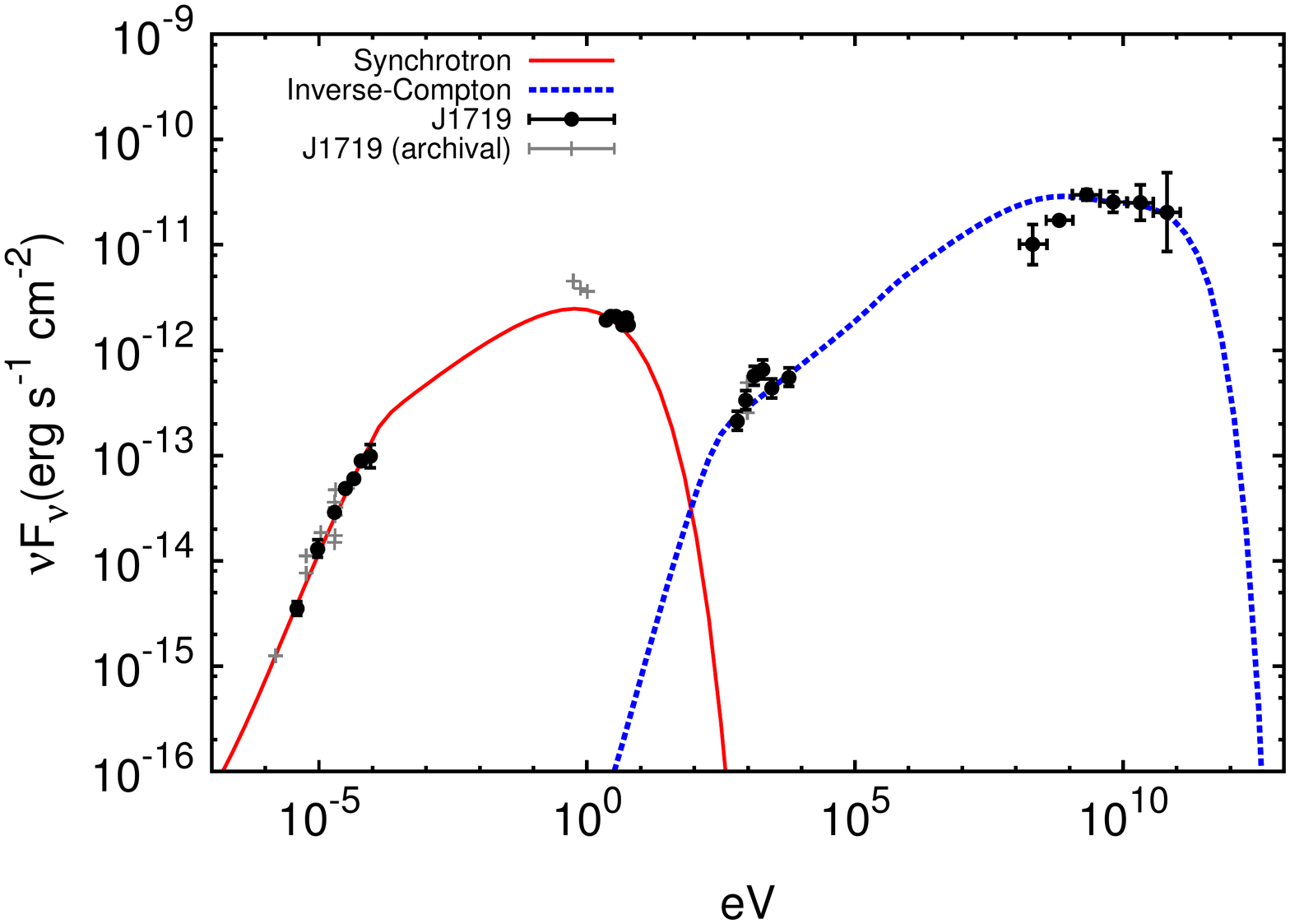} }
		\subfloat[J1751]{ \includegraphics[height=8cm, clip=true, trim=1.5cm 1cm 1cm 3.5cm,angle=-90]{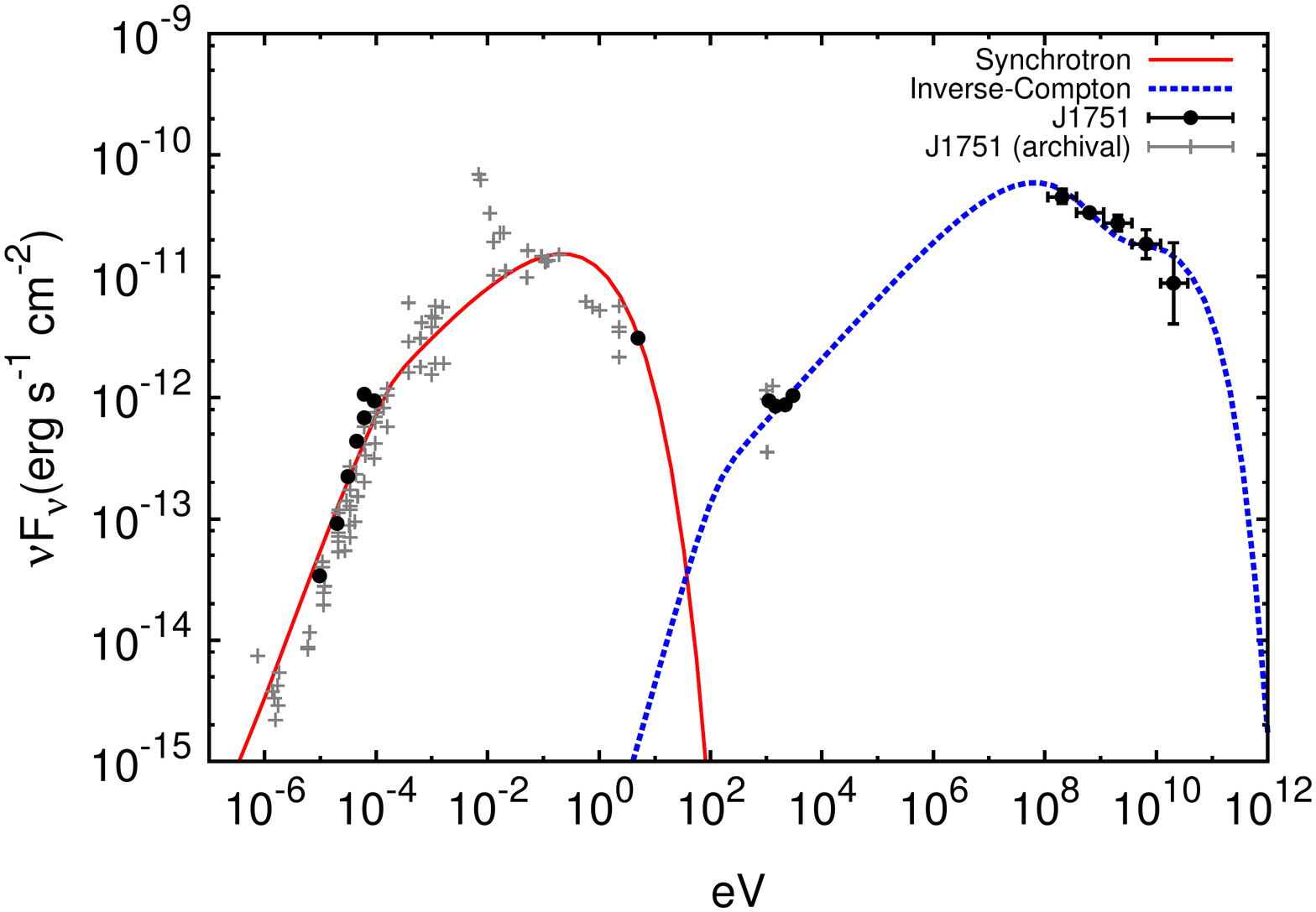} }
\\
		\subfloat[J1849]{ \includegraphics[height=8cm, clip=true, trim=1.5cm 1cm 1cm 3.5cm,angle=-90]{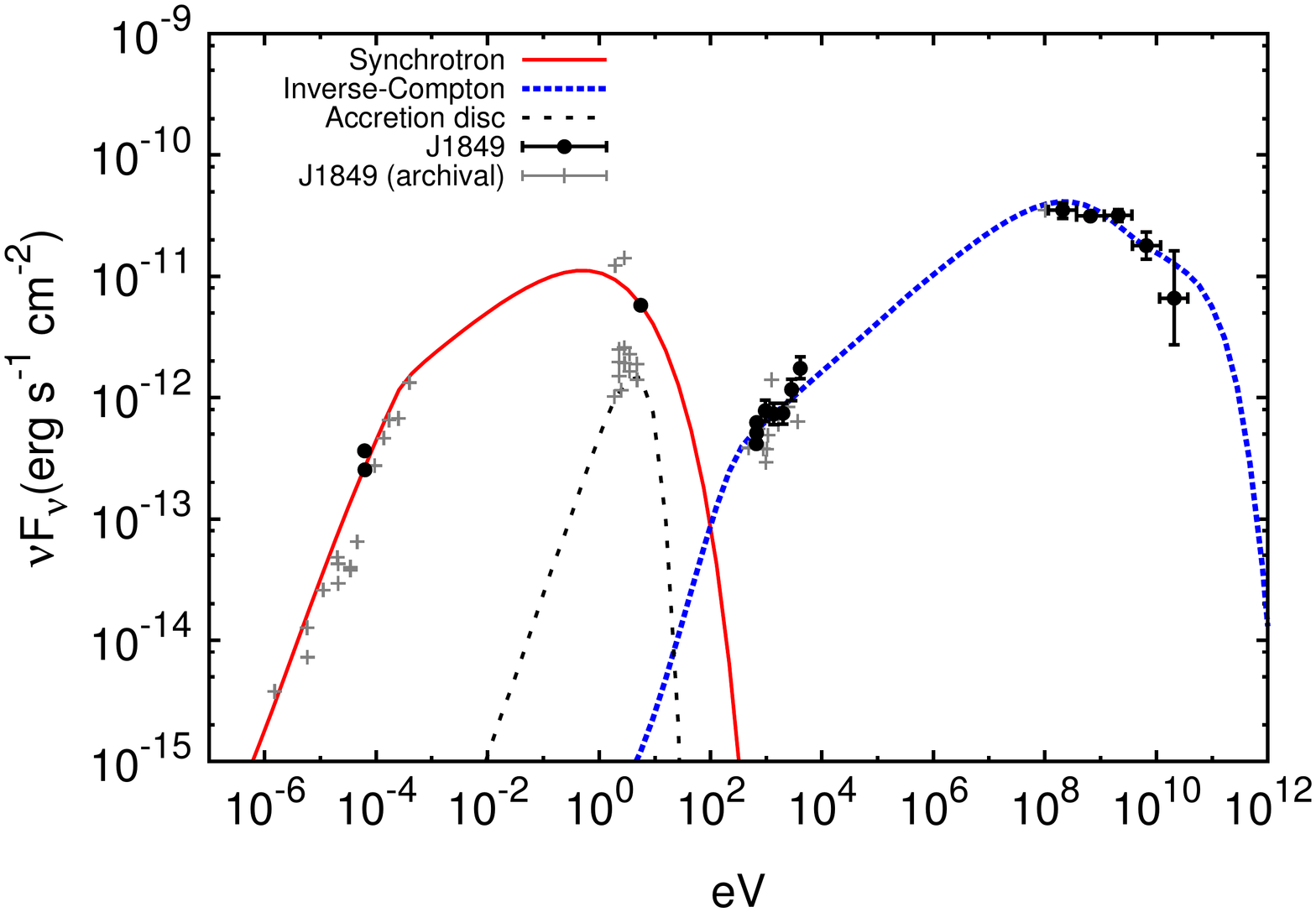} }				
		\subfloat[J2000]{ \includegraphics[height=8cm, clip=true, trim=1.5cm 1cm 1cm 3.5cm,angle=-90]{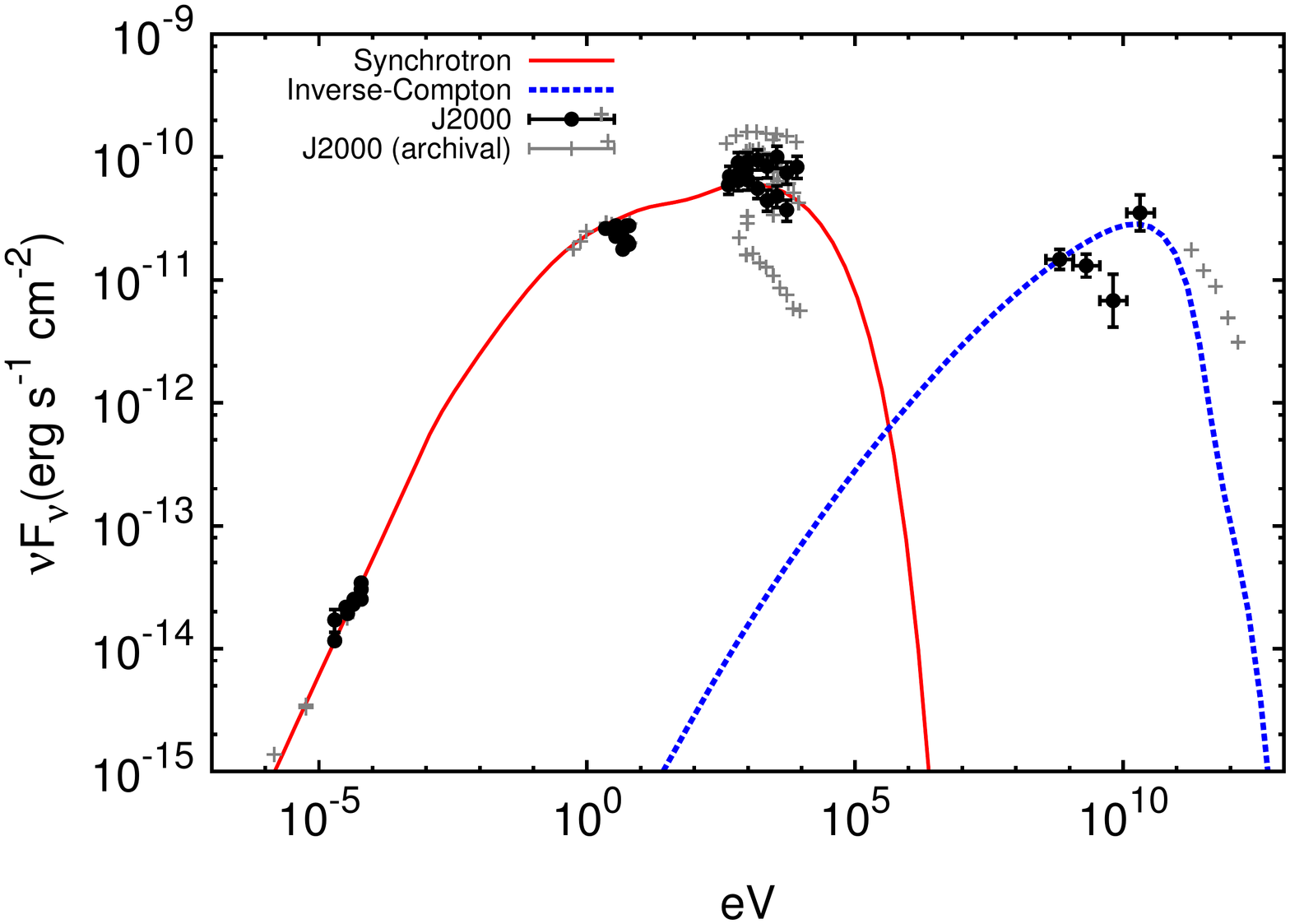} }

\end{figure*}

\begin{figure*}
	\centering
		\subfloat[J2143]{ \includegraphics[height=8cm, clip=true, trim=1.5cm 1cm 1cm 3.5cm,angle=-90]{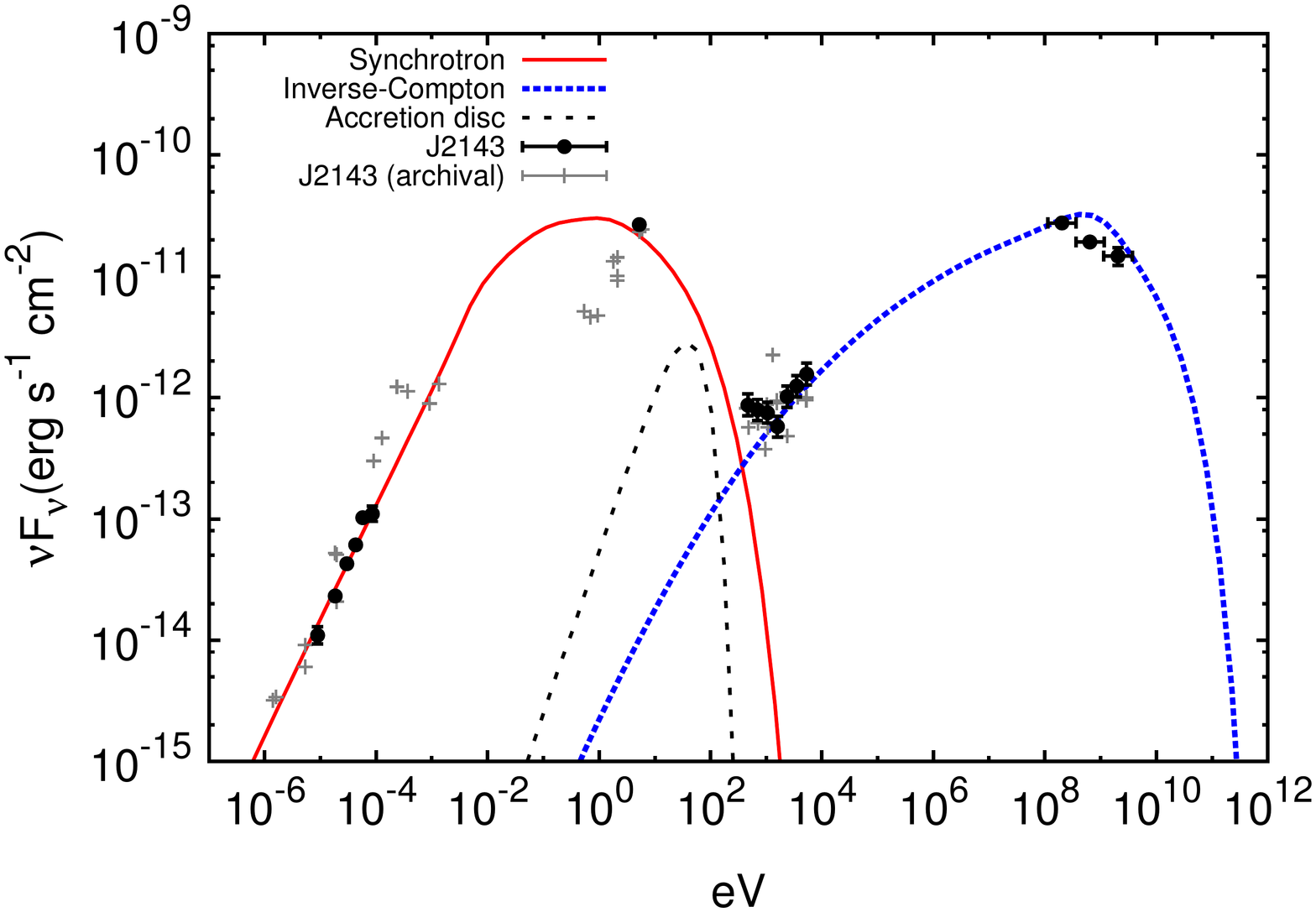} }	
		\subfloat[J2158]{ \includegraphics[height=8cm, clip=true, trim=1.5cm 1cm 1cm 3.5cm,angle=-90]{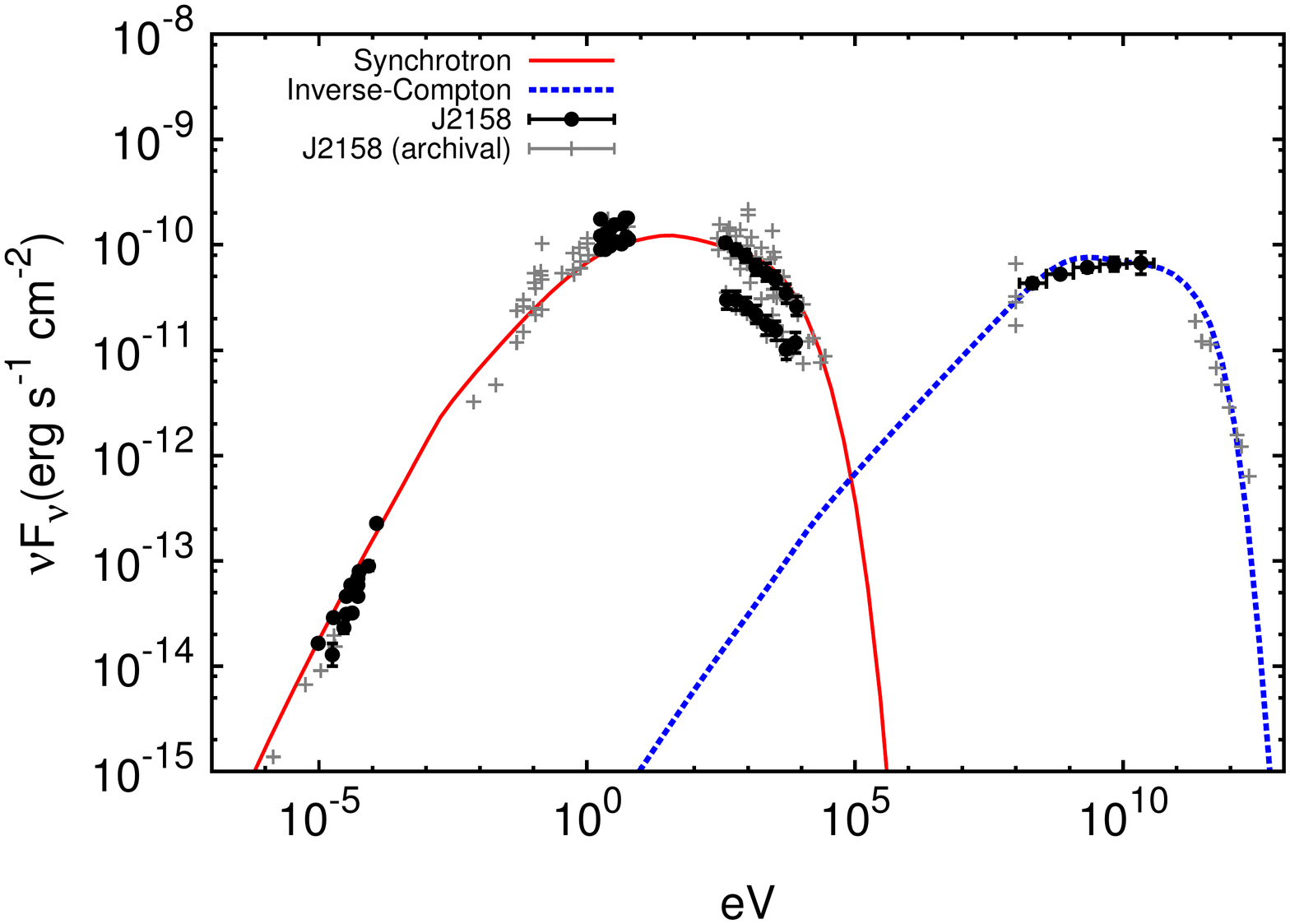} }	
\\
		\subfloat[BLLac]{ \includegraphics[height=8cm, clip=true, trim=1.5cm 1cm 1cm 3.5cm,angle=-90]{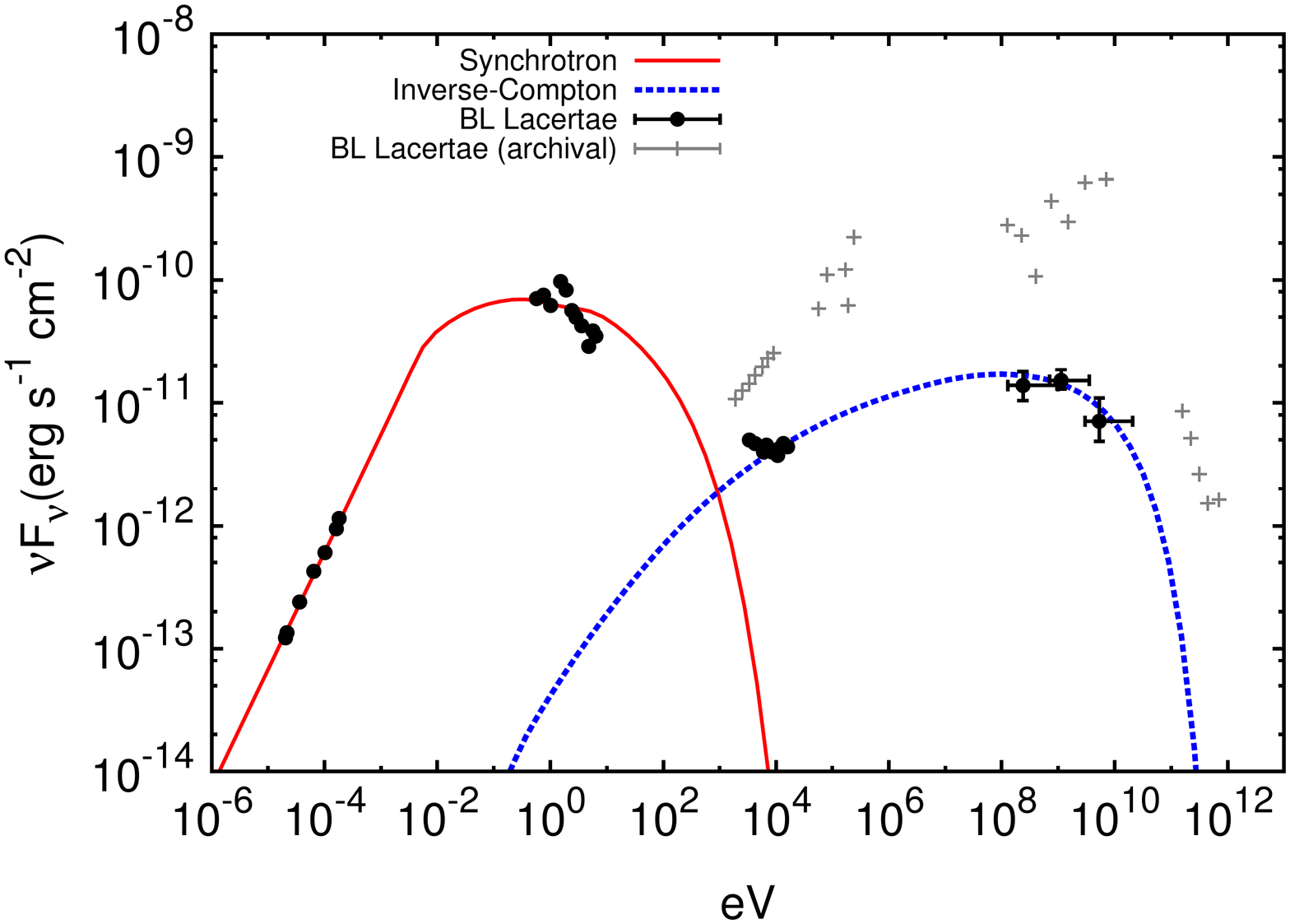} }	
		\subfloat[J2254]{ \includegraphics[height=8cm, clip=true, trim=1.5cm 1cm 1cm 3.5cm,angle=-90]{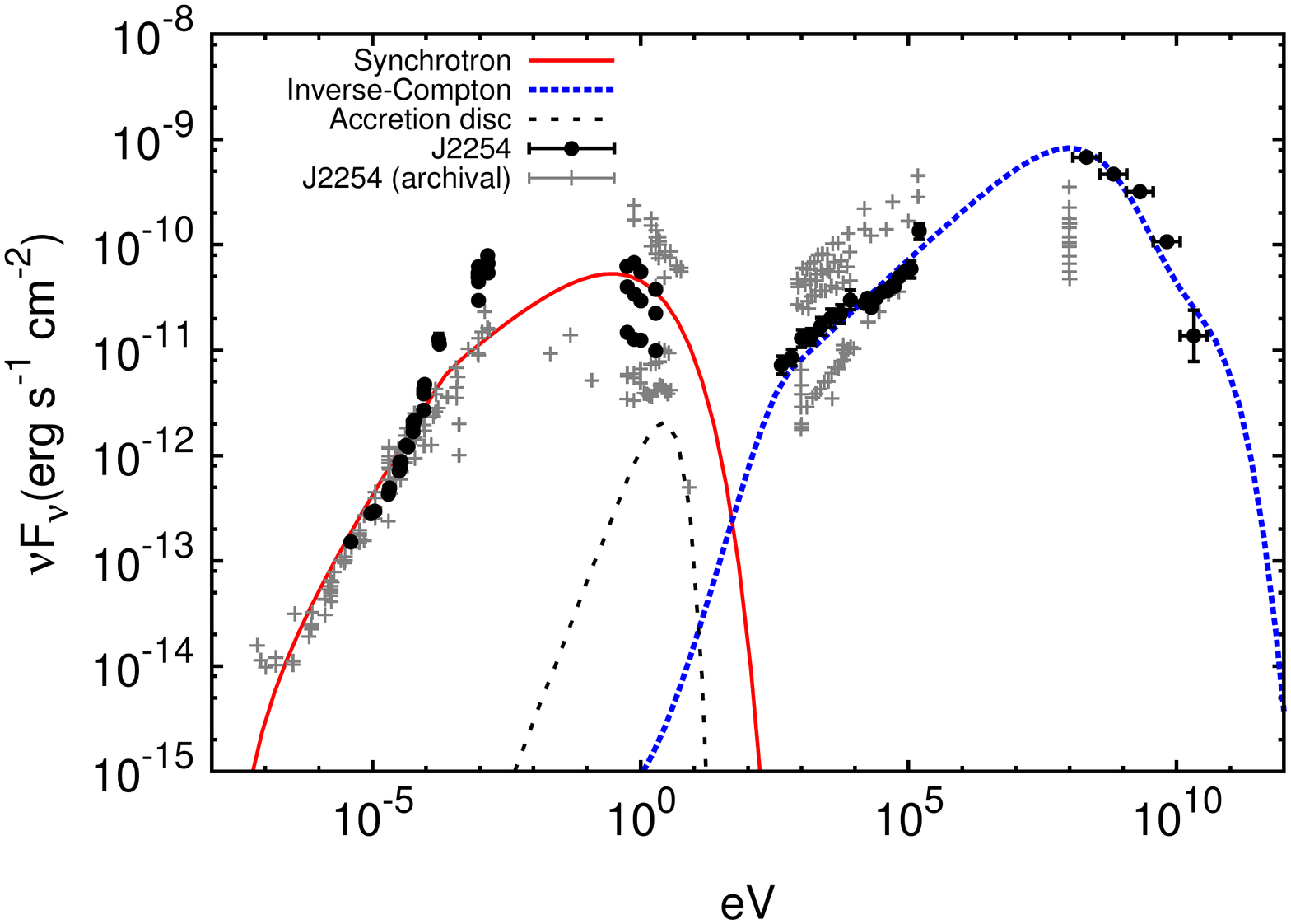} }	
\\
		\subfloat[J2327]{ \includegraphics[height=8cm, clip=true, trim=1.5cm 1cm 1cm 3.5cm,angle=-90]{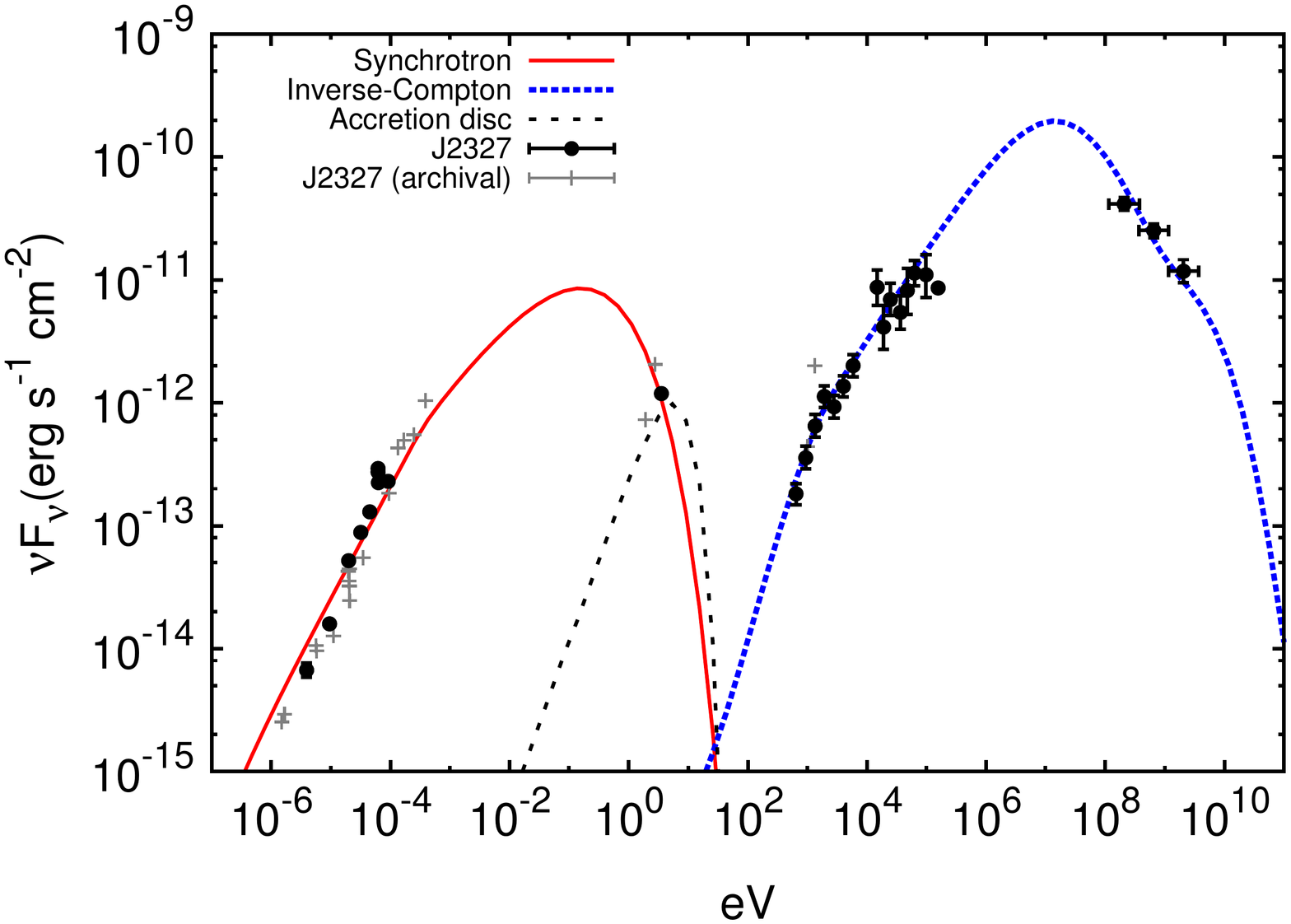} }	
		\subfloat[J2345]{ \includegraphics[height=8cm, clip=true, trim=1.5cm 1cm 1cm 3.5cm,angle=-90]{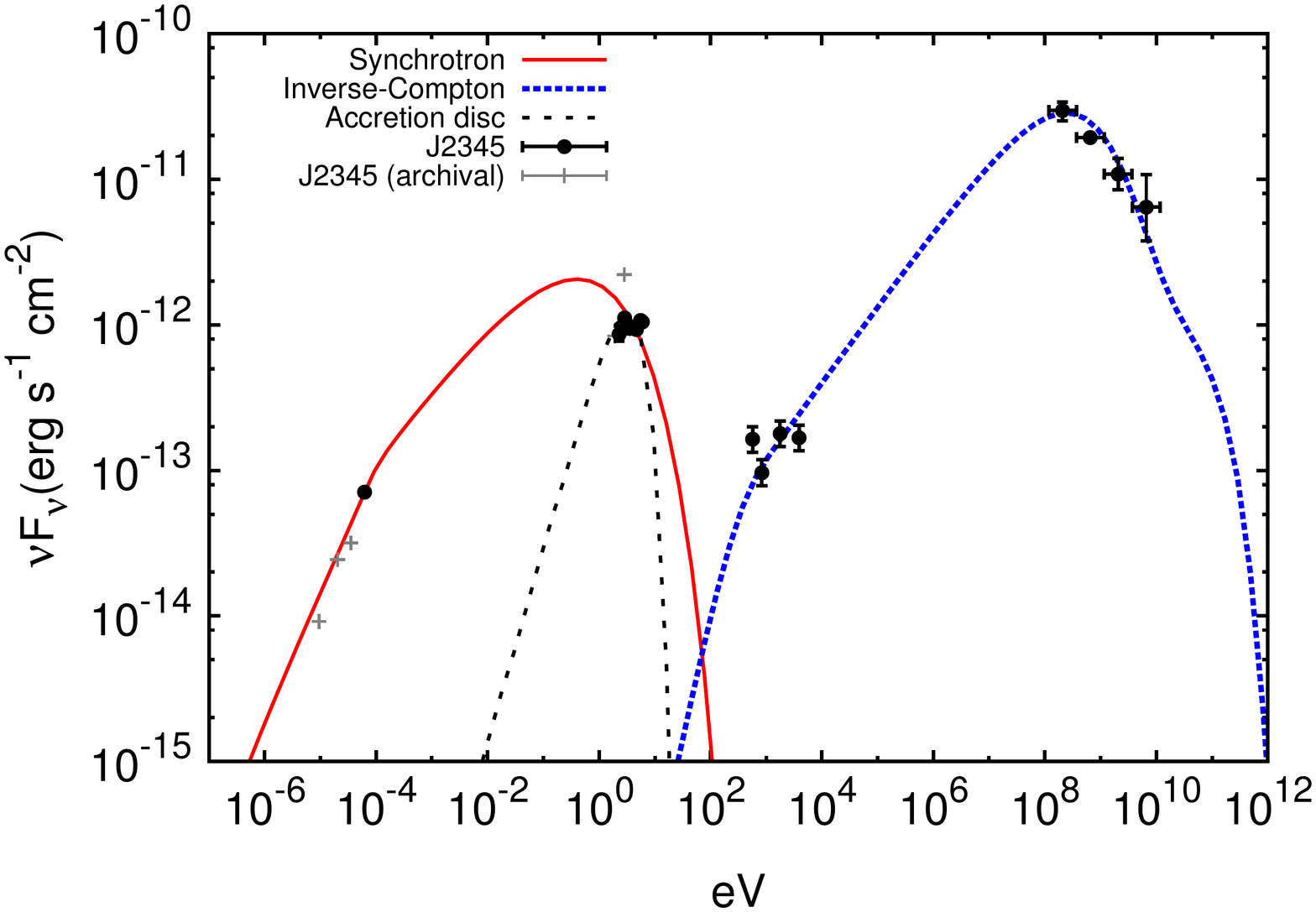} }	
\end{figure*}

\begin{figure*}
          \centering
          \includegraphics[width=22cm, angle=90]{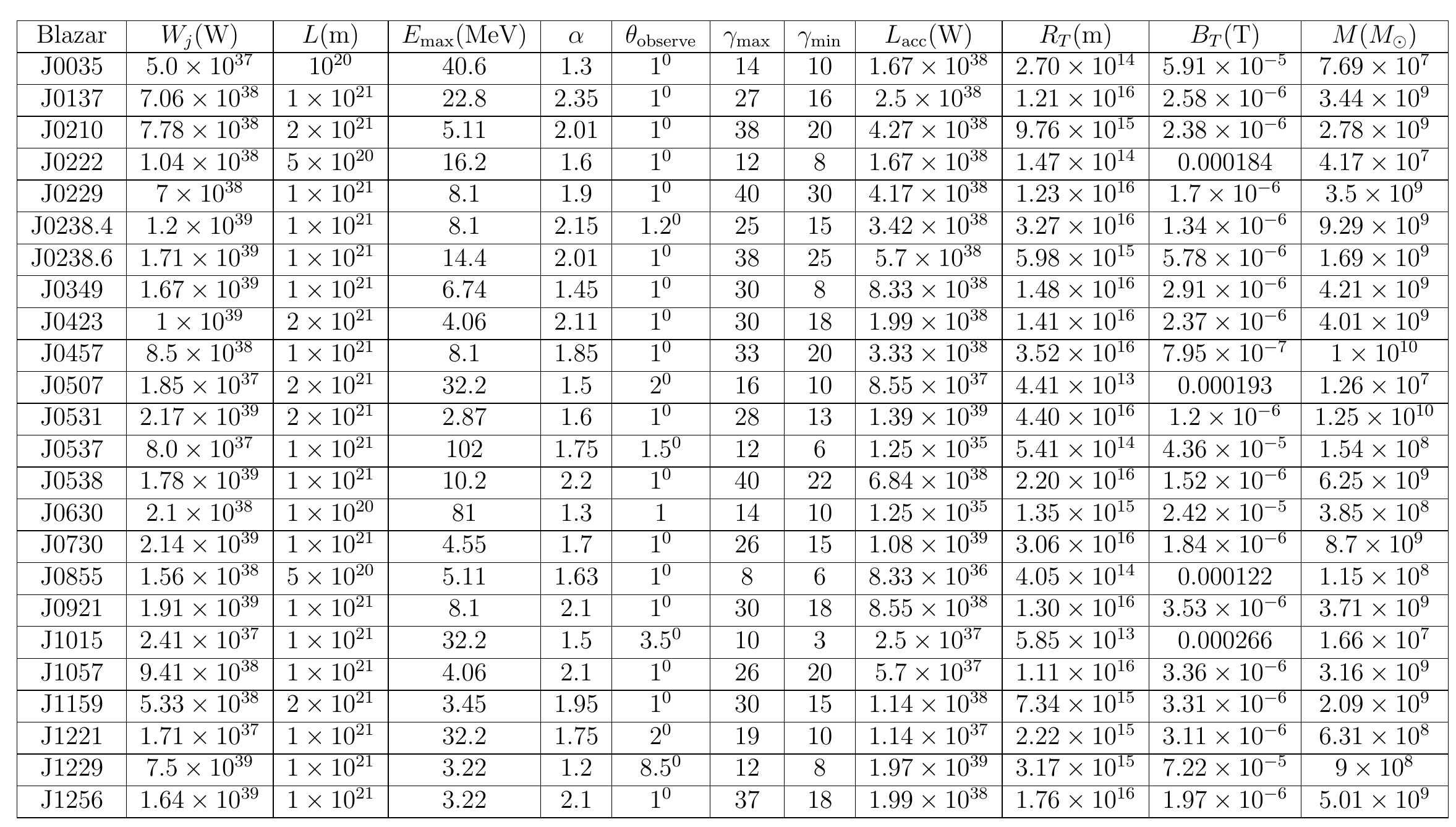}
          	\caption{Fitting parameters for all 42 blazars used in Figs. \ref{spectra} and \ref{corr}. Columns from left to right: Fermi blazar name (RA only), initial jet power, jet length, electron cutoff energy, injected electron power law index, jet misalignment angle, bulk Lorentz factor at transition region, bulk Lorentz factor at end of jet, accretion disc luminosity, radius of transition region, magnetic field strength at transition region and effective black hole mass (if the jet transition region were to occur at $10^{5}r_{s}$ as in M87) also used as the black hole mass for the thin accretion disc fit.}
\label{table}
\end{figure*}
          
\begin{figure*}
          \centering
          \includegraphics[width=22cm, angle=90]{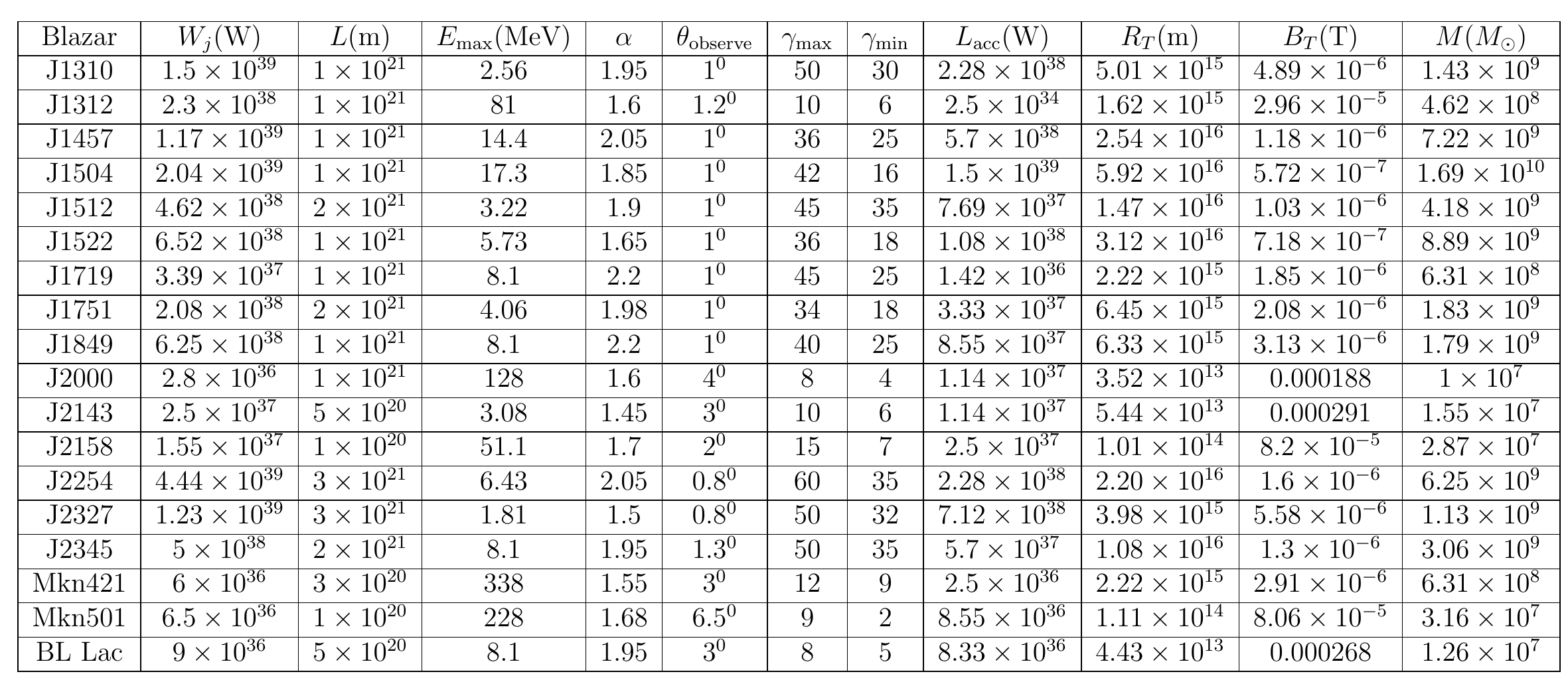}
\end{figure*}

\label{lastpage}

\end{document}